\documentclass{jfm}
\usepackage{graphicx}
\usepackage{epstopdf, epsfig}

\usepackage{xcolor}
\usepackage{amsmath}
\usepackage{wasysym}%
\usepackage{amssymb}
\usepackage{subcaption}
\usepackage[percent]{overpic}

\newcommand{\pop}[2]{\frac{\partial #1}{\partial #2}}
\newcommand{\cpop}[3]{\frac{\partial^2 #1}{\partial #2 \partial #3}}

\definecolor{darkgreen}{RGB}{0, 100, 0}

\def\genbox#1#2#3#4#5#6{
    \leavevmode\raise#4bp\hbox to#5bp{\vrule height#5bp depth0bp width0bp
    \pdfliteral{q .5 w \csname #2COLOR\endcsname\space RG
                       \csname #3PDF\endcsname{#5}{#6} S Q
             \ifx1#1 q \csname #2COLOR\endcsname\space rg 
                       \csname #3PDF\endcsname{#5}{#6} f Q\fi}\hss}}

\def\sqbox      #1#2{\genbox{#1}{#2}  {sq}       {0}   {4.5}  {2.25}}
\def\trianbox   #1#2{\genbox{#1}{#2}  {trian}    {0}   {5}    {2.5}}
\def\uptrianbox #1#2{\genbox{#1}{#2}  {uptrian}  {0}   {5}    {2.5}}
\def\circbox    #1#2{\genbox{#1}{#2}  {circ}     {0}   {5}    {2.5}}
\def\diabox     #1#2{\genbox{#1}{#2}  {dia}      {-.5} {6}    {3}}

\shorttitle{Analysis of wall-pressure fluctuation sources from DNS of
  turbulent channel flow} \shortauthor{S. Anantharamu and K. Mahesh}

\title{Analysis of wall-pressure fluctuation sources from DNS of
  turbulent channel flow}

\author{Sreevatsa Anantharamu
 \and Krishnan Mahesh
\corresp{\email{kmahesh@umn.edu}}}

\affiliation{Aerospace Engineering and Mechanics, University of Minnesota - Twin Cities,
Minneapolis, MN 55455, USA}

\begin{document}

\maketitle



\begin{abstract}
  The sources of wall-pressure fluctuations in turbulent channel flow are studied using a novel framework. The wall-pressure power spectral density (PSD) $(\phi_{pp}(\omega))$ is expressed as an integrated contribution from all wall-parallel plane pairs, $\phi_{pp}(\omega)=\int_{-\delta}^{+\delta}\int_{-\delta}^{+\delta}\Gamma(r,s,\omega)\,\mathrm{dr}\,\mathrm{ds}$, using the Green's function. Here, $\Gamma(r,s,\omega)$ is termed the net source cross spectral density (CSD) between two wall-parallel planes, $y=r$ and $y=s$ and $\delta$ is the half channel height. Direct Numerical Simulation (DNS) data at friction Reynolds number of $180$ and $400$ are used to compute $\Gamma(r,s,\omega)$. Analysis of the net source CSD, $\Gamma(r,s,\omega)$ reveals that the location of dominant sources responsible for the premultiplied peak in the power spectra at $\omega^+\approx 0.35$ \citep{hu2006wall} and the wavenumber spectra at $\lambda^+\approx 200$ \citep{panton2017correlation} is in the buffer layer at $y^+\approx 16.5$ and $18.4$ for $Re_{\tau}=180$ and $400$, respectively. The contribution from a wall-parallel plane (located at distance $y^+$ from the wall) to wall-pressure PSD is log-normal in $y^+$ for $\omega^+>0.35$. A dominant inner-overlap region interaction of the sources is observed at low frequencies. Further, the decorrelated features of the wall-pressure fluctuation sources are analyzed using spectral Proper Orthogonal Decomposition (POD). Instead of the commonly used $L^2$ inner product, we require the modes to be orthogonal in an inner product with a symmetric positive definite kernel. Spectral POD supports the case that the net source is composed of two components - active and inactive. The dominant spectral POD mode that comprises the active part contributes to the entire wall-pressure PSD. The suboptimal spectral POD modes that constitute the inactive portion do not contribute to the PSD. Further, the active and inactive parts of the net source are decorrelated because they stem from different modes. The structure represented by the dominant POD mode at the premultiplied wall-pressure PSD peak inclines in the downstream direction. At the low-frequency linear PSD peak, the dominant mode resembles a large scale vertical pattern. Such patterns have been observed previously in the instantaneous contours of rapid pressure fluctuations by \cite{abe2005dns}.
\end{abstract}


\section{Introduction}


In a turbulent flow, wall-pressure fluctuations excite flexible
structures. The fluctuations' spatio-temporal features determine their
relation to the far-field sound radiation resulting from the
structural excitation. Pressure fluctuations in an incompressible flow
are governed by the Poisson equation,
\begin{equation} \label{eqn:pflucpe}
  -\cpop{p}{x_i}{x_i}=2\rho\pop{U_i}{x_j}\pop{u'_j}{x_i}+\rho\cpop{}{x_i}{x_j}\left(u'_iu'_j-\overline{u'_iu'_j}\right),
\end{equation}
with appropriate boundary conditions. Here, $p$ is the fluctuating pressure, $\rho$ is the constant fluid density, and $U_i$ and $u'_i$ are the mean and fluctuating component of the flow velocities, respectively. The linear and quadratic (in fluctuation) source terms in the above equations are called the rapid and slow terms, respectively \citep{pope2001turbulent}. The Poisson equation implies that the pressure fluctuation is a global quantity, meaning that the velocity at every point in the domain affects $p$ at every point. This makes it harder to use arguments that are based on local length and velocity scale (that work {\color{black}reasonably} well for local mean and fluctuating velocities) to analyse pressure fluctuations.

Several experiments (\cite{willmarth1962measurements},
\cite{corcos1964structure}, \cite{blake1970turbulent},
\cite{farabee1991spectral}. \cite{gravante1998characterization},
\cite{tsuji2007pressure}, \cite{klewicki2008statistical}) and
numerical simulations (\cite{kim1989structure}, \cite{choi1990space},
{\color{black}\cite{kim1993propagation}}, \cite{chang1999relationship},
{\color{black}\cite{abe2005dns}}, \cite{hu2006wall},
{\color{black}\cite{jimenez2008turbulent}} \cite{sillero2013one},
\cite{park2016space}, \cite{panton2017correlation}) have studied the
spatio-temporal features of wall-pressure fluctuation in turbulent
boundary layer and channel flows at different Reynolds
numbers. Reviews by \cite{willmarth1975pressure}, \cite{bull1996wall}
and \cite{blake2017mechanics} summarize the features of wall-pressure
fluctuations in wall-bounded flows.


\cite{farabee1991spectral} measublack wall-pressure fluctuations in a
boundary layer at friction Reynolds numbers
$Re_{\tau}=u_{\tau}\delta/\nu$ ranging from $1000-2000$, where
$u_{\tau}=\sqrt{\tau_w/\rho}$ is the friction velocity, $\delta$ is
the boundary layer thickness, $\nu$ is the kinematic viscosity of the
fluid, $\tau_w$ is the wall-shear stress and $\rho$ is the density of
the fluid. Non-dimensionalization of the power spectral density (PSD)
based on $\rho,U_o$ and $\delta^*$, where $\delta^*$ is the
displacement thickness of the boundary layer, yielded collapse of the
low frequency region ($\omega\delta/u_{\tau}<5$). The mid frequency
($5<\omega\delta/u_{\tau}<100$) region showed collapse with outer flow
variables ($u_{\tau},\delta,\tau_w$), but the high frequency region
($\omega\delta/u_{\tau}>0.3Re_{\tau}$) collapsed with inner flow
variables ($u_{\tau},\nu,\tau_w$). An overlap region
($100<\omega\delta/u_{\tau}<0.3Re_{\tau}$) showed collapse with both
outer and inner flow variables. Based on the wall-normal location
associated with the corresponding non-dimensional variable group,
\cite{farabee1991spectral} hypothesized the dominant contribution to
the low, mid and high frequency regions of the wall-pressure PSD to be
from the unsteady potential region (above the boundary layer), outer
region and inner region of the boundary layer, respectively.

\cite{chang1999relationship} analyzed the contribution of individual
source terms to wall-pressure fluctuation PSD using Green's function
formulation for $Re_{\tau}=180$ channel flow. The contributions from
the viscous sublayer, buffer, logarithmic and the outer region to
wall-pressure fluctuation wavenumber spectra were investigated by
computing partial pressures from sources located in the corresponding
regions. The buffer region contribution was seen to be the most
dominant for both slow and rapid terms over most of the wavenumber
range. The logarithmic region was seen to contribute to the low
wavenumbers through the rapid term. The viscous region was observed to
contribute only to the high wavenumbers through both rapid and slow
terms.

\cite{panton2017correlation} investigated wall-pressure fluctuations
using DNS datasets of turbulent channel flow at $Re_{\tau}$ ranging
from $180-5200$. The premultiplied wall-pressure streamwise wavenumber
spectra showed a peak around $\lambda_1^+\approx 200-300$.  Here,
$\lambda_1^+$ is the non-dimensional streamwise wavelength based on
inner units.  Because the peak wavenumber scaled with inner units,
\cite{panton2017correlation} believed the location of the
corresponding velocity sources to be in the inner region of the
channel. Further, with increasing Reynolds number, the low wavenumber
contribution was observed to increase in magnitude and separate from
the high wavenumber contribution. Since the dominant low wavenumbers
did not scale with inner units, the corresponding velocity sources
were believed to be in the outer region of the channel. Hence, the
outer region contribution to wall-pressure becomes important at very
high Reynolds numbers.


We investigate the decorrelated features of wall-pressure fluctuation
sources in the turbulent channel using Spectral Proper Orthogonal
Decomposition (spectral POD). Spectral POD was originally introduced
by \cite{lumley2007stochastic} and recently analyzed by
\cite{towne2018spectral} for its relation to Dynamic Mode
Decomposition and Resolvent analysis. It involves the
eigendecomposition of the cross spectral density of the quantity of
interest. The technique has been used previously
\citep{schmidt2018spectral} as a post-processing tool to infer
wavepackets in axsymmetric jets. We use this technique to obtain the
decorrelated contribution from each wall-parallel plane to
wall-pressure fluctuation PSD. To our knowledge, this is the first
work that uses spectral POD to analyze wall-pressure fluctuation
sources.

Unlike the methodology of \cite{chang1999relationship}, the proposed
method takes into account the wall-normal cross correlation of the
source terms and accounts for the phase relationships between
different wall-parallel planes. The contribution of cross-correlation
between sources in any two wall-parallel planes to wall-pressure PSD
is quantified as a function of frequency. Also, the collapse of the
frequency and wavenumber spectrum based on inner and outer flow
variables as carried out in \cite{farabee1991spectral} and
\cite{panton2017correlation} do not yield such information on the
wall-normal distribution, insight into which can be obtained from the
proposed analysis. A `net source distribution function' (also termed
as `net source' for brevity) is defined which yields the integrated
effect of all sources in a particular wall-parallel plane. The cross
spectral density (CSD) of the net source function is computed from the
generated DNS database. The net source CSD when doubly integrated in
the wall-normal direction yields the wall-pressure PSD and, when
singly integrated yields the CSD between wall-pressure fluctuation and
the net source. In addition to the spectral features, spectral POD is
used to identify the decorrelated contribution from each wall-parallel
plane. We present a parallel implementation of the analysis framework
that is streaming, thus enabling processing of large data sets.

The paper is organized as follows. We discuss the DNS simulation
details in section \ref{sec:dnssimdetails}. The theory and
implementation of the proposed analysis framework to investigate
wall-pressure sources is discussed in sections
\ref{subsec:analysistheory} and \ref{subsec:analysisimpl},
respectively. Finally, in section \ref{sec:results}, we discuss the
spectral features of the net source function, the spectral POD results
and its relevance to wall-pressure fluctuation PSD using DNS data at
$Re_{\tau}=180$ and $400$.

\section{DNS simulation details} \label{sec:dnssimdetails}

The incompressible Navier-Stokes equations are solved using the
collocated finite volume method of \cite{mahesh2004numerical} in a
frame of reference moving with the bulk velocity of the fluid as done
by \cite{bernardini2013turbulent}. Better pblackiction of the convection
velocities and high wavenumber component of the streamwise velocity
fluctuations was observed by \cite{bernardini2013turbulent} in the
moving frame of reference. We observed a slightly better pblackiction of
high frequency component of the wall-pressure frequency spectra with
the moving frame of reference formulation. The method is second-order
accurate in space. We use the Crank-Nicholson time integration scheme
to ensure second-order accuracy in time and to allow for larger
timesteps. The method uses a least-square cell-centeblack pressure
gradient reconstruction to ensure discrete kinetic energy conservation
in space. This ensures stability at large Reynolds number without
adding numerical dissipation.

We define the subscripts $x, y$ and $z$ to be the streamwise,
wall-normal and spanwise directions. The computational domain is a
Cartesian box with side lengths $L_x=6\pi\delta, L_y=2\delta$ and
$L_z=2\pi\delta$. A long streamwise domain was chosen to include large
scale contribution within the domain. Also, the long domain eliminates
periodicity effects otherwise seen in low-frequency streamwise
wavenumber frequency spectra (not shown). The spurious high levels of
low wavenumber region observed in the results of \cite{choi1990space}
at low frequencies is not present in the current simulation results
(not shown). Table \ref{tab:grid} shows the grid sizes
($N_x, N_y, N_z$) for $Re_{\tau}=180$ and $400$. The mesh is uniform
in streamwise and spanwise directions, and a hyperbolic tangent
spacing is used in the wall-normal direction with a stretching factor
of $2.07$ for both $Re_{\tau}$. The mesh spacing in viscous units
($\Delta x^+, \Delta z^+, \Delta y_w^{+}, \Delta y_c^{+}$) is given in
table \ref{tab:grid}, where $\Delta y_w^{+}, \Delta y_c^{+}$ is the
wall-normal mesh spacing at the wall and at the centerline
respectively. A superscript of $+$ indicates non-dimensionalization
with respect to inner layer variables $u_{\tau}$ and $\nu$
respectively. The resolution is sufficient enough to resolve the near
wall fine scale features. The velocity of the moving frame of
reference ($U_{bref}^+$) is chosen to be $15.8$ and $17.8$ for
$Re_{\tau}=180$ and $400$ respectively. These values are close to the
actual bulk velocity in the stationary frame of
reference. {\color{black} A non-dimensional body force
  ($f_x\delta/\rho u_{\tau}^2$) of $1$ is applied in the streamwise
  direction throughout the domain}. A slip velocity equal to the
negative of the frame velocity is applied at the walls. Periodic
boundary conditions are used in the streamwise and spanwise
directions. A timestep of $5\times 10^{-4}\delta/u_{\tau}$ is used for
both the simulations. The flow is initially transient and subsequently
reaches a statistically stationary state when the discharge starts to
oscillate around a mean value. The total simulation time for both
$Re_{\tau}=180$ and $400$ cases is $8\delta/u_{\tau}$ after the
initial transient period. {\color{black} We sample the data every
  timestep to compute wall-pressure statistics.}

\begin{table}
  \begin{center}
    \def~{\hphantom{0}}
    \begin{tabular}{lccccccc}
      $Re_{\tau}$  & $N_x\times N_y\times N_z$ & $\Delta x^+$ & $\Delta z^+$ & $\Delta y_w^{+}$ & $\Delta y_c^{+}$ & $U_{bref}^+$ & $Tu_{\tau}/\delta$\\[3pt]
      180   & $720\times 176\times 330$ & $4.7$ & $3.4$ & $0.27$ & $4.4$ & $15.8$ & $8$\\
      400   & $1388\times 288\times 660$ & $5.4$ & $3.8$ & $0.37$ & $5.9$ & $17.8$ & $8$\\
    \end{tabular}
    \caption{Grid sizes, mesh spacing and velocity of the moving frame
      of reference used in the DNS simulation.}
    \label{tab:grid}
  \end{center}
\end{table}

\section{Analysis framework} \label{sec:analysis}

\subsection{Theory} \label{subsec:analysistheory}

We first write the solution to equation \ref{eqn:pflucpe} using the
Green's function formulation. The streamwise and spanwise extents are
taken to be infinite and the frame of reference is assumed to be
stationary. We use zero normal derivative of pressure fluctuation as
the boundary condition at the top and bottom walls. The Stokes
component of pressure arising from the non-zero wall-normal derivative
of wall-pressure fluctuation at the top and bottom wall has been shown
to be negligible when compablack to the rapid and slow terms for high
Reynolds number flows
{\color{black}\citep{hoyas2006scaling,gerolymos2013wall}}. The
wall-normal coordinates of the top and bottom wall are $y=-\delta$ and
$y=+\delta$ respectively. The Fourier transform is defined as
\begin{equation}
  g(t)=\int_{-\infty}^{\infty}\hat{g}(\omega)e^{i\omega t}\mathrm{d\omega} ;\, \hat{g}(\omega)=\frac{1}{2\pi}\int_{-\infty}^{\infty}g(t)e^{-i\omega t}\mathrm{dt}.
\end{equation}
where $\hat{g}(\omega)$ is the Fourier transform of $g(t)$. The
pressure fluctuation
\begin{equation} \label{eqn:greenfuncsoln}
  \begin{split}
    p(x,y,z,t)&=\iint_{-\infty}^{\infty}\hat{p}(k_1,y,k_3,t)\mathrm{e}^{i\left(k_1x+k_3z\right)}\mathrm{dk}_1\mathrm{dk}_3 \\
    \hat{p}(k_1,y,k_3,t)&=\int_{-\delta}^{+\delta}G(y,y',k_1,k_3)\hat{f}(k_1,y,k_3,t)\mathrm{dy'},\\
    \hat{f}(k_1,y,k_3,t)&=\frac{1}{\left(2\pi\right)^2}\iint_{-\infty}^{\infty}f(x,y,z,t)\mathrm{e}^{-i(k_1x+k_3z)}\mathrm{dx}\,\mathrm{dz},
  \end{split}
\end{equation}
where $f(x,y,z,t)$ is the right hand side source term in the Poisson
equation (equation \ref{eqn:pflucpe}), $\hat{p}(k_1,y,k_3,t)$ and
$\hat{f}(k_1,y,k_3,t)$ denote the Fourier transform in the spanwise
and streamwise directions of $p(x,y,z,t)$ and $f(x,y,z,t)$
respectively and the Fourier transform $\hat{p}(k_1,y,k_3,t)$ is
defined similar to $\hat{f}(k_1,y,k_3,t)$ in the above equation. The
Green's function $G(y,y',k_1,k_3)$ can be shown to be
\begin{equation} \label{eqn:greenfn}
  \begin{split}
    G(y,y',k_1,k_3)&=\begin{cases}
      \frac{\rm{cosh}(k(y'-\delta))\rm{cosh}(k(y+\delta))}{2k \rm{sinh}(k\delta)\rm{cosh}(k\delta)},  y\leq y',\\
      \frac{\rm{cosh}(k(y'+\delta))\rm{cosh}(k(y-\delta))}{2k \rm{sinh}(k\delta)\rm{cosh}(k\delta)}, y > y',
    \end{cases}\\
    k&=\sqrt{k_1^2+k_3^2},
  \end{split}
\end{equation}
for all combinations of $k_1,k_3$ except when both $k_1=0$ and
$k_3=0$, for which we can obtain
\begin{equation}
  G(y,y',k_1,k_3)=\begin{cases}
    \frac{1}{2}(y-y'),  y\leq y',\\
    \frac{1}{2}(y'-y),  y > y'.
  \end{cases}
\end{equation}
In order to ensure uniqueness of the Green's function when $k=0$, we
have made use of the condition that the instantaneous average of the
top and bottom wall-pressure fluctuation is zero. The above Green's
function has been previously used by \cite{kim1989structure} to obtain
wall-pressure fluctuations from the \cite{kim1987turbulence}
simulation.

The wall-pressure fluctuation of a point $(x,z)$ on the bottom wall is
\begin{equation} \label{eqn:phatxzomega}
  \begin{split}
    p(x,-\delta,z,t)&=\iint_{-\infty}^{\infty}p(k_1,-\delta,k_3,t)e^{i\left(k_1x+k_3z\right)}\mathrm{dk_1}\mathrm{dk_3},\\
    &=\iint_{-\infty}^{\infty}\int_{-\delta}^{+\delta}G(-\delta,y,k_1,k_3)f(k_1,y,k_3,t)\mathrm{dy}e^{i\left(k_1x+k_3z\right)}\mathrm{dk_1}\mathrm{dk_3},\\
    &=\int_{-\delta}^{+\delta}\iint_{-\infty}^{\infty}G(-\delta,y,k_1,k_3)f(k_1,y,k_3,t)e^{i\left(k_1x+k_3z\right)}\mathrm{dk_1}\mathrm{dk_3}\mathrm{dy},\\
    &=\int_{-\delta}^{+\delta}f_G(x,y,z,t)\mathrm{dy},
  \end{split}
\end{equation}
where $f_G(x,y,z,t)$ is termed as the `net source' because it includes
contribution from all sources in a wall-parallel plane and the Green's
function. It includes the contribution from all streamwise and
spanwise wavenumbers. The Green's function essentially assigns a
weight to each wavenumber $(k_1,k_3)$ component of the source in the
wall-parallel plane. Note that the function $f_G(x,y,z,t)$ is
homogenous in the streamwise and spanwise directions.

In order to characterize the features of the net source function
$f_G(x,y,z,t)$, the net source CSD $\Gamma(r,s,\omega)$ is defined as
\begin{equation}
  \Gamma(r,s,\omega)=\frac{1}{2\pi}\int_{-\infty}^{\infty}\langle f^*_G(x,r,z,t) f_G(x,s,z,t+\tau)\rangle e^{-i\omega\tau}\mathrm{d\tau}.
\end{equation}
It can be related to the five-dimensional CSD
$\varphi_{ff}(r,s,k_1,k_3,\omega)$ of the pressure Poisson source
terms as
\begin{equation} \label{eqn:gammadefn} \Gamma(r,s,\omega)=\iint
  \limits_{-\infty}^{+\infty}G^*(0,r,\mathrm{k}_1,\mathrm{k_3})G(0,s,\mathrm{k}_1,\mathrm{k_3})\varphi_{ff}(r,s,k_1,k_3,\omega)\mathrm{dk}_1\,\mathrm{dk}_3,
\end{equation}
where
\begin{equation} \label{eqn:phiff}
  \begin{split}
    &\varphi_{ff}(r,s,k_1,k_3,\omega) = \\
    &\frac{1}{\left(2\pi\right)^3}\iiint
    \limits_{-\infty}^{+\infty}\langle f^*(x,r,z,t)
    f(x+\xi_1,s,z+\xi_3,t+\tau) \rangle \mathrm{e}^{-i\left(
        k_1\xi_1+k_3\xi_3+\omega\tau\right)}
    \mathrm{d\xi}_1\,\mathrm{d\xi}_3\,\mathrm{d\tau}.
  \end{split}
\end{equation}

The PSD of the spatially homogenous wall-pressure fluctuation
$\phi_{pp}(\omega)$ is related to the net source CSD.
\begin{equation} \label{eqn:phippgammareln}
  \phi_{pp}(\omega)=\int_{-\delta}^{+\delta}\int_{-\delta}^{+\delta}
  \Gamma(r,s,\omega)\,\mathrm{dr}\,\mathrm{ds}.
\end{equation}
In order to analyze the contribution from a particular wall-parallel
plane at $y=r$, we include its cross-correlation with every other
wall-normal location $y'=s$ by integrating $\Gamma(r,s,\omega)$ along
$s$.
\begin{equation} \label{eqn:psidefn} \Psi(r,\omega) =
  \int_{-\delta}^{+\delta}\Gamma(r,s,\omega)\,\mathrm{ds}.
\end{equation}
The resulting function $\Psi(r,\omega)$ can be shown to be the CSD of
the wall-pressure fluctuation and the net source at $r$, i.e.,
\begin{equation}
  \Psi(r,\omega)=\frac{1}{2\pi}\int_{-\infty}^{\infty}\langle f^*_G(x,r,z,t)p(x,-1,z,t+\tau)\rangle e^{-i\omega\tau}\mathrm{d\tau}.
\end{equation}
We will call $\Psi(r,\omega)$ as the wall-pressure fluctuation
{\color{black}-} net source CSD. The wall-pressure PSD can be
expressed in terms of $\Psi(r,\omega)$.
\begin{equation} \label{eqn:phipppsireln}
  \phi_{pp}(\omega)=\int_{-\delta}^{+\delta}\Psi(r,\omega)\,\mathrm{dr}.
\end{equation}

{\color{black} Next, we identify decorrelated features in the dataset that contribute the most to the wall-pressure PSD using $\Gamma(r,s,\omega)$. To accomplish this, we use the Poisson inner product defined in equation \ref{eqn:geninnerprod} to enforce the orthonormality of the modes instead of the commonly used $L^2$ inner product. We decompose $\Gamma(r,s,\omega)$ as
  \begin{equation} \label{eqn:spodgamma}
    \Gamma(r,s,\omega)=\sum_{i=1}^{\infty}\lambda_i(\omega)\Phi_i(r,\omega)\Phi_i^*(s,\omega),
  \end{equation}
  where $\{\lambda_i(\omega),\Phi_i(r,\omega)\}_{i=1}^{\infty}$ are
  the spectral POD eigenvalue and mode pairs. The mode
  $\Phi_i(r,\omega)$ relates to the eigenfunction
  $\bar{\Phi}_i(r,\omega)$ of $\Gamma(r,s,\omega)$ through the
  relation
  \begin{equation}
    \Phi_i(r,\omega)=\left(-(1-\beta)\frac{\partial^2}{\partial y^2}
      +\beta\right)\bar{\Phi}_i(r,\omega),
  \end{equation}
  where $\beta$ is a real number satisfying $0<\beta\leq 1$ and the
  eigenfunctions are assumed to satisfy zero Neumann boundary
  conditions $\bar{\Phi}_i(r,\omega)$ at $r=-\delta$ and $r=+\delta$. The
  eigenvalue problem for $\bar{\Phi}_i(r,\omega)$ and
  $\lambda_i(\omega)$ is
  \begin{equation} \label{eqn:eigenvalueproblem}
    \int_{-\delta}^{+\delta}\Gamma(r,s,\omega)\bar{\Phi}_i(s,\omega)\,\mathrm{ds}=\lambda_i(\omega)\left(-(1-\beta)\frac{\partial^2}{\partial
        y^2} +\beta\right)\bar{\Phi}_i(r,\omega).
  \end{equation}
  The spectral POD eigenvalues are arranged in decreasing order. The
  eigenfunctions $\bar{\Phi_i}(r,\omega)$ satisfy the orthonormality
  condition
  \begin{equation}\label{eqn:geninnerprod}
    \begin{split}
      &=\int_{-\delta}^{+\delta}\bar{\Phi}^*_i(r,\omega)\left(-(1-\beta)\frac{\partial^2}{\partial
          y^2} +\beta\right)\bar{\Phi}_j(r,\omega) \,\mathrm{dr}\\
      &=\delta_{ij},
    \end{split}
  \end{equation} 
  where $\delta_{ij}$ is the Kroenecker delta. We will call the inner product above `the Poisson inner product' because the kernel $\left(-(1-\beta)\frac{\partial^2}{\partial y^2} +\beta\right)$ can be related to the Poisson equation. If we choose $\beta=1$, then the Poisson inner product is the standard $L^2$ inner product.

  The contribution of each spectral POD mode to wall-pressure PSD can
  be obtained by integrating equation \ref{eqn:spodgamma} in $r$ and
  $s$,
  \begin{equation} \label{eqn:phippwrtgamma}
    \phi_{pp}(\omega)=\sum_{i=1}^{\infty}\gamma_i(\omega);\,\gamma_i(\omega)=\lambda_i(\omega)|\int_{-\delta}^{+\delta}\Phi_i(r,\omega)\,\mathrm{dr}|^2;\,i=1,\dots,\infty.
  \end{equation}
  In the above equation, the wall-pressure PSD is expressed as sum of
  positive contributions $\{\gamma_i(\omega)\}_{i=1}^{\infty}$ from
  each spectral POD mode. We will use the quantities
  $\{\gamma_i(\omega)\}_{i=1}^{\infty}$ to identify the spectral POD
  modes that are the dominant contributors to wall-pressure PSD.

  The spectral POD modes and eigenvalues depend on the parameter $\beta$. For a chosen value of $\beta$, we will have the corresponding set of spectral POD modes $\{\Phi_i(y,\omega)\}_{i=1}^{\infty}$ and eigenvalues $\{\lambda_i(\omega)\}_{i=1}^{\infty}$. However, irrespective of the chosen $\beta$, the component of the net source Fourier transform ($\hat{f}_G(x,y,z,\omega)$) along the spectral POD modes will be decorrelated, i.e.,
  \begin{equation} \label{eqn:netsourceexpansion}
    \begin{split}
      f_G(x,y,z,t)&=\int_{-\infty}^{+\infty}\hat{f}_G(x,y,z,\omega)e^{i\omega t}\,\mathrm{d\omega},\\
      \hat{f}_G(x,y,z,\omega)&=\sum_{j=1}^{\infty}\alpha_j(x,z,\omega)\Phi_j^*(y,\omega),\\
      \langle\alpha_i(x,z,\omega)\alpha^*_j(x,z,\omega_o)\rangle
      &=\lambda_i(\omega)\delta_{ij}\delta(\omega-\omega_o),
    \end{split}
  \end{equation}
  where $\{\alpha_j(x,z,\omega)\}_{j=1}^{\infty}$ are the
  coefficients, $\langle \cdot \rangle$ denotes ensemble average and
  $\delta$ is the Dirac delta function.

  On the other hand, choosing the $L^2$ inner product ($\beta=1$) to enforce the orthonormality of the modes will also optimally decompose the wall-normal integral of the PSD $\Gamma(r,r,\omega)$. Substituting $s=r$ in equation \ref{eqn:spodgamma} and integrating in $r$, we obtain
  \begin{equation}\label{eqn:intgammadecomp}
    \begin{split}
      \Gamma(r,r,\omega)&=\sum_{j=1}^{\infty}\lambda_j(\omega)|\Phi_i(r,\omega)|^2,\\
      \int_{-\delta}^{+\delta}\Gamma(r,r,\omega)\,\mathrm{dr}&=\sum_{j=1}^{\infty}\lambda_j(\omega).
    \end{split}
  \end{equation}

  However, the dominant spectral POD modes obtained with the $L^2$ inner product do not necessarily isolate the main contribution to the wall-pressure PSD. i.e., the value of $\gamma_i(\omega)$. That is, the wall-pressure PSD can be distributed over a large number of modes that each individually contribute a small fraction. This makes it difficult to identify the few dominant decorrelated source patterns. Further, the single dominant wall-pressure mode (mode with largest $\gamma_i(\omega)$) does not necessarily contain any useful information about the source because it contributes only a small fraction to the wall-pressure PSD. This was observed at low frequencies (see figure \ref{fig:spodeigwt}). 

Our goal is to identify useful decorrelated features of the wall pressure source, not to optimally decompose the integrated net source PSD (as done by the $L^2$ inner product). Therefore, we use the parameter $\beta$ to our advantage and select a suitable value for $\beta$.

For $0<\beta<1$, it can be shown that the Poisson inner product optimally decomposes $\iint_{-\delta}^{+\delta}{{G}(s,r,\beta/(1-\beta),0)}/({1-\beta})\Gamma(r,s,\omega)\,\mathrm{dr}\,\mathrm{ds}$ into the sum of spectral POD eigenvalues,
  \begin{equation} \label{eqn:sumofeig}
    \iint_{-\delta}^{+\delta}\frac{{G}(s,r,\frac{\beta}{1-\beta},0)}{1-\beta}\Gamma(r,s,\omega)\,\mathrm{dr}\,\mathrm{ds}=\sum_{j=1}^{\infty}\lambda_j(\omega),
  \end{equation}
  where the Green's function $G$ is given in equation \ref{eqn:greenfn}. We can observe that as $\beta$ approaches $0$, the Green's function $G(r,s,\beta/(1-\beta),0)$ becomes flatter and approaches a function that is constant in $r$ and $s$. Thus, the left hand side in the above equation approaches the wall-pressure PSD $\phi_{pp}(\omega)=\iint_{-1}^{+1}\Gamma(r,s,\omega)\,\mathrm{dr}\,\mathrm{ds}$ (up to a scaling). As we decrease $\beta$, we can therefore expect the dominant spectral POD modes to be the dominant contributors to wall-pressure PSD. Therefore, the Poisson inner product in equation \ref{eqn:geninnerprod} identifies the few dominant features of wall-pressure sources that are decorrelated.

  The Poisson inner product defined in equation \ref{eqn:geninnerprod} does not fall into the category presented by \cite{towne2018spectral}. They requiblack the eigenfunctions to be orthonormal in a weighted $L^2$ inner product. Here, we use the Poisson inner product (\ref{eqn:geninnerprod}) that has a symmetric positive definite kernel.

  The set of spectral POD modes obtained with any $\beta$ is complete. Therefore, we can relate the POD modes obtained with two different values of $\beta$ to each other through a linear transformation. That is, if $\{\hat{\Phi}_i(y,\omega)\}_{i=1}^{\infty}$ and $\{\tilde{\Phi}_i(y,\omega)\}_{i=1}^{\infty}$ are the two sets of spectral POD modes obtained with two different values of $\beta$, then
  \begin{equation}\label{eqn:cdefn}
    \hat{\Phi}_i(y,\omega)=\sum_jC^*_{ij}(\omega)\tilde{\Phi}_j(y,\omega),
  \end{equation}
  where the matrix $C(\omega)=[C_{ij}(\omega)]$ is the linear transformation. Further, we show in Appendix \ref{app:orthoreln} that the linear transformation $C(\omega)$ is indeed orthogonal with an appropriate row scaling, i.e.,
  \begin{equation}\label{eqn:corthoreln}
    \left({\hat{\Lambda}^{1/2}(\omega)}C(\omega)\right)^{H}{\hat{\Lambda}^{1/2}(\omega)}C(\omega)=\tilde{\Lambda}(\omega),
  \end{equation}
where $\hat{\Lambda}(\omega)$ and $\tilde{\Lambda}(\omega)$ are the diagonal matrices of eigenvalues of the set of modes $\{\hat{\Phi}_i(y,\omega)\}_{i=1}^{\infty}$ and $\{\tilde{\Phi}_i(y,\omega)\}_{i=1}^{\infty}$, respectively.


}

We can show that
\begin{equation} \label{eqn:magphi}
  |\int_{-\delta}^{+\delta}\Phi_i(y,\omega)\mathrm{dy}|=\int_{-\delta}^{+\delta}|\Phi_i(y,\omega)|cos\left(\angle\Phi_i(y,\omega)-\angle
    \Phi^n_i(\omega)\right)\mathrm{dy},
\end{equation}
where
$\angle \Phi_i^n(\omega)=\angle\left(
  \int_{-\delta}^{+\delta}\Phi_i(y,\omega)\mathrm{dy}\right)$ and $\angle$
denotes the phase of the complex number that follows it. Using
equation \ref{eqn:magphi} in equation \ref{eqn:phippwrtgamma}, we
obtain
\begin{equation} \label{eqn:gammamagphase}
  \gamma_i(\omega)=\lambda_i(\omega)\left(\int_{-\delta}^{+\delta}|\Phi_i(r,\omega)|cos\left(\angle\Phi_i(r,\omega)-\angle\Phi_i^n(\omega)\right)\mathrm{dr}\right)^2;\,i=1,\dots,\infty.
\end{equation}
From the above equation, we can observe that the eigenvalue, magnitude
and phase of the spectral POD mode, all play a role in determining its
contribution to wall-pressure PSD. Sources contained in wall-normal
regions where the phase is in the range
$|\angle \Phi_i(y,\omega)-\angle \Phi^n_i(\omega)|<\pi/2$ undergo
destructive interference with the sources contained in the region
where $\pi/2<|\angle \Phi_i(y,\omega)-\angle
\Phi^n_i(\omega)|<\pi$. Therefore, the interference of the sources
from different wall-normal regions represented by a spectral POD mode
plays a role in determining the net contribution to wall-pressure PSD
from the mode.

\subsection{Implementation} \label{subsec:analysisimpl}

The five-dimensional CSD $\varphi_{ff}(r,s,k_1,k_3,\omega)$ defined in
equation \ref{eqn:phiff} contains all pertinent information on
velocity field sources from cross-correlation of two wall-normal
locations. However, computing the function is extremely memory
intensive. For the $Re_{\tau}=400$ case, assuming $2000$ frequencies,
we would need $\approx 1220TB$ to store $\varphi_{ff}$. We use a
streaming parallel implementation procedure to compute the net source
CSD $\Gamma(r,s,\omega)$ that makes the computation feasible.

The source term in equation \ref{eqn:pflucpe} is computed and stoblack
from the DNS. The stoblack data is divided into multiple chunks to
compute the ensemble average in equation \ref{eqn:phiff}. For a given
chunk, the source terms are first converted to stationary frame of
reference and then Fourier transformed in $x$, $z$ and $t$. The
Fourier transforms are then used to update the net source CSD.
Details of the parallel implementation are provided in the Appendix.

A total of 16000 timesteps are used to obtain the net source CSD
$\Gamma(r,s,\omega)$ for both $Re_{\tau}$. {\color{black}We sample
  the data ever timestep}. The number of timesteps in each chunk is
2000 and $50\%$ overlap is used in time to increase statistical
convergence.  {\color{black} The frequency resolution of the analysis is
  $\Delta\omega\delta/u_{\tau}=2\pi$.}

\section{Results and discussion} \label{sec:results}

First, we discuss the spectral features of the wall-pressure
fluctuations obtained from the finite volume solver. Then, the wall
pressure net source cross spectral density (wall-pressure fluctuation - net source CSD) and the
dominant decorrelated net source patterns obtained using spectral POD
are discussed. {\color{black}For validation of the current DNS, we
  refer the reader to appendix \ref{sec:valid}}.

\subsection{DNS wall-pressure fluctuations}

\begin{figure}
  \centering
  \begin{subfigure}{0.5\textwidth}
    \begin{overpic}[width=\linewidth]{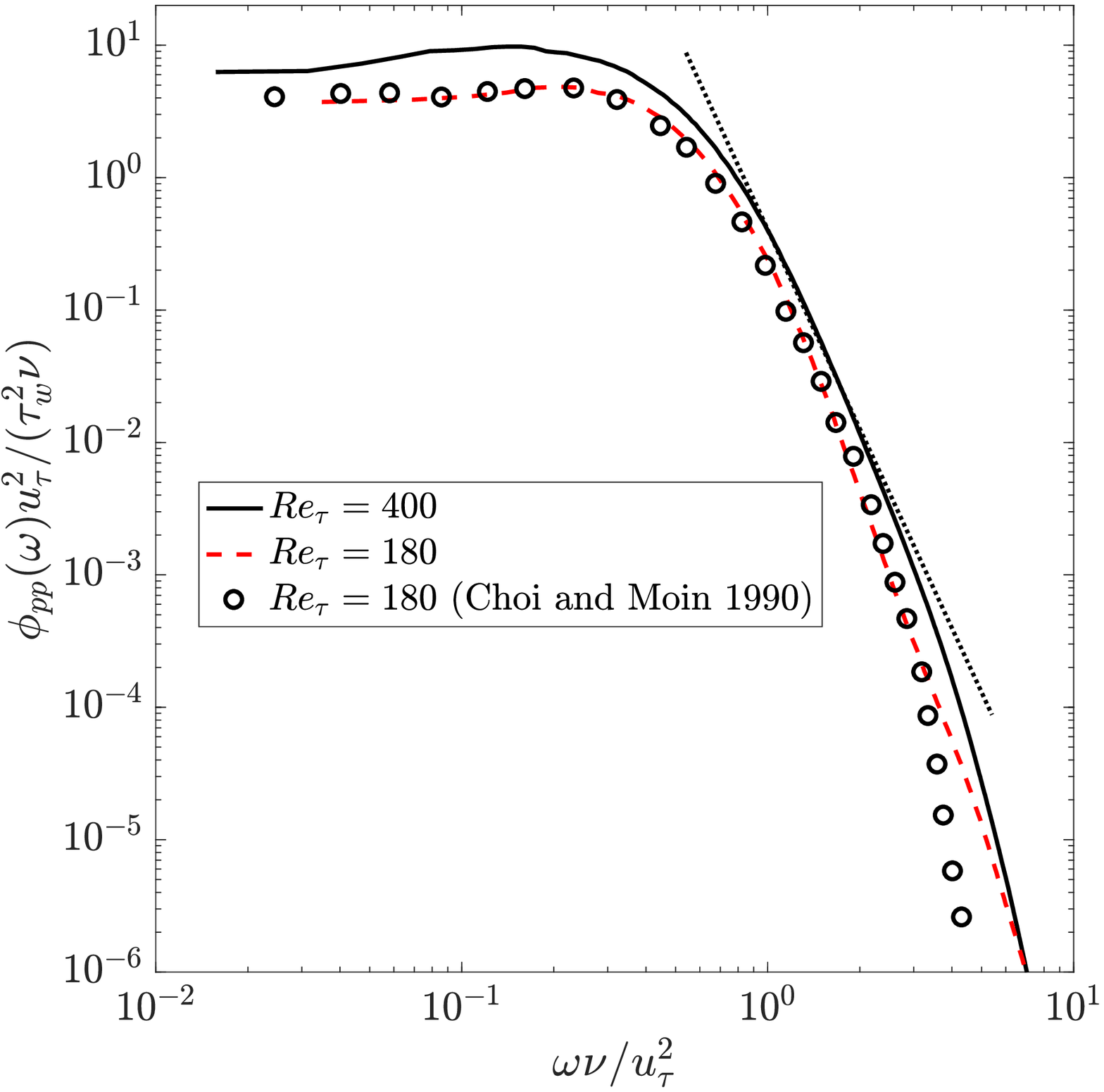}
      \put(1,95){$(a)$} \put(70,65){$-5$}
    \end{overpic}
  \end{subfigure}%
  \begin{subfigure}{0.5\textwidth}
    \begin{overpic}[width=\linewidth]{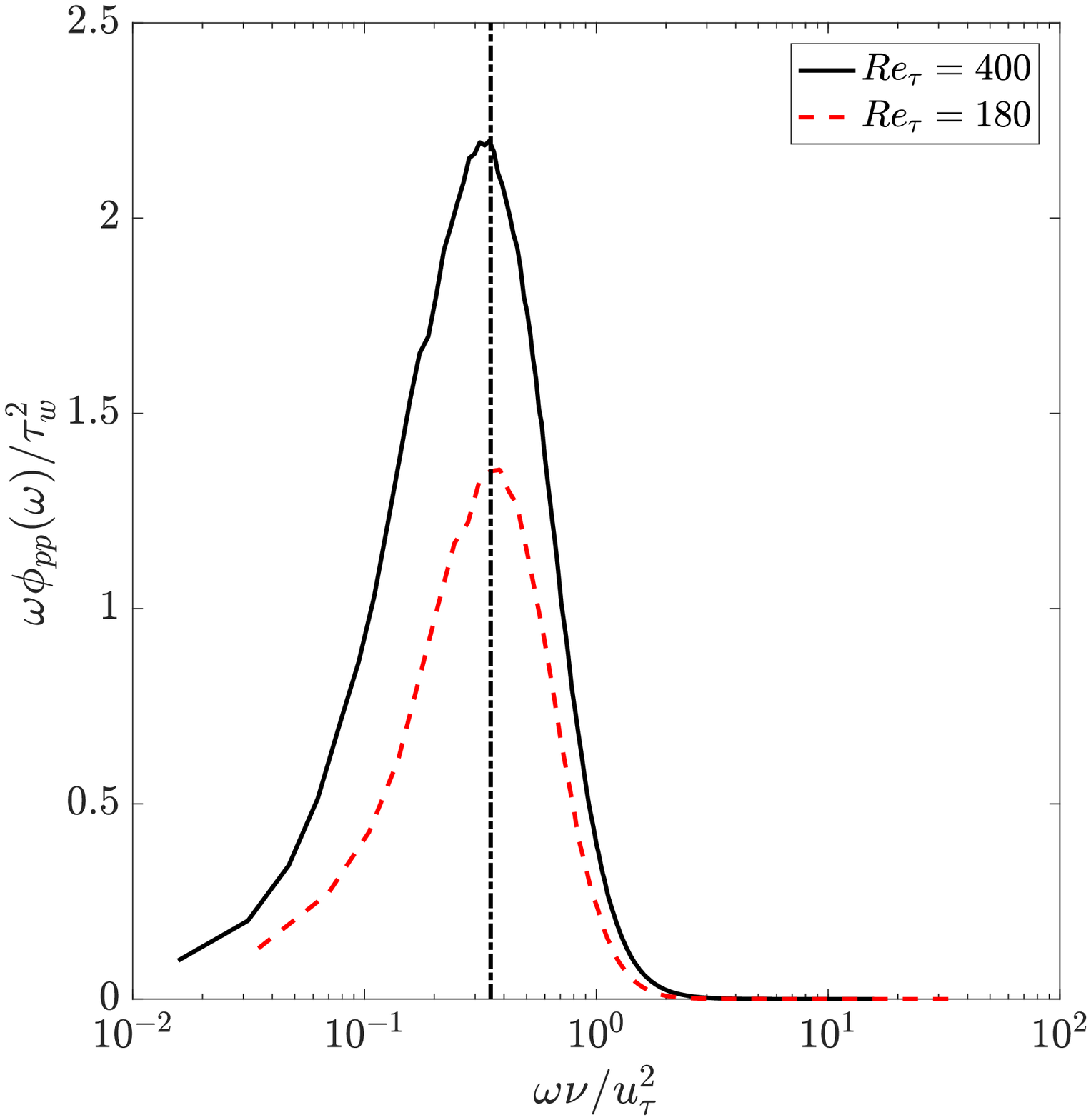}
      \put(1,95){$(b)$} \put(38,5){$0.35$}
    \end{overpic}
  \end{subfigure}
  \caption{Wall-pressure fluctuation power spectra in $(a)$ inner
    units $(b)$ premultiplied form with inner units on x-axis.}
  \label{fig:wallprespowerspec}
\end{figure}

\begin{figure}
  \centering
  \begin{subfigure}{0.5\textwidth}
    \begin{overpic}[width=\linewidth]{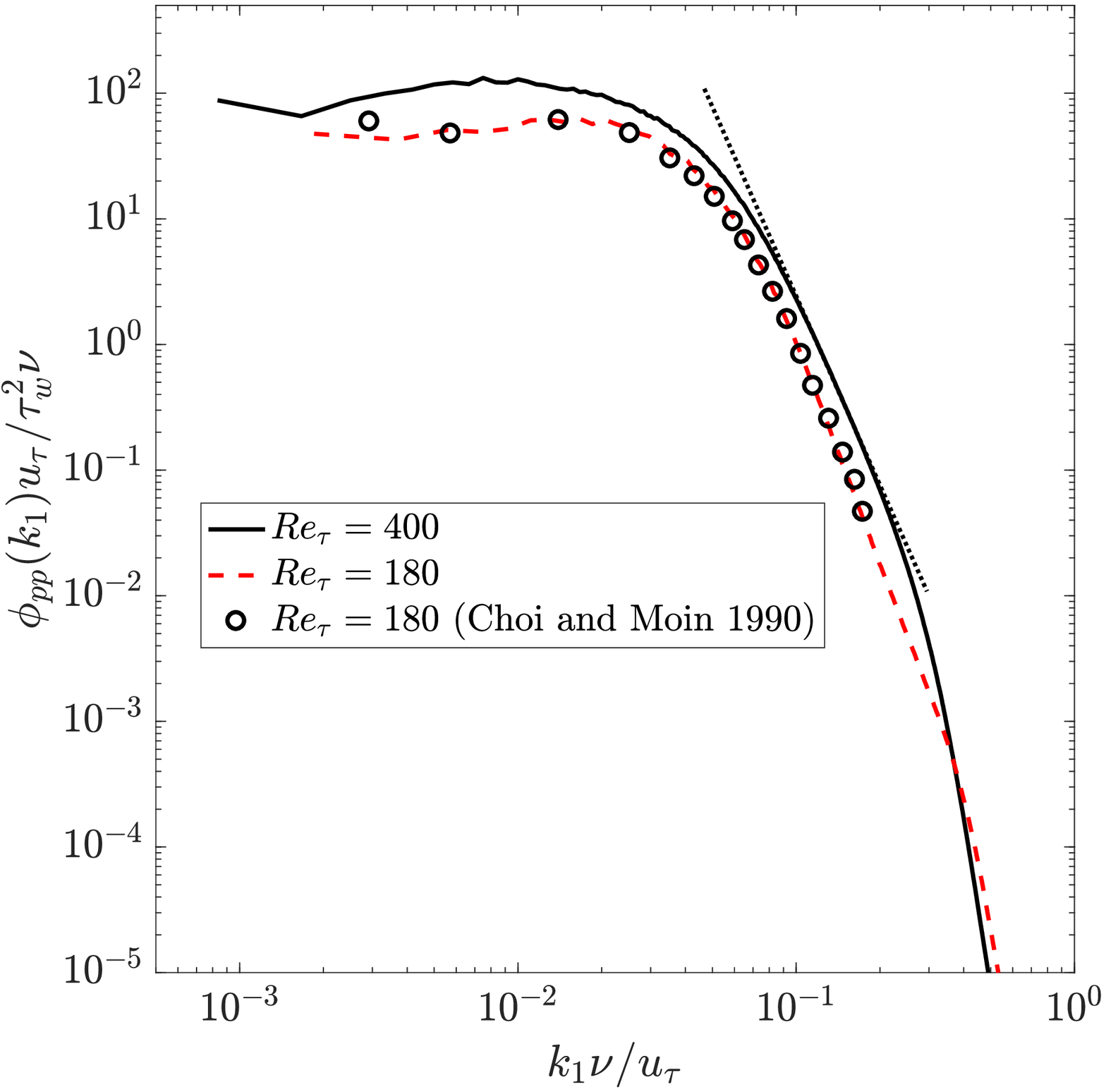}
      \put(1,95){$(a)$} \put(70,65){$-5$}
    \end{overpic}
  \end{subfigure}%
  \begin{subfigure}{0.5\textwidth}
    \begin{overpic}[width=\linewidth]{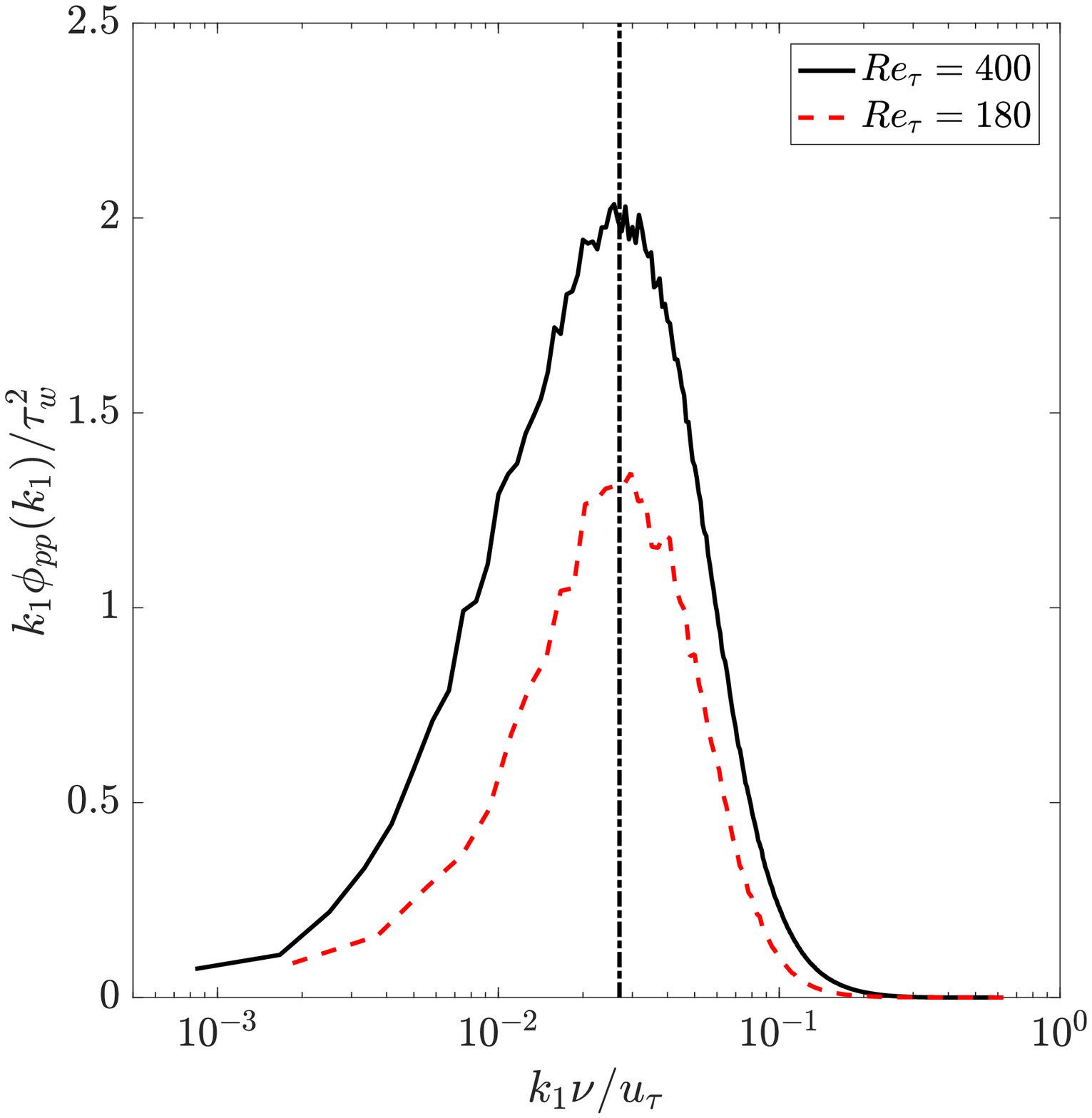}
      \put(1,95){$(b)$} \put(48,6){$0.027$}
    \end{overpic}
  \end{subfigure}
  \caption{Wall-pressure fluctuation streamwise wavenumber spectra in
    $(a)$ inner units $(b)$ pre-multiplied form with inner units on
    x-axis.}
  \label{fig:wallpresk1spec}
\end{figure}

The one-sided PSD of the obtained $Re_{\tau}=180$ and $400$
wall-pressure fluctuations scaled with inner variables is shown in
figure \ref{fig:wallprespowerspec}a. The streamwise wavenumber spectra
of the fluctuations at the two $Re_{\tau}$ are shown in figure
\ref{fig:wallpresk1spec}a. Both the PSD and wavenumber spectra at
$Re_{\tau}=180$ agree well with the results of
\cite{choi1990space}. The high frequency region with
$\omega^+=\omega \nu/u_{\tau}^2 > 1$, shows a {\color{black}small
  region of} $-5$ decay for the higher Reynolds number
($Re_{\tau}=400$). The high wavenumber region of the wavenumber
spectra plotted in figure \ref{fig:wallpresk1spec}a also shows a
{\color{black}small region of $-5$} decay in the region
$k_1^+=k_1\nu/u_{\tau}>0.1$, for the $Re_{\tau}=400$ case. The
premultiplied power spectra plotted in figure
\ref{fig:wallprespowerspec}b for both $Re_{\tau}$ shows a peak at
$\omega_p^+=0.35$. This peak at the same frequency has been previously
observed by \cite{hu2006wall} for $Re_{\tau}$ upto $1440$. Similar to
the power spectra, the premultiplied streamwise wavenumber spectra in
figure \ref{fig:wallpresk1spec}b also shows a peak at
$k_1^+=k_1\nu/u_{\tau}$ at $k_p^+=0.027$. This peak has also been
previously observed by \cite{panton2017correlation} for $Re_{\tau}$
over the range $180$ to $5000$. The wall-pressure fluctuation PSD
computed from the net source CSD using equation
\ref{eqn:phippgammareln} agrees with that obtained directly from the
solver (figure \ref{fig:wallprespowerspec}a) for both $Re_{\tau}$ (not
shown). We will investigate the distribution of the net
sources that give rise to this premultiplied PSD peak in the next
section.

\begin{figure}
  \centering
  \begin{subfigure}{0.5\textwidth}
    \begin{overpic}[width=\linewidth]{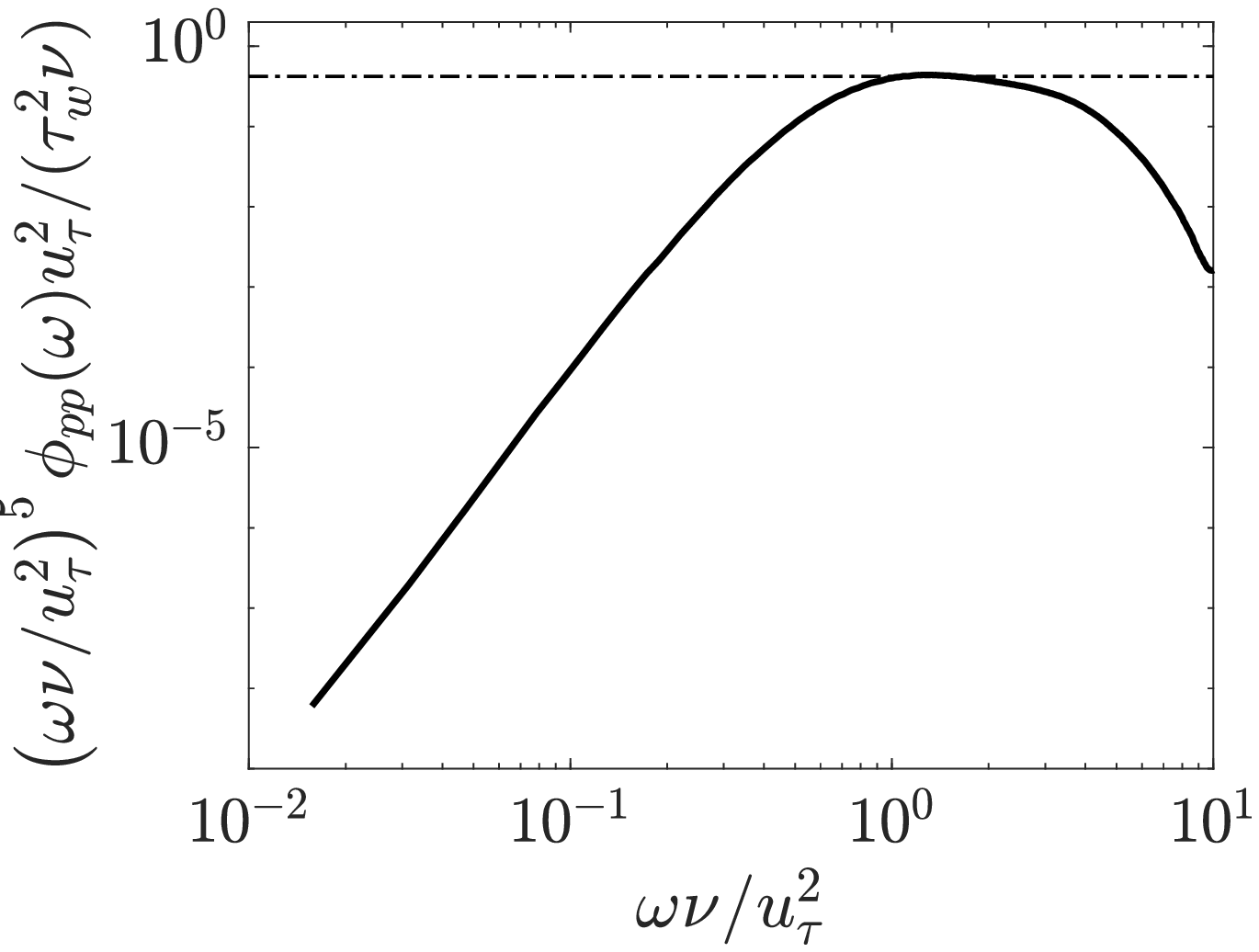}
      \put(1,72){$(a)$} 
    \end{overpic}
  \end{subfigure}%
  \begin{subfigure}{0.5\textwidth}
    \begin{overpic}[width=\linewidth]{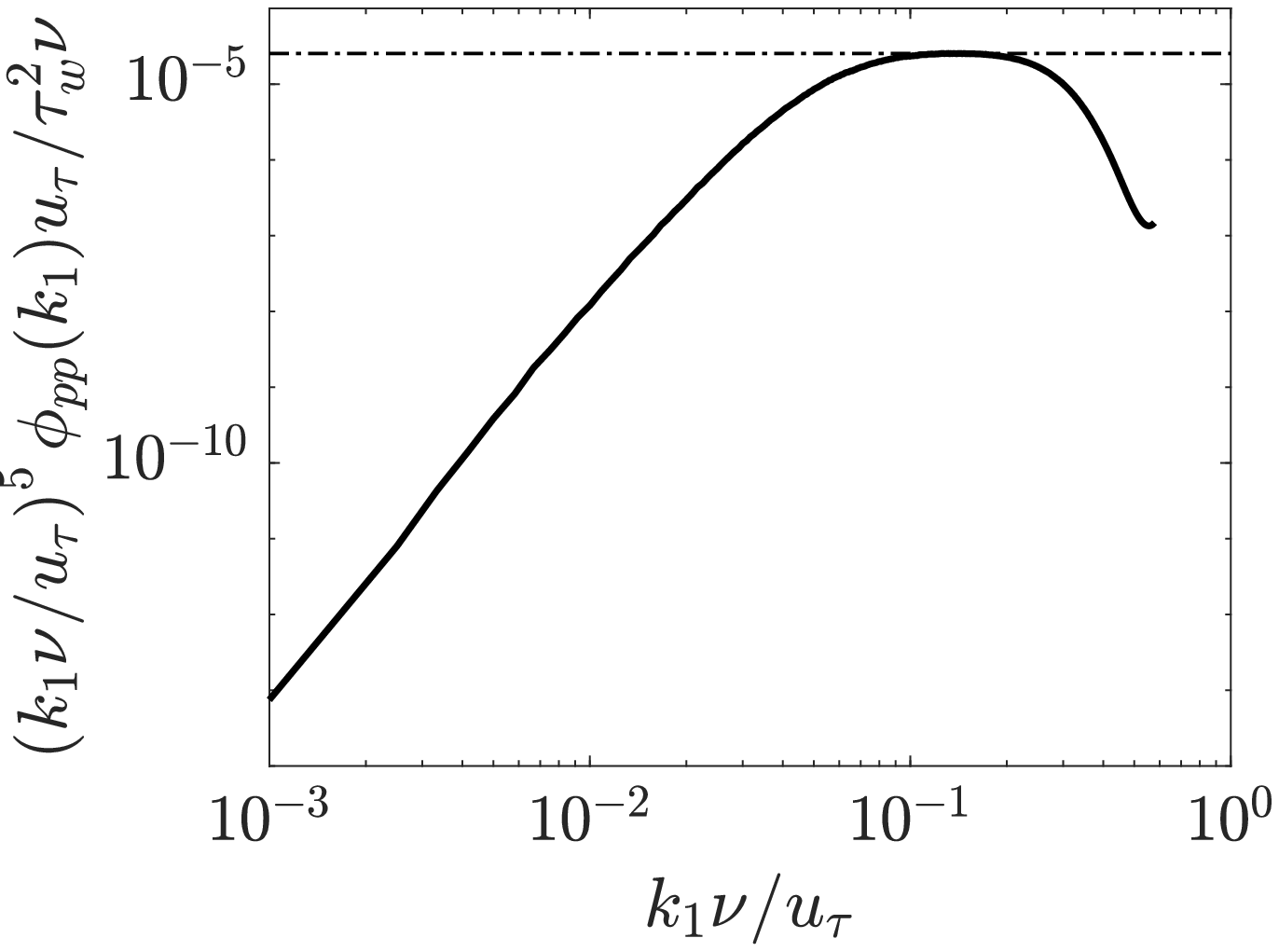}
      \put(1,72){$(b)$}
    \end{overpic}
  \end{subfigure}
  \caption{Diagnostic function to verify region of $-5$ slope in the $Re_{\tau}=400$ a) power and b) streamwise wavenumber spectrum. The dashed-dotted horizontal line in figures a and b indicates constant values of $0.42$ and $2.53\times 10^{-5}$, respectively.}
  \label{fig:diagnostic}
\end{figure}

{\color{black} To identify the range of $-5$ decay in the power and
  streamwise wavenumber spectrum, we plot the diagnostic functions
  $\left(\omega\nu/u_{\tau}^2\right)^5\phi_{pp}(\omega)u_{\tau}^2/\left(\tau_w^2\nu\right)$
  and
  $\left(k_1\nu/u_{\tau}\right)^5\phi_{pp}(k_1)u_{\tau}/\left(\tau_w^2\nu\right)$
  in figures \ref{fig:diagnostic}a and b, respectively.  The function
  is constant in the range of $-5$ slope. The diagnostic function does
  not return a significant range of frequency and wavenumbers that
  show $-5$ decay. We observe a constant value (indicated by
  dashed-dotted horizontal line) for only a very small range of
  frequencies and wavenumbers. To observe the decay in a significant
  range, we require higher Reynolds numbers.}

\begin{figure}
  \centering
  \begin{subfigure}{0.45\textwidth}
    \begin{overpic}[width=\linewidth]{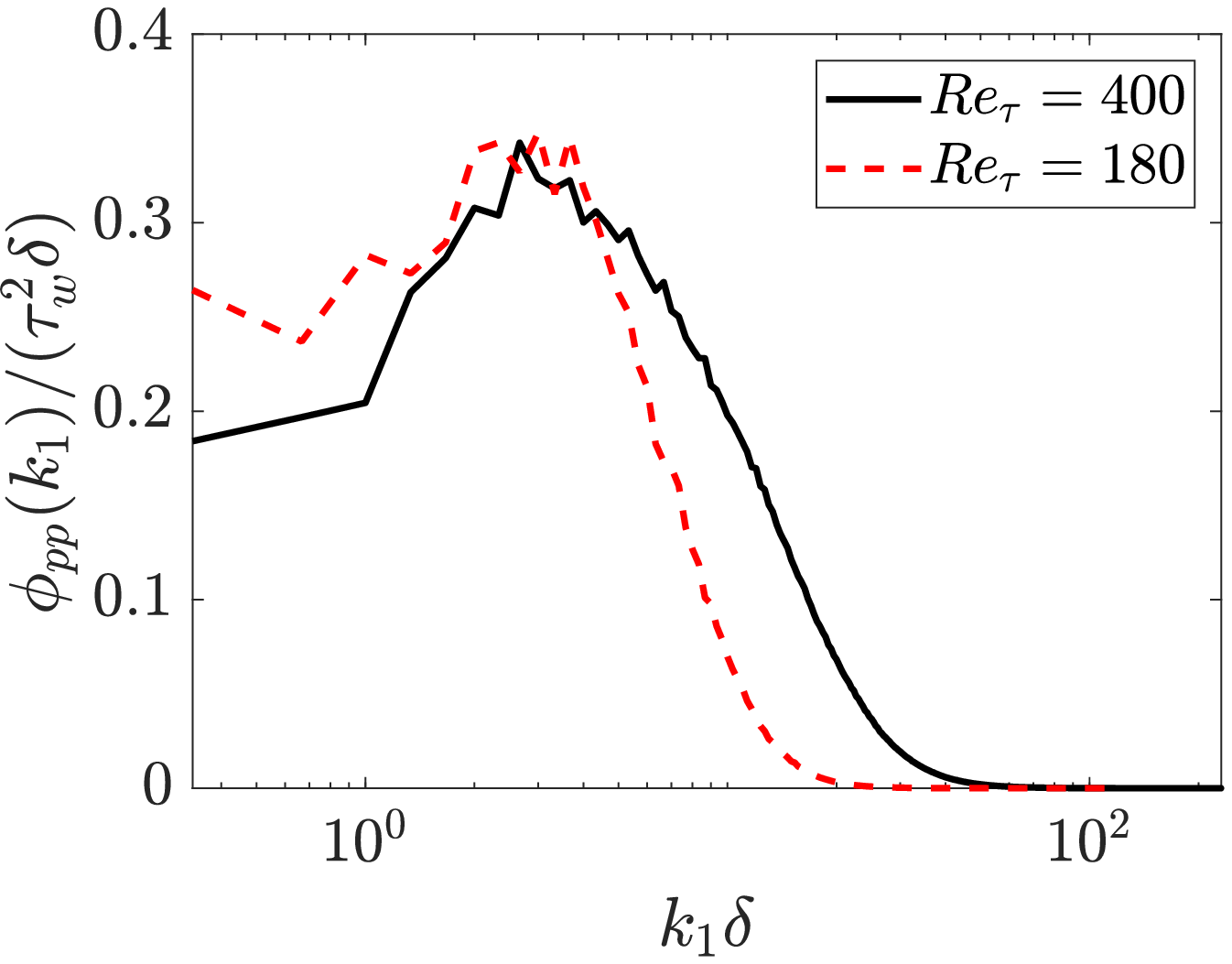}
      \put(1,75){$(a)$}
    \end{overpic}
    \begin{overpic}[width=\linewidth]{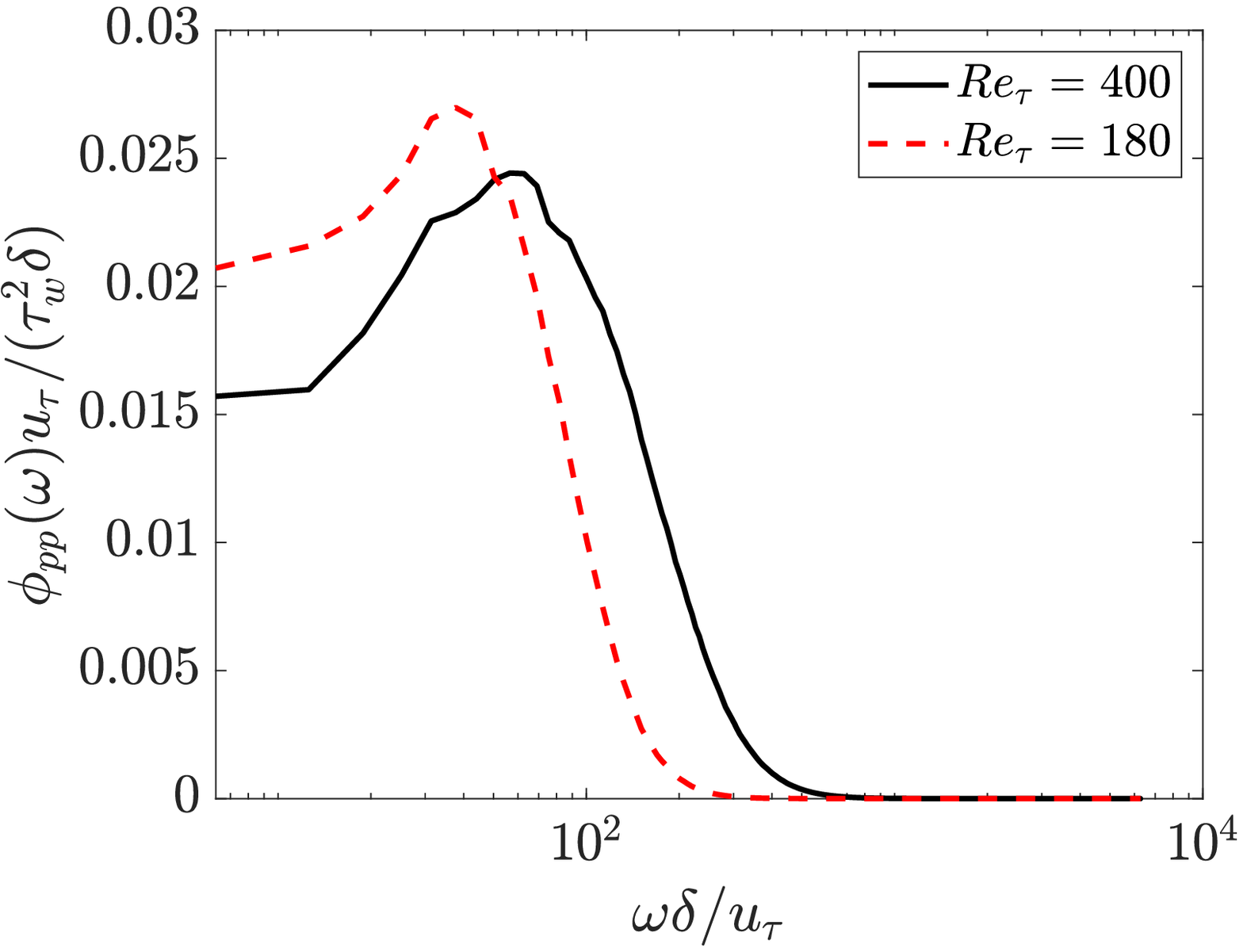}
      \put(1,75){$(b)$}
    \end{overpic}
  \end{subfigure}
  \begin{subfigure}{0.5\textwidth}
    \begin{overpic}[width=\linewidth]{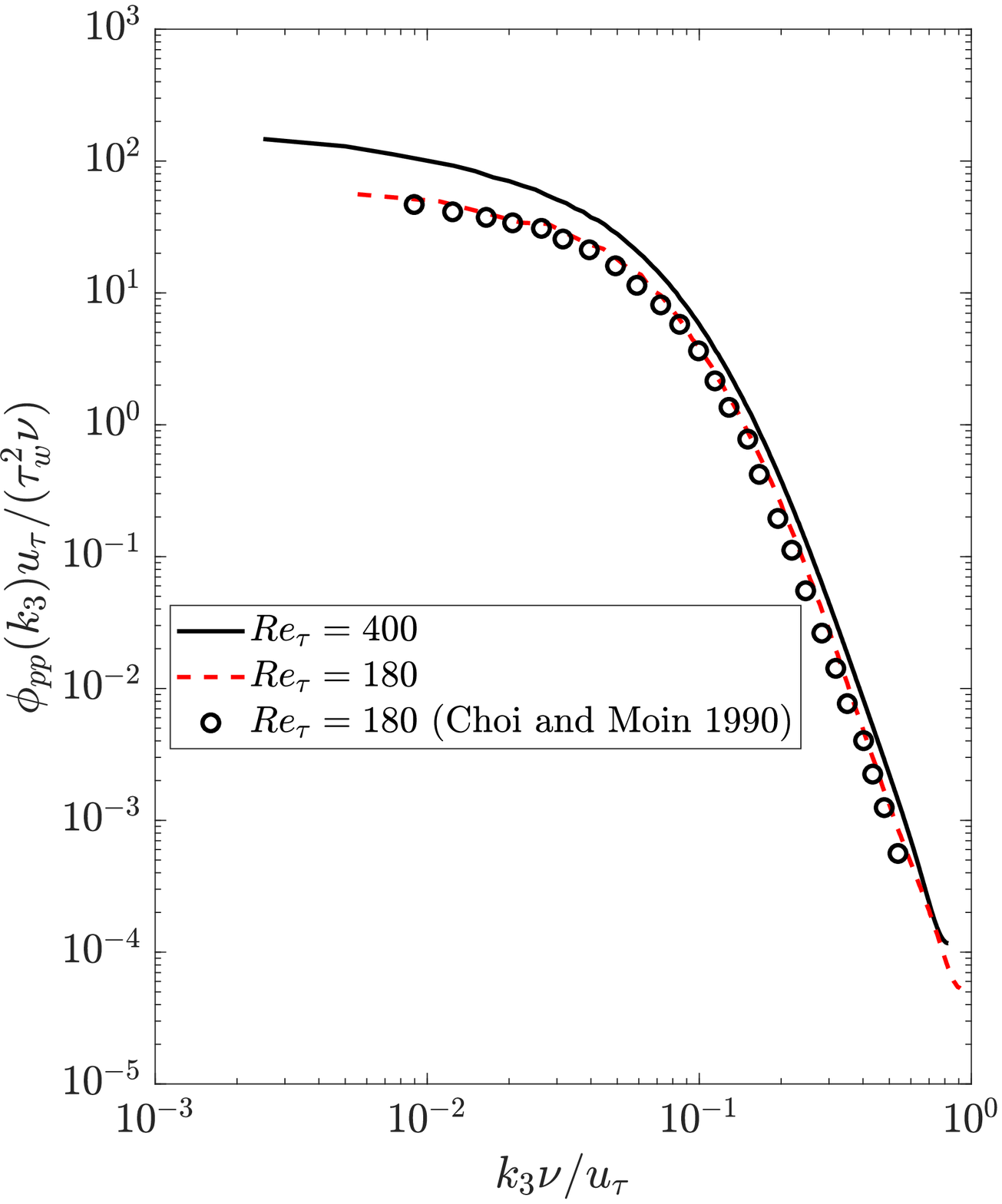}
      \put(1,95){$(c)$}
    \end{overpic}
  \end{subfigure}%
  \caption{$(a)$ Wall-pressure fluctuation streamwise wavenumber
    spectrum with linear y-axis. $(b)$ Wall-pressure fluctuation PSD
    with linear y-axis. $(c)$ Wall-pressure fluctuation spanwise
    wavenumber spectrum in inner units. }
  \label{fig:wallpresk3spec}
\end{figure}

{\color{black} The wall-pressure wavenumber spectra shows a low
  wavenumber peak around $k_1\delta\approx 3$ for both $Re_{\tau}=180$
  and $400$, respectively when the y-axis is plotted in linear
  coordinates (figure \ref{fig:wallpresk3spec}a). This corresponds to
  streamwise wavelengths $\lambda_1/\delta$ of $\sim 2$. Such low
  wavenumber peaks in the range $k_x\delta\approx 2.5-3.4$
  ($\lambda_x/\delta\approx 1.8-2.4$) have been previously observed by
  \cite{abe2005dns} and \cite{panton2017correlation} in turbulent
  channel for friction Reynolds numbers ranging from $180$ to
  $5000$. We observe the corresponding low frequency peak in the
  wall-pressure PSD at $\omega\delta/u_{\tau}=37.6$ and $50.2$ for
  $Re_{\tau}=180$ and $400$, respectively (shown in figure
  \ref{fig:wallpresk3spec}b). Later, we identify the decorrelated
  fluid sources responsible for this low frequency peak in the PSD
  using spectral POD and relate to the observations of
  \cite{abe2005dns}.}

{\color{black}Figure \ref{fig:wallpresk3spec}c shows the spanwise
  wavenumber spectrum of the wall-pressure fluctuations in inner
  units. The spectrum at $Re_{\tau}=180$ agrees well with
  \cite{choi1990space}. Therefore, the spanwise resolution is
  sufficient enough to resolve the fine scale spanwise features of
  wall-pressure fluctuations.}


\subsection{Wall-pressure source distribution analysis}

\begin{figure}
  \centering
  \begin{subfigure}{.5\textwidth}
    \begin{overpic}[width=\linewidth]{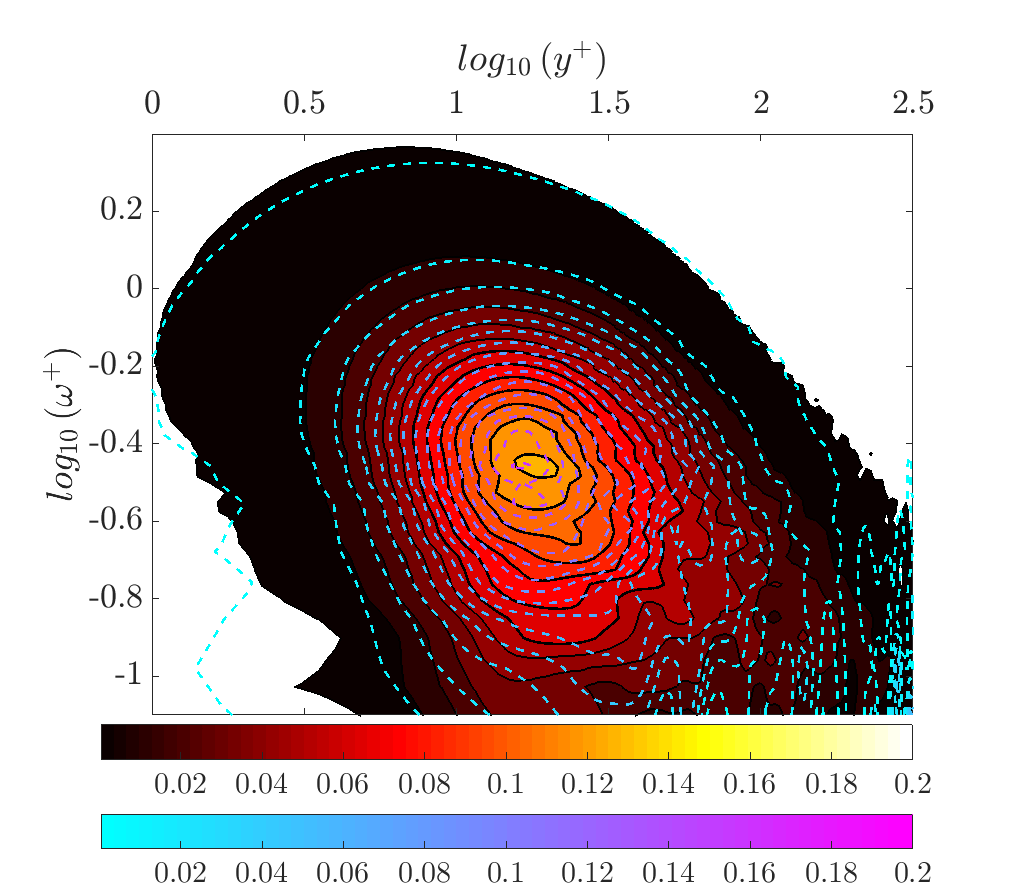}
      \put(2,4){$C_2$} \put(2,13){$C_1$} \put(2,80){\scriptsize$(a)$}
    \end{overpic}
  \end{subfigure}%
  \begin{subfigure}{.5\textwidth}
    \begin{overpic}[width=\linewidth]{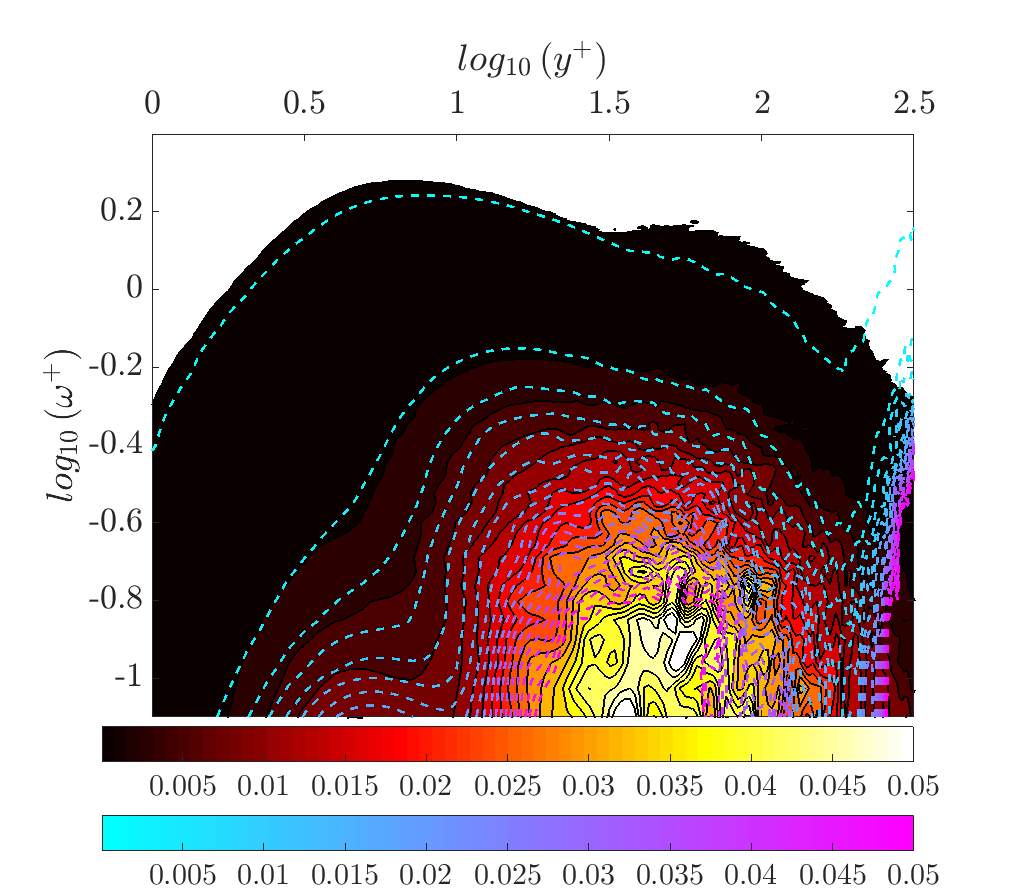}
      \put(2,4){$C_2$} \put(2,13){$C_1$} \put(2,80){\scriptsize$(b)$}
    \end{overpic}
  \end{subfigure}%
  \caption{a) Real part of premultiplied wall-pressure fluctuation - net source CSD
    ($y^+\omega^+Re(\Psi(y^+,\omega^+))/\langle p^2\rangle$) for
    $Re_{\tau}=400$ (black solid lines with filled contours with
    colormap $C_1$) and $180$ (line contours with colormap
    $C_2$). Contour lines are 20 equally spaced values between 4e-4
    and 2e-1. b) Premultiplied net source PSD
    $\left(y^+\omega^+\Gamma(y^+,y^+,\omega^+)/\langle
      \Gamma^2\rangle\right)$ for $Re_{\tau}=400$ (black solid lines
    with filled contours with colormap $C_1$) and $180$ (line contours
    with colormap $C_2$). Contour lines are 20 equally spaced values
    between 4e-5 and 5e-2}
  \label{fig:fracpremult}
\end{figure}
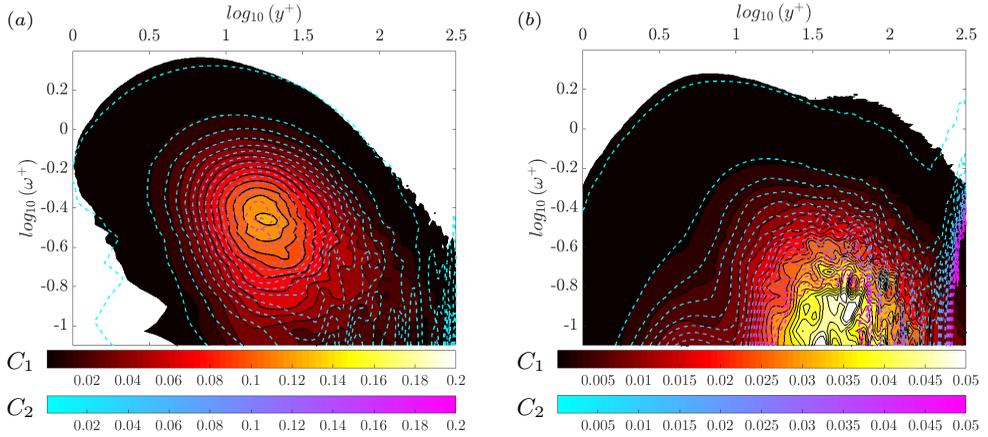

The wall-parallel plane that contributes the most to the wall-pressure PSD can be determined from the real part of the wall-pressure fluctuation - net source CSD $\Psi(y^+,\omega^+)$ (defined in section \ref{subsec:analysistheory}). Figure \ref{fig:fracpremult}a shows the wall-pressure fluctuation - net source CSD in premultiplied form normalized by the root mean square (RMS) of the wall-pressure fluctuations ($y^+\omega^+Re(\Psi^+(y^+,\omega^+))/\langle p^2\rangle^+$). $y^+$ is the distance from the wall in viscous units. The coordinates {\color{black}$(\omega_p^+,y_p^+)$} of the peak value in the contours occur at $(0.35,16.5)$ and $(0.35,18.4)$ for $Re_{\tau}=180$ and $400$, respectively. The frequency coordinate of the peak in the contour levels ($\omega^+=0.35$) is same as the premultiplied power spectra peak location shown in figure \ref{fig:wallprespowerspec}b. Therefore, the corresponding wall-normal coordinate yields the location of the wall-parallel plane that contributes the most to the premultiplied power spectra peak. Specifically, it is the cross-correlations with this dominant plane that contributes the most. This coincidence is not surprising since integrating figure \ref{fig:fracpremult}a in wall-normal direction yields figure \ref{fig:wallprespowerspec}b (normalized by $\langle p^2\rangle$). The wall-normal coordinate of the peak indicates that it is the correlations with the buffer region that contribute the most to the wall-pressure PSD at the Reynolds numbers consideblack. 

{\color{black}Even though the peak location differs slightly in inner
  units for the two $Re_{\tau}$, the main implication of this result
  is that the peak lies in the buffer region. Further, we cannot
  expect the same location of the peak for both $Re_{\tau}$. This is
  because the real part of the peak wall-pressure fluctuation - net source CSD includes the
  contribution from the correlations with the rest of the channel
  (since
  $\Psi(y_p^+,\omega)=\int_{-\delta}^{+\delta}\Gamma(y_p^+,y',\omega)\,\mathrm{dy'}$)
  and not just the inner layer. Therefore, the peak need not
  neccessarily scale in inner units. We believe that changing the
  Reynolds number would not affect this main finding. We expect the
  peak value of wall-pressure fluctuation - net source CSD to still occur in the buffer
  layer.} 

The phase difference between the wall-pressure and the dominant net
source obtained from the argument of $\Psi(y_p^+,\omega_p^+)$ is
$0.013\pi$ and $0.016\pi$ for $Re_{\tau}=180$ and $400$ respectively,
is very small. Hence, the dominant net sources and the wall-pressure
fluctuation are in phase with each other. The contour levels of the
normalized wall-pressure fluctuation - net source CSD plotted in figure \ref{fig:fracpremult}a almost
overlap in the range $\omega^+>0.3\sim 10^{-0.5}$. This indicates that
the high frequency contribution to the RMS scales in inner
units. However, in the near wall region ($y^+<10$), the overlap in the
contours is observed for a much larger frequency range
$\omega^+>0.16 \sim 10^{-0.8}$. This implies that for most of the
frequency range, the contribution to wall-pressure PSD from the
near-wall region scales in inner units.

\begin{figure}
  \centering
  \begin{subfigure}{.5\textwidth}
    \begin{overpic}[width=\linewidth]{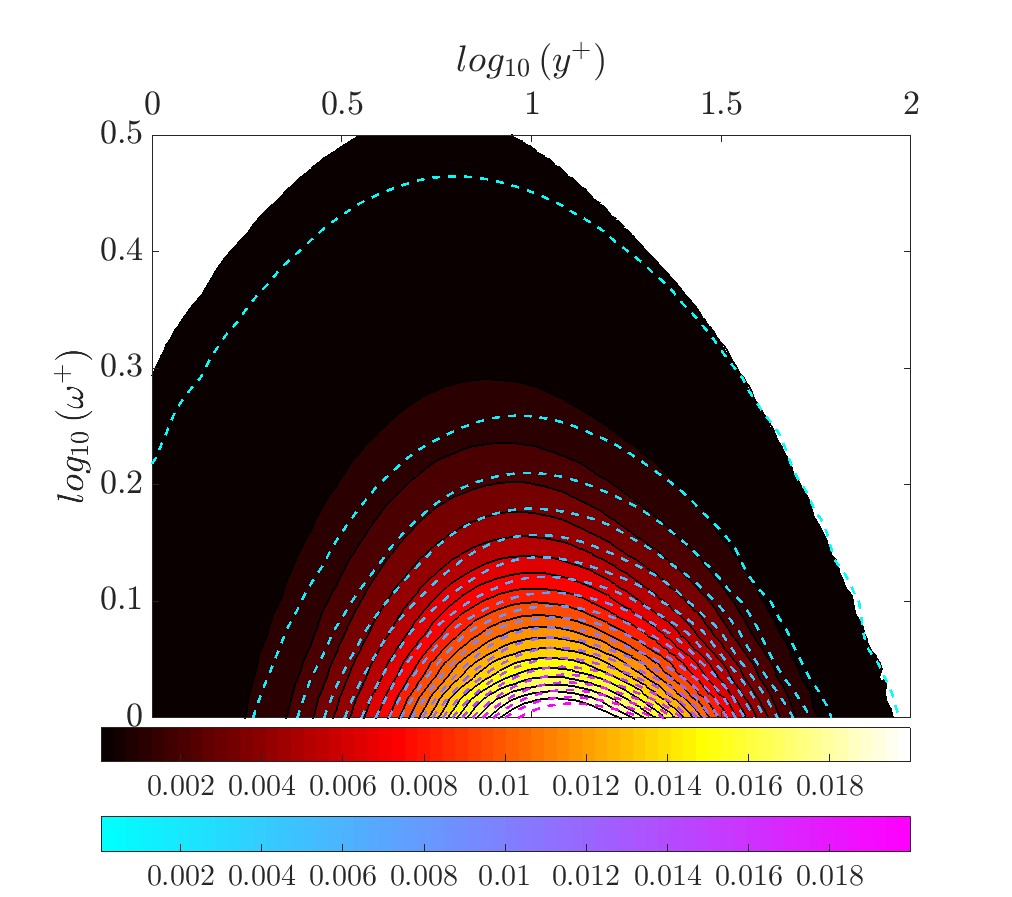}
      \put(2,4){$C_2$} \put(2,13){$C_1$} \put(2,80){\scriptsize$(a)$}
    \end{overpic}
  \end{subfigure}%
  \begin{subfigure}{.5\textwidth}
    \begin{overpic}[width=\linewidth]{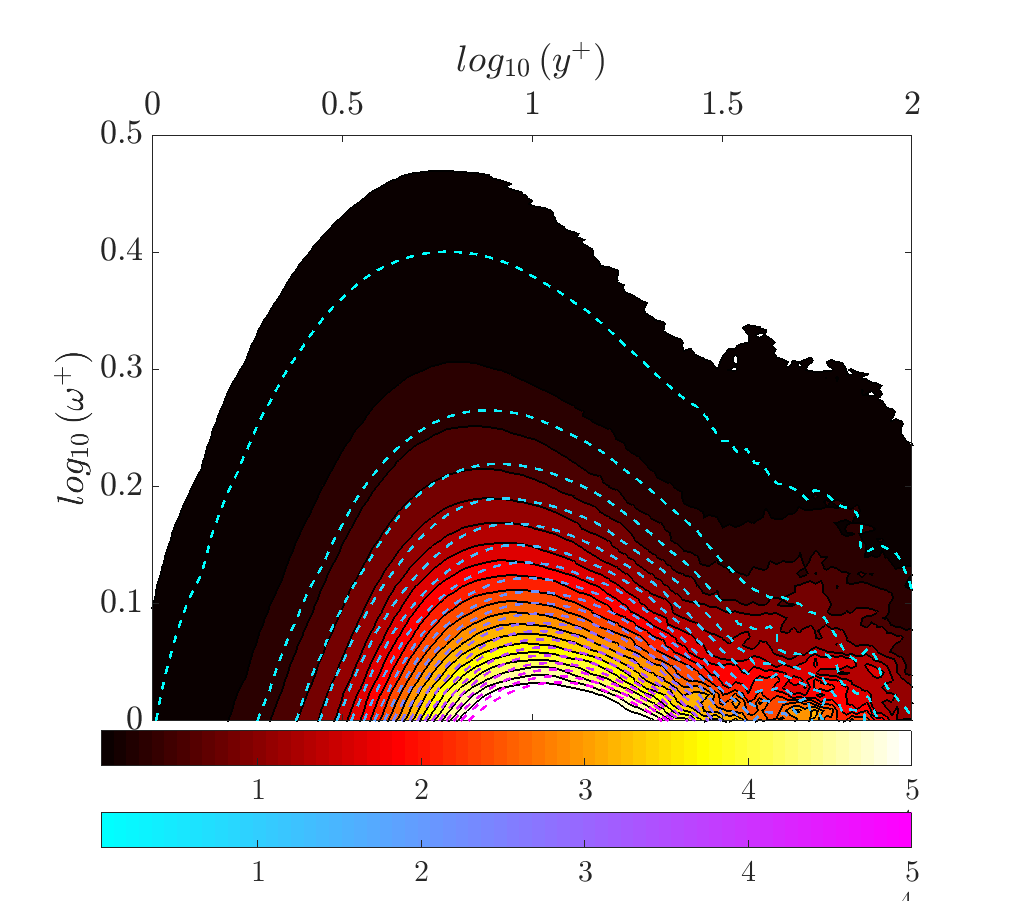}
      \put(2,4){$C_2$} \put(2,13){$C_1$} \put(2,80){\scriptsize$(b)$}
      \put(92,2){\scriptsize$\times 10^{-4}$}
      \put(92,11){\scriptsize$\times 10^{-4}$}
    \end{overpic}
  \end{subfigure}%
  \caption{a) Real part of the high frequency premultiplied wall-pressure fluctuation - net source CSD
    ($y^+\omega^+Re(\Psi(y^+,\omega^+))/\langle p^2\rangle$) for
    $Re_{\tau}=400$ (black solid lines with filled contours with
    colormap $C_1$) and $180$ (line contours with colormap
    $C_2$). Contour lines are 20 equally spaced values between 4e-5
    and 2e-2. b) High frequency premultiplied fractional net source
    power spectral density
    $y^+\omega^+\Gamma(y^+,y^+,\omega^+)/\langle \Gamma^2\rangle$ for
    $Re_{\tau}=400$ (black solid lines with filled contours with
    colormap $C_1$) and $180$ (line contours with colormap
    $C_2$). Contour lines are 20 equally spaced values between 4e-6
    and 2e-2}
  \label{fig:fracpremulthighfreq}
\end{figure}

\begin{figure}
  \centering
  \begin{subfigure}{.5\textwidth}
    \begin{overpic}[width=\linewidth]{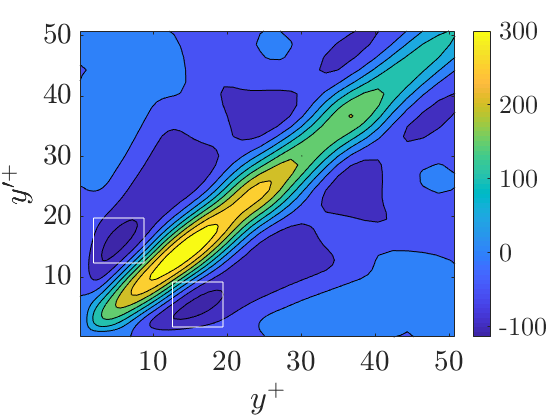}
    \end{overpic}
  \end{subfigure}%
  \begin{subfigure}{.5\textwidth}
    \begin{overpic}[width=\linewidth]{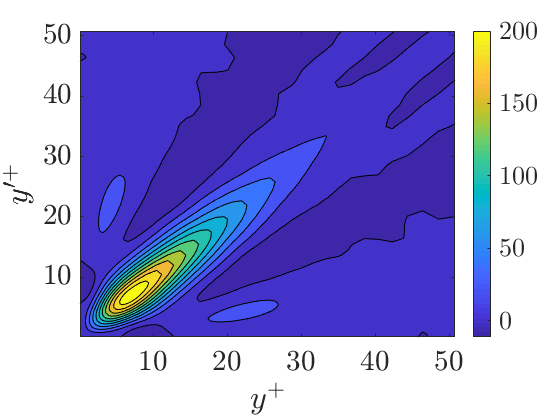}
    \end{overpic}
  \end{subfigure}%
  \caption{Real part of $\Gamma(r,s,\omega)$ (normalized by
    $\phi_{pp}(\omega)$) at a) $\omega^+\approx 0.35$ and b)
    $\omega^+\approx 1$ for $Re_{\tau}=180$.}
  \label{fig:diff_psi_psd}
\end{figure}

Next, we investigate whether the net source PSD can be used to infer
the location of the dominant source of wall-pressure fluctuation
instead of the wall-pressure fluctuation - net source CSD. Figure \ref{fig:fracpremult}b shows the
contours of the premultiplied net source PSD
$\Gamma(y^+,y^+,\omega^+)$ in fractional form for both
$Re_{\tau}$. The main contribution to the net source PSD is seen to be
from the region around $y^+\approx 30$ and at frequencies much lower
than $\omega^+\approx 0.35$. There is no signature of the distinct
premultiplied peak observed in figure \ref{fig:fracpremult}a. From
visual inspection at low frequencies ($\omega^+<1$), the shape of the
contours in figure \ref{fig:fracpremult}b do not have similar shape to
those in figure \ref{fig:fracpremult}a. However, at high frequencies
$\omega^+>1$, we observe from figure \ref{fig:fracpremulthighfreq}a
and \ref{fig:fracpremulthighfreq}b that the contour shapes near the
wall ($y^+<30$) are almost identical.  Therefore, the net source PSD
$\Gamma(y^+,y^+,\omega^+)$ is a good proxy for wall-pressure fluctuation - net source CSD
$\Psi(y^+,\omega^+)$ at high frequencies to obtain the pattern of the
net sources. {\color{black} The reason for this behavior can be
  understood from the near wall contours of the real part of the net
  source CSD shown in figure \ref{fig:diff_psi_psd}. Figures
  \ref{fig:diff_psi_psd}a and \ref{fig:diff_psi_psd}b show the
  contours at frequencies $\omega^+=0.35$ and $\omega^+=1$,
  respectively for $Re_{\tau}=180$. Clearly, the low frequency
  ($\omega^+=0.35$) contours show a large negatively cross correlated
  region around $(y^+,y'^+)=(5,15)$ (shown by white boxes). These
  dominant negative regions found at low frequencies contribute to the
  wall-pressure fluctuation - net source CSD (equation \ref{eqn:psidefn}) leading to different shapes
  compablack to net source PSD.  However, such negative regions are not
  present at the higher frequency $\omega^+=1$. Therefore, the wall-pressure fluctuation - net source
  CSD and the net source CSD have similar shapes near to the wall at
  high frequencies.}




\begin{figure}
  \centering
  \begin{subfigure}{.5\textwidth}
    \begin{overpic}[width=\linewidth]{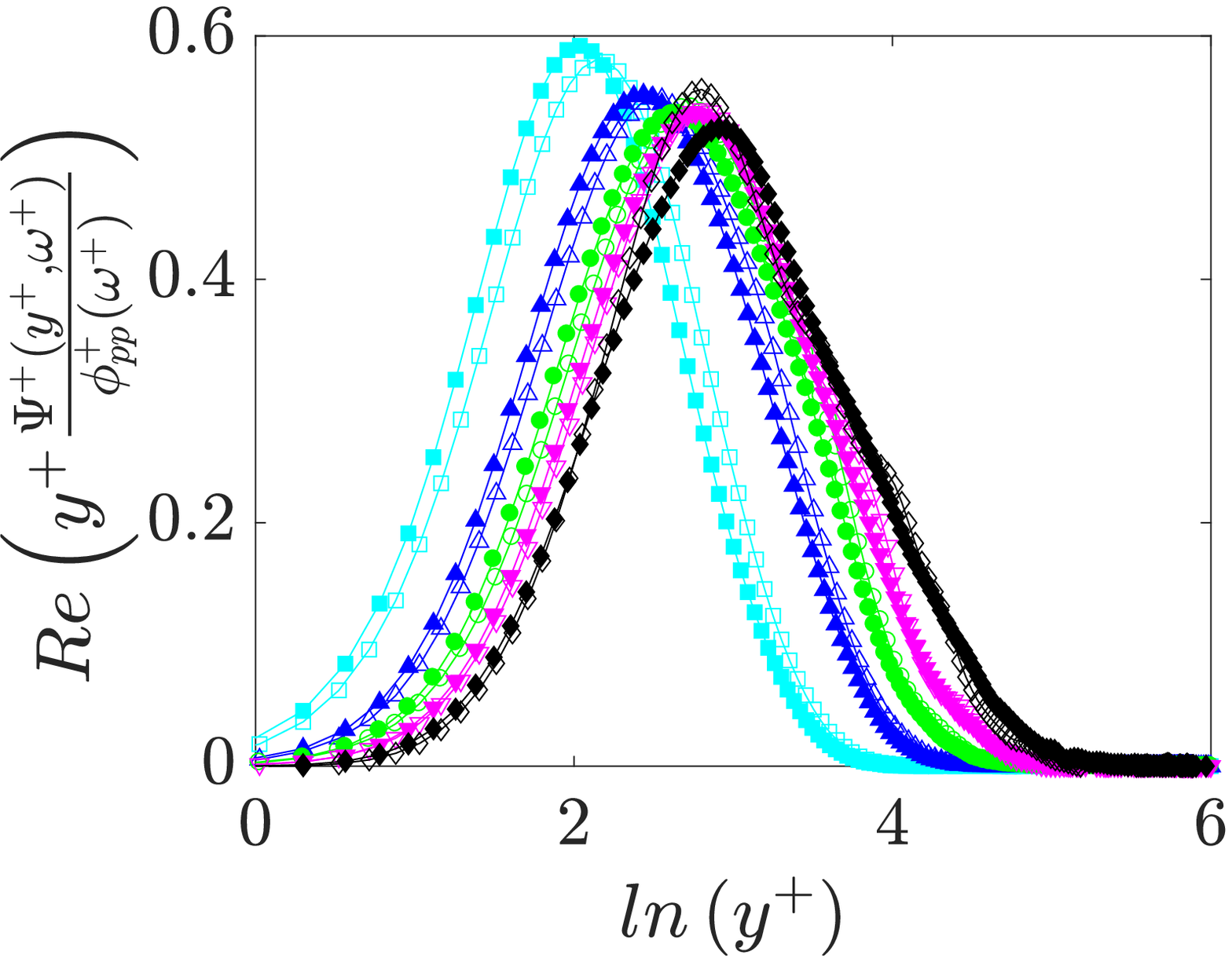}
      \put(1,70){$(a)$}
    \end{overpic}
  \end{subfigure}%
  \begin{subfigure}{.5\textwidth}
    \begin{overpic}[width=\linewidth]{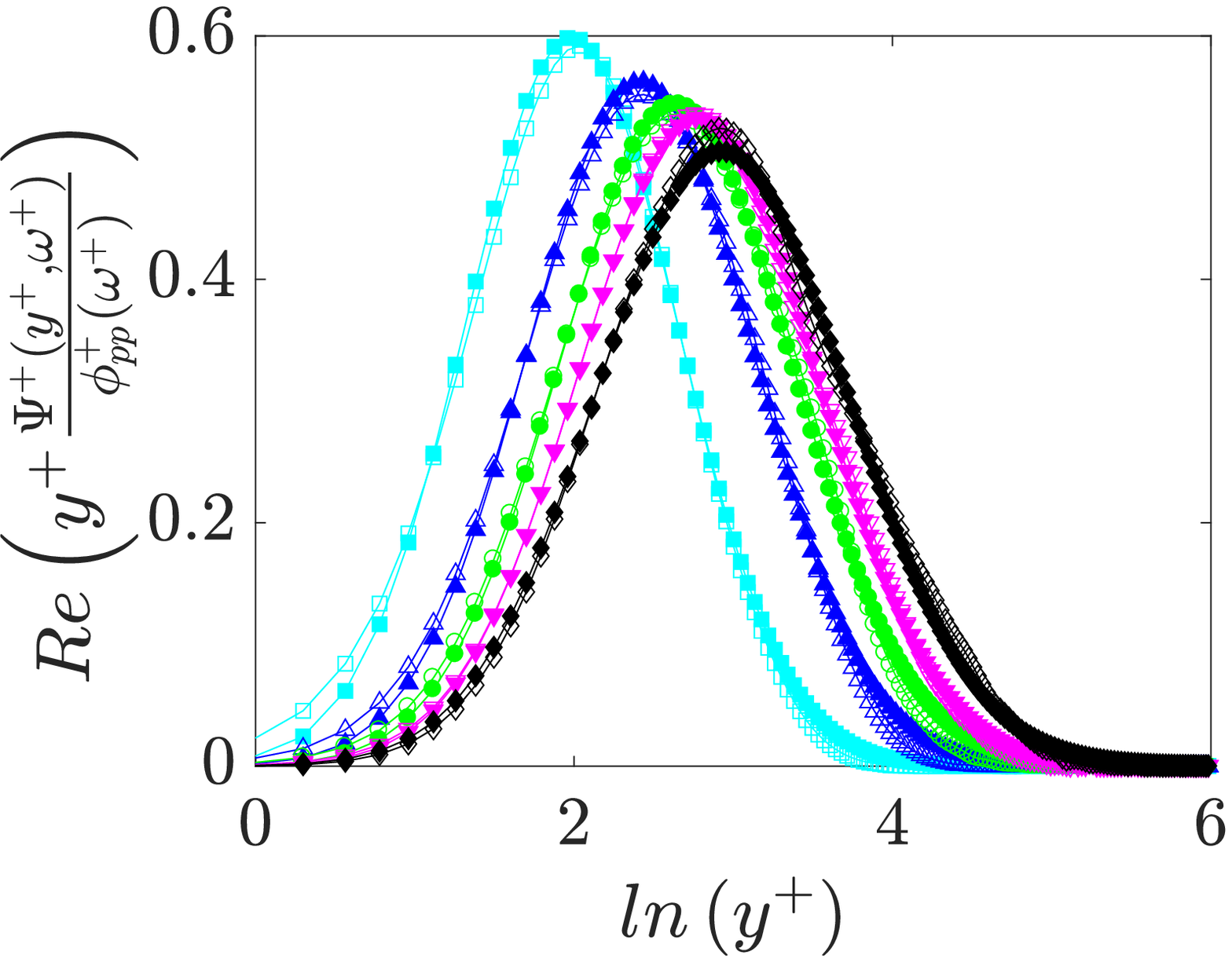}
      \put(1,70){$(b)$}
    \end{overpic}
  \end{subfigure}%

  \begin{subfigure}{.5\textwidth}
    \begin{overpic}[width=\linewidth]{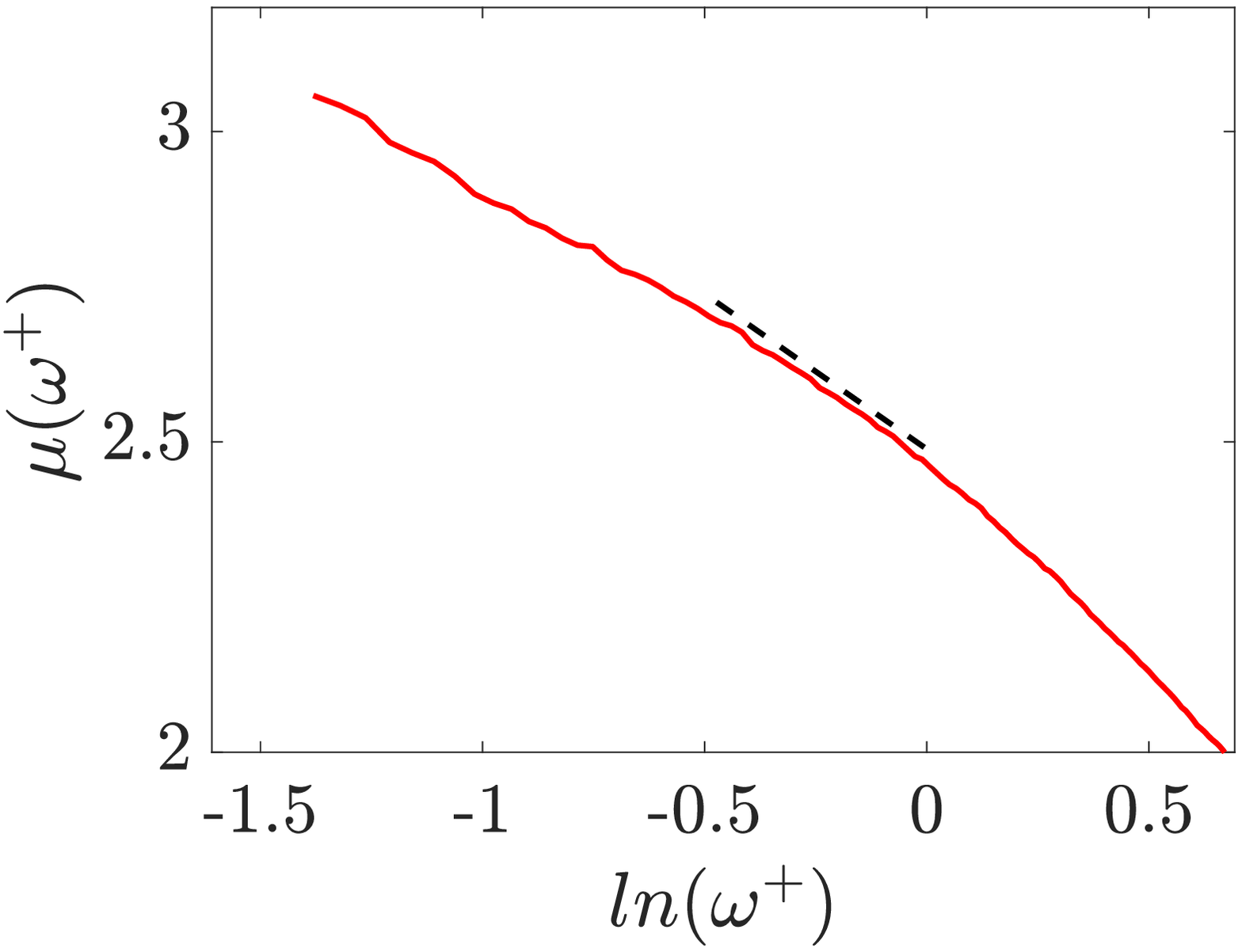}
      \put(1,70){$(c)$} \put(60,45){$-0.5$}
    \end{overpic}
  \end{subfigure}%
  \begin{subfigure}{.5\textwidth}
    \begin{overpic}[width=\linewidth]{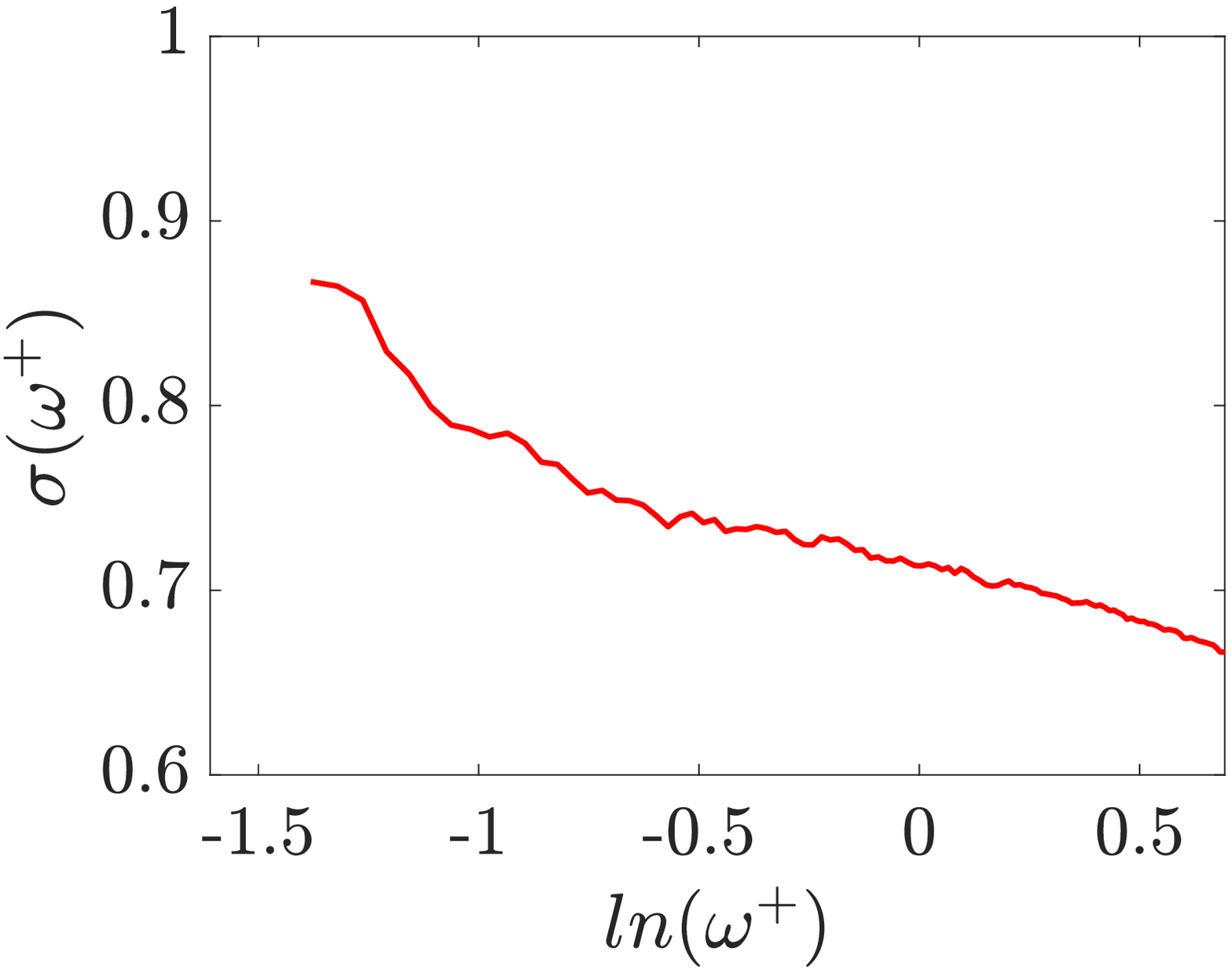}
      \put(1,70){$(d)$}
    \end{overpic}
  \end{subfigure}%
  \caption{a) Premultiplied wall-pressure fluctuation - net source CSD
    ($y^+\Psi^+(y^+,\omega^+)/\phi_{pp}^+(\omega^+)$) at different
    frequencies for $Re_{\tau}=180$ (empty markers) and $400$ (filled
    markers). b) Comparison of fitted log-normal Gaussians (filled
    markers) to the wall-pressure fluctuation - net source for $Re_{\tau}=400$ (empty markers). c)
    Variation of mean $\mu(\omega^+)$ and d) standard deviation
    $\sigma(\omega^+)$ of the fitted log-normal profile for
    $Re_{\tau}=400$. $\diabox0{cblack},\diabox1{cblack}$ :
    $\omega^+\approx 0.35$,
    $\uptrianbox0{cmagenta},\uptrianbox1{cmagenta}$ :
    $\omega^+\approx 0.5$, $\circbox0{cgreen},\circbox1{cgreen}$ :
    $\omega^+\approx 0.7$, $\trianbox0{cblack},\trianbox1{cblack}$ :
    $\omega^+\approx 1$, $\sqbox0{ccyan},\sqbox1{ccyan}$ :
    $\omega^+\approx 2$.}
  \label{fig:lognormal}
\end{figure}

The wall-pressure fluctuation - net source CSD (normalized with wall-pressure PSD) is plotted in
premultiplied form for selected frequencies between $\omega^+=0.35$
and $\omega^+=2$ in figure \ref{fig:lognormal}a. Due to the
normalization, each profile has unit area under it. From the figure,
we can observe that the curves for $Re_{\tau}=180$ and $400$ are very
close to each other for the different frequencies plotted. Further,
visual inspection shows that we can model the profiles using
log-normal function in $y^+$. Therefore, normalized log-normal
profiles of the form
\begin{equation}
  f(y^+,\omega^+)=\frac{1}{\sigma\left(\omega^+\right)\sqrt{2\pi}}exp\left(-\left(\frac{ln(y^+)-\mu\left(\omega^+\right)}{\sqrt{2}\sigma\left(\omega^+\right)}\right)^2\right)
\end{equation}
are fitted to the $Re_{\tau}=400$ data for different $\omega^+$ using
nonlinear least squares fit and plotted in figure
\ref{fig:lognormal}b.

The mean and standard deviation of the fitted log-normal curves
characterize the location and the width of the dominant net source
respectively as a function of frequency. The correlation between the
planes contained in this width have a sizeable contribution to
wall-pressure PSD. Figure \ref{fig:lognormal}c and
\ref{fig:lognormal}d show the mean ($\mu(\omega^+)$) and standard
deviation ($\sigma(\omega^+)$) as a function of frequency
respectively. We define the location of the dominant net source
$y^+_p(\omega^+)$ as $y^+_p(\omega^+)=exp(\mu(\omega^+))$. From figure
\ref{fig:lognormal}c, we observe that the location of the dominant net
source moves closer to the wall with increase in frequency through a
power law dependence $y^+_p\sim \left(\omega^+\right)^{m}$. The value
of $m$ depends on the frequency range. In the low
$(-1.5<ln(\omega^+)<0.5)$, mid ($-0.5<ln(\omega^+)<0$) and high
$(\omega^+>1)$ frequency range, the value of the exponent $m$ is
larger than $-0.5$, equal to $-0.5$ and smaller than $-0.5$
respectively.

\begin{figure}
  \centering
  \begin{subfigure}{.5\textwidth}
    \begin{overpic}[width=\linewidth]{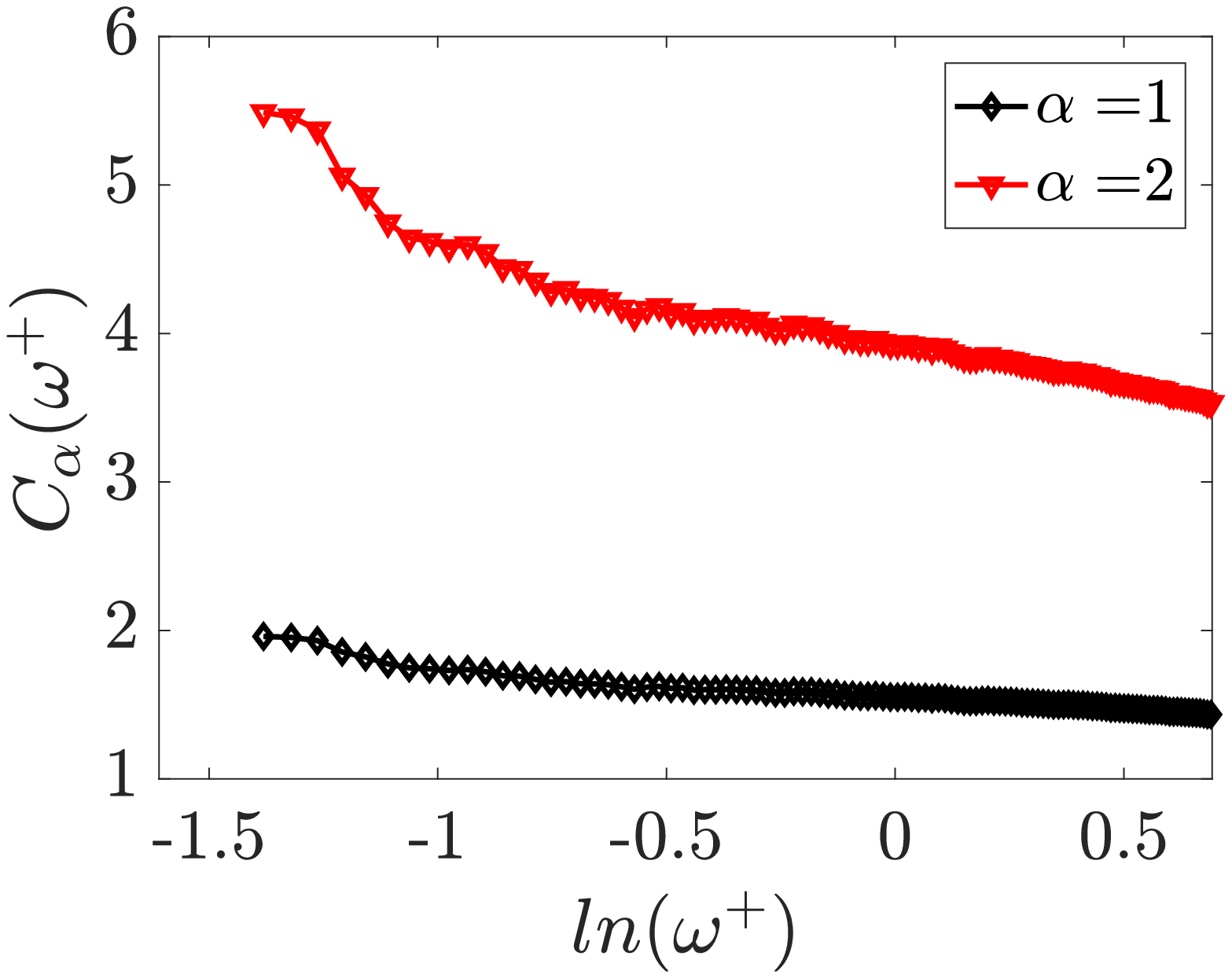}
      \put(1,70){\scriptsize$(a)$}
    \end{overpic}
  \end{subfigure}%
  \begin{subfigure}{.5\textwidth}
    \begin{overpic}[width=\linewidth]{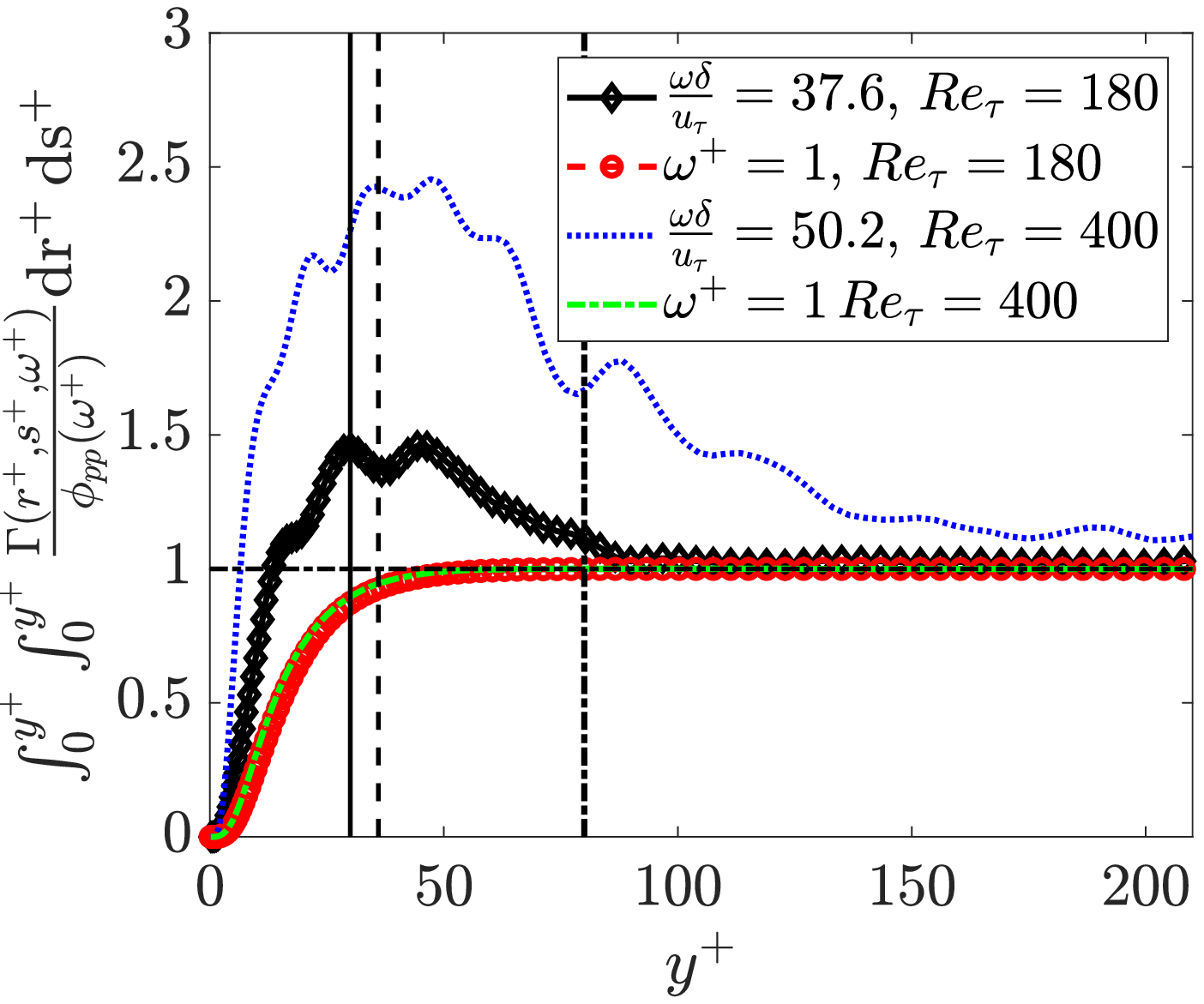}
      \put(1,70){\scriptsize$(b)$}
    \end{overpic}
  \end{subfigure}%
  \caption{a) Variation of $C_{\alpha}(\omega^+)$ for
    $Re_{\tau}=400$. b) Partial wall-pressure fluctuation spectra from
    sources that extend from the wall to a particular $y^+$ for
    $Re_{\tau}=180$ and $400$ in near wall region. The vertical solid,
    dashed, dash-dotted lines denote $y^+=30$, $y/\delta=0.2$ for
    $Re_{\tau}=180$ and $y/\delta=0.2$ for $Re_{\tau}=400$,
    respectively. The horizontal dash-dotted line denotes partial
    contribution equal to $1$.}
  \label{fig:partial_gamma}
\end{figure}

Figure \ref{fig:lognormal}d shows that the standard deviation of the
log-normal profiles decreases with increasing frequency. We use the
standard deviation profile to show that for $\omega^+>e^{-1}$, the
width of the dominant net source is proportional to its location. We
define the wall-normal width of the net source
$\Delta y^+(\omega^+;\alpha,\sigma)$ as
\begin{equation}
  \begin{split}
    \Delta
    y^+(\omega^+;\alpha,\sigma)=y^+_{max}(\omega^+;\alpha,\sigma)-y^+_{min}(\omega^+;\alpha,\sigma),
  \end{split}
\end{equation}
where $y^+_{max},y^+_{min}$ and $y^+_p$ are related as
\begin{equation}
  \begin{split}
    ln\left(y^+_{max}(\omega^+;\alpha,\sigma)\right)-ln\left(y^+_{min}(\omega^+;\alpha,\sigma)\right)
    = 2\alpha\sigma(\omega^+),\\
    ln\left(y^+_{p}(\omega^+)\right)-ln\left(y^+_{min}(\omega^+;\alpha,\sigma)\right)
    = \alpha\sigma(\omega^+).
  \end{split}
\end{equation}
The parameter $\alpha$ is the proportion of the standard deviation
used to define the width of the net source. Using the above
expressions, the width $\Delta y^+(\omega^+;\alpha,\sigma)$ can be
shown to be
\begin{equation}
  \begin{split}
    \Delta y^+(\omega^+;\alpha,\sigma)&=C(\omega^+;\alpha,\sigma) y^+_{p}(\omega^+),\\
    C(\omega^+;\alpha,\sigma)&=\left(e^{\alpha\sigma(\omega^+)}-e^{-\alpha\sigma(\omega^+)}\right).
  \end{split}
\end{equation}
The variation of $C_{\alpha}(\omega^+)$ for $\alpha=1,2$ using
$Re_{\tau}=400$ data is shown in figure \ref{fig:partial_gamma}a. The
proportionality constant is observed to vary slowly for
$\omega^+>ln(-1)$. Hence, in this frequency range, the width of the
dominant net source is proportional to its location.

The contribution of the interaction between the net sources in the
inner and overlap/outer region to wall-pressure PSD can be
investigated using the wall-pressure fluctuation - net source CSD. Figure \ref{fig:partial_gamma}b
shows the partial contribution (normalized by the wall-pressure PSD)
$\int_0^{y^+}\int_0^{y^+}\Gamma(r^+,s^+,\omega^+)/\phi_{pp}(\omega^+)\mathrm{dr^+}\mathrm{ds^+}$
from the net sources contained between the wall and a given $y^+$ for
two selected frequencies. {\color{black}At the low frequency
  wall-pressure linear PSD peak (which is $\omega\delta/u_{\tau}=37.6$
  and $\omega\delta/u_{\tau}=50.2$ for $Re_{\tau}=180$ and $400$,
  respectively) }, we observe that the partial contribution first
increases and then decreases. However, a monotonically increasing
behavior is observed for the high frequency. In order to investigate
the implication of the non-monotonic low frequency behavior, we split
the domain {\color{black}$0<y/\delta<1$} into an inner region
$0<y^+<30$, an outer/overlap region $30<y^+<Re_{\tau}$. The
contribution to wall-pressure PSD from sources within
{\color{black}$y/\delta=1$} can then be accordingly split as
\begin{align}
  \int_{0}^{Re_{\tau}}\int_{0}^{Re_{\tau}}\frac{\Gamma(r^+,s^+,\omega^+)}{\phi_{pp}(\omega^+)}\,\mathrm{dr^+}\,\mathrm{ds^+}=\int_{0}^{30}\int_{0}^{30}\frac{\Gamma(r^+,s^+,\omega^+)}{\phi_{pp}(\omega^+)}\,\mathrm{dr^+}\,\mathrm{ds^+}+\\ \nonumber
  \int_{30}^{Re_{\tau}}\int_{30}^{Re_{\tau}}\frac{\Gamma(r^+,s^+,\omega^+)}{\phi_{pp}(\omega^+)}\,\mathrm{dr^+}\,\mathrm{ds^+}+\\ \nonumber
  2Re\left(\int_{0}^{30}\int_{30}^{Re_{\tau}}\frac{\Gamma(r^+,s^+,\omega^+)}{\phi_{pp}(\omega^+)}\,\mathrm{dr^+}\,\mathrm{ds^+}\right).
\end{align}
From figure \ref{fig:partial_gamma}b, we observe that at the lower
frequency, this contribution from sources within
{\color{black}$y/\delta=1$} is smaller than the inner region
contribution
$\int_{0}^{30}\int_{0}^{30}\frac{\Gamma(r^+,s^+,\omega^+)}{\phi_{pp}(\omega^+)}\,\mathrm{dr^+}\,\mathrm{ds^+}$. Therefore,
\begin{equation}
  \int_{30}^{Re_{\tau}}\int_{30}^{Re_{\tau}}\frac{\Gamma(r^+,s^+,\omega^+)}{\phi_{pp}(\omega^+)}\,\mathrm{dr^+}\,\mathrm{ds^+} \leq -2Re\left(\int_{0}^{30}\int_{30}^{Re_{\tau}}\frac{\Gamma^+(r^+,s^+,\omega^+)}{\phi_{pp}(\omega^+)}\,\mathrm{dr^+}\,\mathrm{ds^+}\right).
\end{equation}
Note the the left hand side of the above inequality is a positive real
number. This indicates that i) the contribution from the
cross-correlations between the inner and the overlap/outer region
dominantes the contribution from the outer/overlap region alone, ii)
the phase difference between the net sources in these two regions is
pblackominantly in the range $\pi/2$ to $\pi$ or $-\pi$ to $-\pi/2$. In
other words, a positive (or negative) low frequency event in the
near-wall region is pblackominantly correlated with a negative (or
positive) low frequency event in the overlap/outer region. Therefore,
the observed non-monotonic behavior at low frequencies implies a
dominant interaction between the net sources the inner and outer
regions of the channel at such frequencies. Such inner-outer
interaction at long streamwise wavelengths has been previously
observed for the streamwise velocity fluctuations by
\cite{del2003spectra}, \cite{morrison2007interaction}, and is the
reason for the mixed scaling \citep{de2000reynolds} of the streamwise
velocity RMS peak in wall-bounded flows.

{\color{black} We further investigate the fractional contribution of the
  wall-pressure sources in the inner ($y^+<30$), overlap
  ($30<y^+<0.2Re_{\tau}$) and outer region ($0.2<y/\delta<1$) and
  their cross-correlations to the wall-pressure PSD by splitting
  $\int_{0}^{Re_{\tau}}\int_{0}^{Re_{\tau}}\frac{\Gamma(r^+,s^+,\omega^+)}{\phi_{pp}(\omega^+)}\,\mathrm{dr^+}\,\mathrm{ds^+}$
  into the sum,
\begin{equation} \label{eqn:frac_cont}
\begin{split}
  &\int_{0}^{Re_{\tau}}\int_{0}^{Re_{\tau}}\frac{\Gamma(r^+,s^+,\omega^+)}{\phi_{pp}(\omega^+)}\,\mathrm{dr^+}\,\mathrm{ds^+}=\\
  &\underbrace{\int_{0}^{30}\int_{0}^{30}\frac{\Gamma(r^+,s^+,\omega^+)}{\phi_{pp}(\omega^+)}\,\mathrm{dr^+}\,\mathrm{ds^+}}_{C_{11}}+\\
  &\underbrace{\int_{30}^{0.2Re_{\tau}}\int_{30}^{0.2Re_{\tau}}\frac{\Gamma(r^+,s^+,\omega^+)}{\phi_{pp}(\omega^+)}\,\mathrm{dr^+}\,\mathrm{ds^+}}_{C_{22}}+\underbrace{\int_{0.2Re_{\tau}}^{Re_{\tau}}\int_{0.2Re_{\tau}}^{Re_{\tau}}\frac{\Gamma(r^+,s^+,\omega^+)}{\phi_{pp}(\omega^+)}\,\mathrm{dr^+}\,\mathrm{ds^+}}_{C_{33}}+\\
  &2Re\bigg(\underbrace{\int_{0}^{30}\int_{30}^{0.2Re_{\tau}}\frac{\Gamma(r^+,s^+,\omega^+)}{\phi_{pp}(\omega^+)}\,\mathrm{dr^+}\,\mathrm{ds^+}}_{C_{12}}+\underbrace{\int_{30}^{0.2Re_{\tau}}\int_{0.2Re_{\tau}}^{Re_{\tau}}\frac{\Gamma(r^+,s^+,\omega^+)}{\phi_{pp}(\omega^+)}\,\mathrm{dr^+}\,\mathrm{ds^+}}_{C_{23}}+\\
&\underbrace{\int_{0.2Re_{\tau}}^{Re_{\tau}}\int_{0}^{30}\frac{\Gamma(r^+,s^+,\omega^+)}{\phi_{pp}(\omega^+)}\,\mathrm{dr^+}\,\mathrm{ds^+}}_{C_{31}}\bigg).
\end{split}
\end{equation}
Table \ref{tab:separatecont} shows the value of each term in the right
hand side of the above equation at the two $Re_{\tau}$ for the same
frequencies chosen in figure \ref{fig:partial_gamma}b. For the lower
frequency, we observe that the magnitude of the contribution from the
cross-correlations between the regions is comparable to the
contribution within the regions. However, at high frequency, the
contribution within each region dominates over the cross-correlation
between the regions. The real part of the cross-correlations is
negative at the lower frequency for $Re_{\tau}=180$. As discussed in
the previous paragraph, this implies that the phase difference of the
wall-pressure sources in the different regions lie in the range
$\pi/2$ to $\pi$ or $-\pi/2$ to $-\pi$. For the higher $Re_{\tau}$,
except the inner and outer region $(2Re(C_{31}))$, the phase
difference between all the other regions lie in the same range as the
lower $Re_{\tau}$. Overall, we observe that the cross-correlation
between the wall-pressure sources present in the inner, overlap and
outer regions are important contributors to the PSD at low frequency
but not at high frequency.

Further, this framework can be used to identify the location of the
dominant sources that lead to the $\omega^{-1}$ behavior of the
wall-pressure PSD in the mid frequency range (observed at very high
Reynolds numbers). \cite{farabee1991spectral} noted that the
$\omega^{-1}$ behavior is responsible for the logarithmic dependence
of the wall-pressure RMS on Reynolds number
\citep{abe2005dns,hu2006wall,jimenez2008turbulent}.

}

\begin{table}
  \begin{center}
    \def~{\hphantom{0}}
    \begin{tabular}{lcccc}
      Term  & \multicolumn{2}{c}{$Re_{\tau}=180$} & \multicolumn{2}{c}{$Re_{\tau}=400$}\\
            & $\frac{\omega\delta}{u_{\tau}}=36.7$ & $\omega^+=1$ & $\frac{\omega\delta}{u_{\tau}}=50.2$ & $\omega^+=1$ \\[3pt]
      $C_{11}$ &  1.45 &  0.87&  2.3 &  0.9 \\
      $C_{22}$ &  0.67 &  0.02 &  3.03 &  0.04\\
      $C_{33}$ &  0.87 &  0.02 &  1.08 &  0.001\\
      $2Re(C_{12})$ &  -0.76&   0.038 & -3.65 & 0.06 \\
      $2Re(C_{23})$ &  -0.44&   0.005 & -1.83 & -0.003  \\
      $2Re(C_{31})$ &  -0.76&   0.039 & 0.09 &  0.0006
    \end{tabular}
    \caption{Fractional contribution of the inner, overlap and outer region to the wall-pressure fluctuation PSD. For definition of $C_{11}$, $C_{22}$, $C_{33}$, $C_{12}$, $C_{23}$ and $C_{31}$, see equation \ref{eqn:frac_cont}.}
    \label{tab:separatecont}
  \end{center}
\end{table}

\subsection{Spectral POD of net source CSD}


{\color{black}Before we investigate the spectral POD modes of the
  net source CSD, we first examine the relevance of the modes to
  wall-pressure fluctuation. We can decompose the wall-pressure
  fluctuation $p(x,0,z,t)$ at a typical point $(x,z)$ on the wall by
  expressing its Fourier transform $\hat{p}(x,0,z,\omega)$ (equation
  \ref{eqn:phatxzomega}) in terms of the spectral POD modes. We have
  \begin{align}
    p(x,0,z,t)=&\int_{-\infty}^{+\infty}\hat{p}(x,0,z,\omega)e^{i\omega t}\mathrm{d}\omega,\\
    p(x,0,z,t)=&\int_{-\infty}^{+\infty}\left(\int_{-\delta}^{+\delta}\hat{f}_{G}(x,y,z,\omega)\mathrm{dy}\right)e^{i\omega t}\mathrm{d}\omega.\nonumber \\
  \end{align}
  We use the decomposition in equation \ref{eqn:netsourceexpansion} to
  express $\hat{f}_G(x,y,z,\omega)$ in terms of the spectral POD modes
  and obtain
  \begin{align}
    p(x,0,z,t)=&\int_{-\infty}^{+\infty}\sum_{j=1}^{\infty}\alpha_i(x,z,\omega)\left(\int_{-\delta}^{+\delta}\Phi^*_j(y,\omega)\mathrm{dy}\right)e^{i\omega t}\mathrm{d\omega}, \label{eqn:spodrellasteqn}
  \end{align}
  Rearranging the integral and writing $\Phi_j(y,\omega)$ as
  $|\Phi_j(y,\omega)|e^{-i\angle\Phi_j(y,\omega)}$, we obtain
  \begin{equation}
    p(x,0,z,t)=\int_{-\delta}^{+\delta}\int_{-\infty}^{+\infty}\sum_{j=1}^{\infty}\alpha_j(x,z,\omega)|\Phi_j(y,\omega)|e^{i\left(-\angle\Phi_j(y,\omega)+\omega t\right)}\mathrm{d\omega}\mathrm{dy}.
  \end{equation}}
The above equation expresses the wall-pressure fluctuation as a
contribution from each spectral POD mode. Recall, that the individual
contributions are decorrelated, i.e.,
$\langle\alpha_i(x,z,\omega)\alpha^*_j(x,z,\omega_o)\rangle=\lambda_i(\omega)\delta_{ij}\delta(\omega-\omega_o)$,
where $\delta$ is the Dirac delta function. Note that since we
integrate over all wavenumbers, the contribution of coherent
structures of all length scales is included.

The wall-normal phase velocity of the net sources represented by
$i^{th}$ spectral POD mode can be quantified as a function of the
wall-normal distance using the phase $\angle\Phi_i(y,\omega)$. We
define a local wall-normal phase velocity $c^+_i(y^+,\omega^+)$ in
viscous units as
\begin{equation} \label{eqn:wallnormalphasespeed}
  c^+_i(y^+,\omega^+)=\omega^+/k_i^+\left(y^+,\omega^+\right),
\end{equation} 
where the local wavenumber $k_i^+(y^+,\omega^+)$ is defined as
$k_i^+(y^+,\omega^+)=\frac{\partial
  \angle\Phi_i(y^+,\omega^+)}{\partial y^+}$. {\color{black}Note
  that negative phase velocity indicates an enclosed wave travelling
  towards the wall and vice versa}. Also, this is similar to
estimating the instantaneous frequency of a temporal signal using
Hilbert transform \citep{huang2014hilbert}.



{\color{black} 

  We found that, for a wide range of frequencies, setting $\beta$ (equation \ref{eqn:geninnerprod}) to $0.1$ gives a dominant spectral POD mode ($\Phi_1$) that contributes to all of the wall-pressure PSD (see figure \ref{fig:spodeigwt_wmat_alpha_0p1}). This observation is consistent with the discussion after equation \ref{eqn:sumofeig}. Note that the lowest frequency in figure \ref{fig:spodeigwt_wmat_alpha_0p1} corresponds to the low-frequency peak in the linear PSD. Also, $\omega^+=0.35$ is the location of the premultiplied PSD peak. Therefore, the dominant modes at these peak frequencies represent the decorrelated source responsible for the peaks.

  Further, the dominant mode represents the active part of the net source Fourier transform ($\hat{f}_G(x,y,z,\omega)$). It is active in the sense that it contributes to the entire PSD. The remaining portion of $\hat{f}_G(x,y,z,\omega)$ is inactive in the sense that it does not contribute to wall-pressure PSD. The suboptimal spectral POD modes comprise this inactive portion. Essentially, the contribution of the suboptimal modes from different wall-normal locations undergo destructive interference resulting in zero net contribution. Since the active and inactive parts of $\hat{f}_G(x,y,z,\omega)$ stem from different modes, they are decorrelated. 

  Separating the active and inactive parts of $\hat{f}_G(x,y,z,\omega)$ in equation \ref{eqn:netsourceexpansion}, we have
  \begin{equation}
    \begin{split}
      \hat{f}_g(x,y,z,\omega)&=\alpha_1(x,z,\omega)\Phi^*_1(y,\omega)+I(x,y,z,\omega),\\
      I(x,y,z,\omega)&=\sum_{j=2}^{\infty}\alpha_j(x,z,\omega)\Phi^*_j(y,\omega),
    \end{split}
  \end{equation}
  where $\alpha_1(x,z,\omega)\Phi^*_1(y,\omega)$ and $I(x,y,z,\omega)$ are the active and inactive portions of $\hat{f}_G(x,y,z,\omega)$, respectively. Correlating the two, we obtain
  \begin{equation}
    \begin{split}
      \langle \alpha^*_1(x,z,\omega)\Phi_1(r,\omega) I(x,s,z,\omega_o)  \rangle &= \sum_{j=2}^{\infty}\langle \alpha^*_1(x,z,\omega)\alpha_j(x,z,\omega) \rangle\Phi_1(r,\omega)\Phi^*_j(s,\omega_o) \\
      &=\sum_{j=2}^{\infty}\lambda_1\delta_{j1}\Phi_1(r,\omega)\Phi^*_j(s,\omega_o)\,(\text{using eqn.}\,\ref{eqn:netsourceexpansion})\\
      &=0.
    \end{split}
  \end{equation}
  Therefore, both parts are decorrelated. Note that $I(x,y,z,\omega)$, the inactive part, is orthogonal to the eigenfunction $\bar{\Phi}^*_1(y,\omega)$ (equation \ref{eqn:eigenvalueproblem}) in the $L^2$ inner product. Decreasing $\beta$ to even smaller values does not affect the mode shape or the eigenvalues for the frequencies in figure \ref{fig:spodeigwt_wmat_alpha_0p1}.

  Since we use a Poisson inner product, the dominant spectral POD mode need not be energetically dominant. In orther words, it need not contribute the most to the integrated net source PSD ($\int_{-\delta}^{+\delta}\Gamma(r,r,\omega)\,\mathrm{dr}$). Figure \ref{fig:spodeigpsdwt_wmat_alpha_0p1} shows this behavior for low frequencies. In the figure, $\bar{\lambda}_j(\omega)=\lambda_j(\omega)|\int_{-\delta}^{+\delta}\Phi_j(y,\omega)\,\mathrm{dy}|^2$ is the contribution of the $j^{th}$ mode to the integrated PSD. We observe that the fractional contribution of the dominant mode increases with frequency. At $\omega^+=1$, the dominant spectral POD mode is the energetically dominant mode.


  \begin{figure}
    \centering
    \begin{subfigure}{.5\textwidth}
      \begin{overpic}[width=\linewidth]{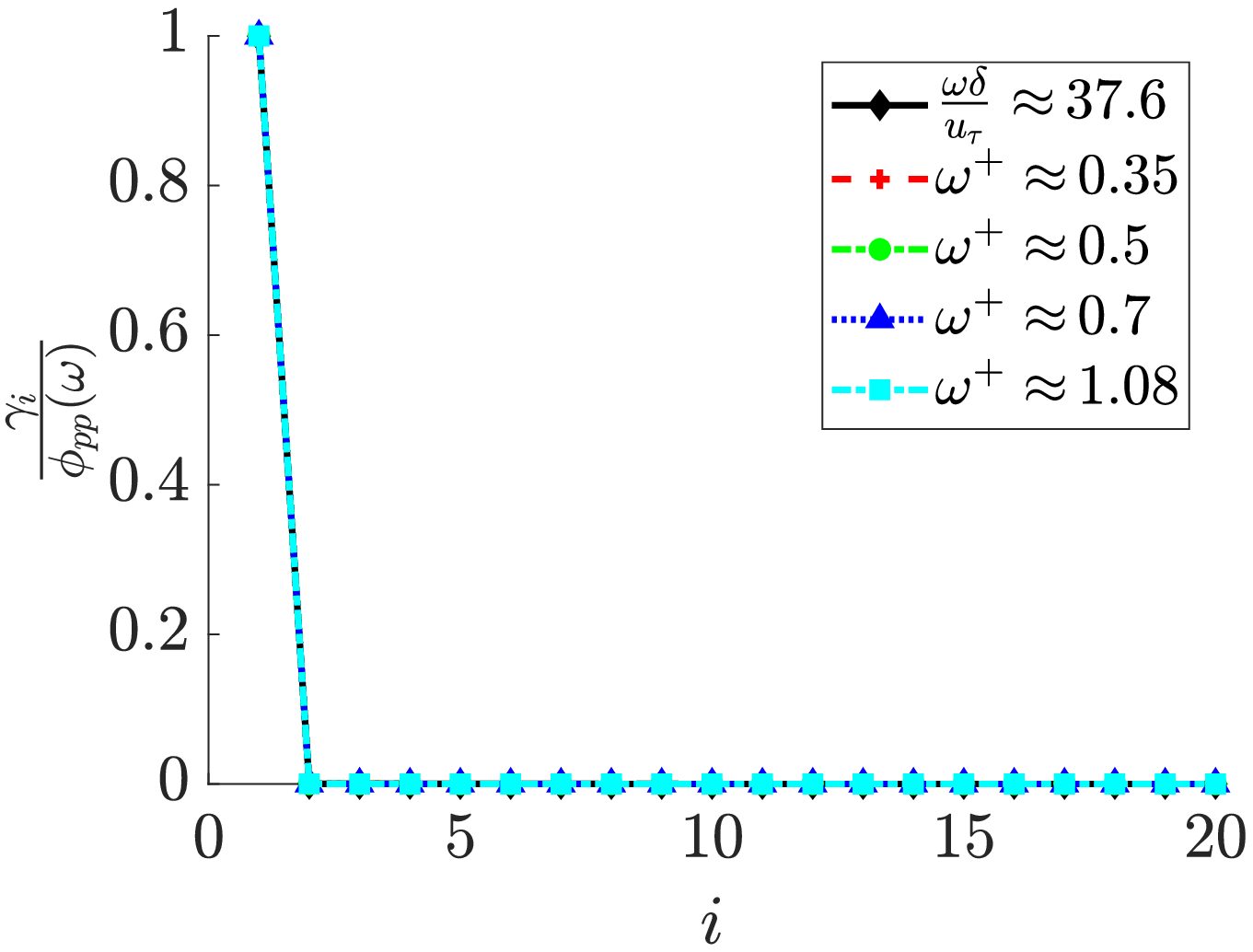}
        \put(1,65){\scriptsize$(a)$}
      \end{overpic}
    \end{subfigure}%
    \begin{subfigure}{.5\textwidth}
      \begin{overpic}[width=\linewidth]{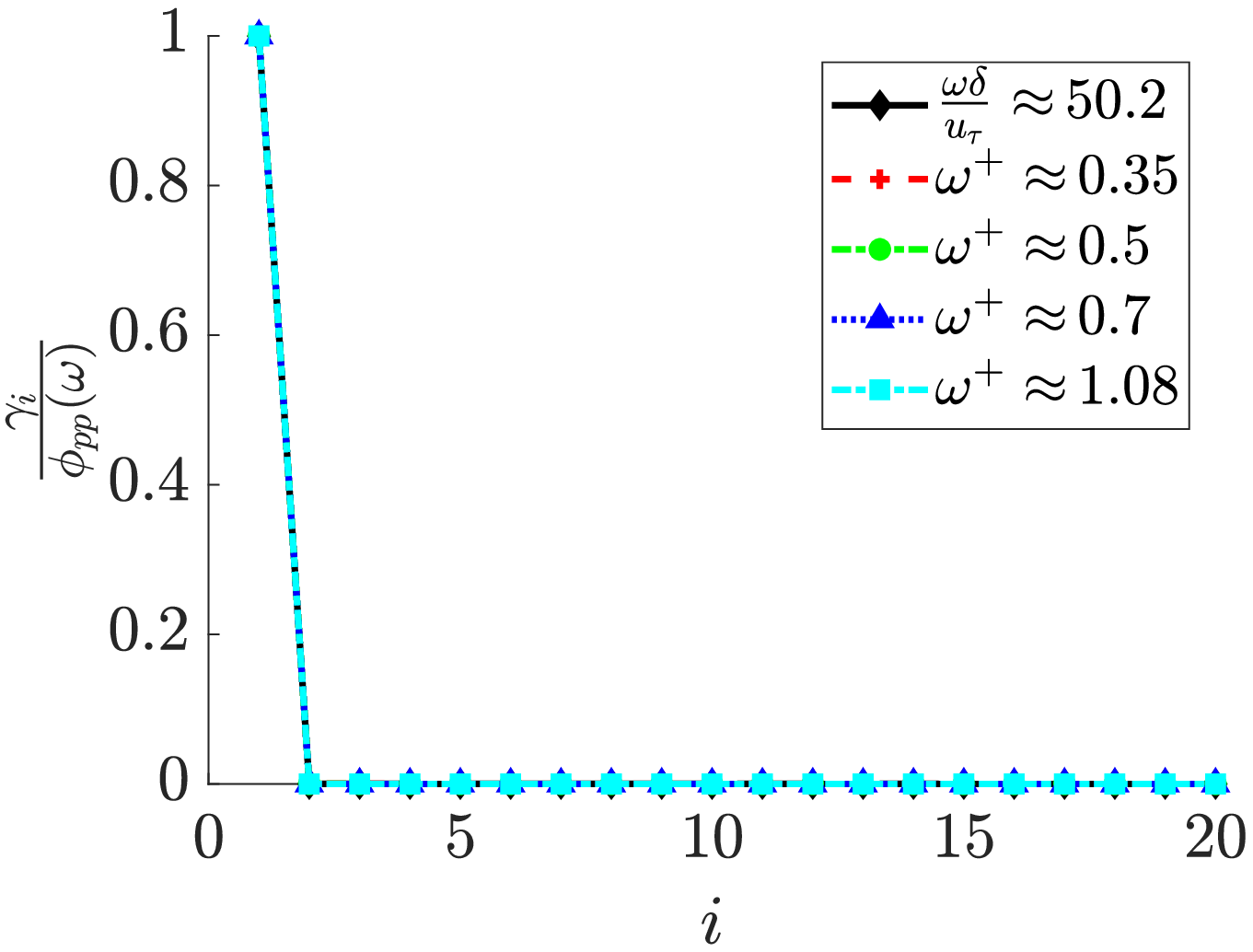}
        \put(1,65){\scriptsize$(b)$}
      \end{overpic}
    \end{subfigure}%
    \caption{Fractional contribution of the first 20 spectral POD
      modes computed using the Poisson inner product ($\beta=0.1$) to
      the wall-pressure PSD for a) $Re_{\tau}=180$ and b)
      $Re_{\tau}=400$ at different frequencies.}
    \label{fig:spodeigwt_wmat_alpha_0p1}
  \end{figure}

  \begin{figure}
    \centering
    \begin{subfigure}{.5\textwidth}
      \begin{overpic}[width=\linewidth]{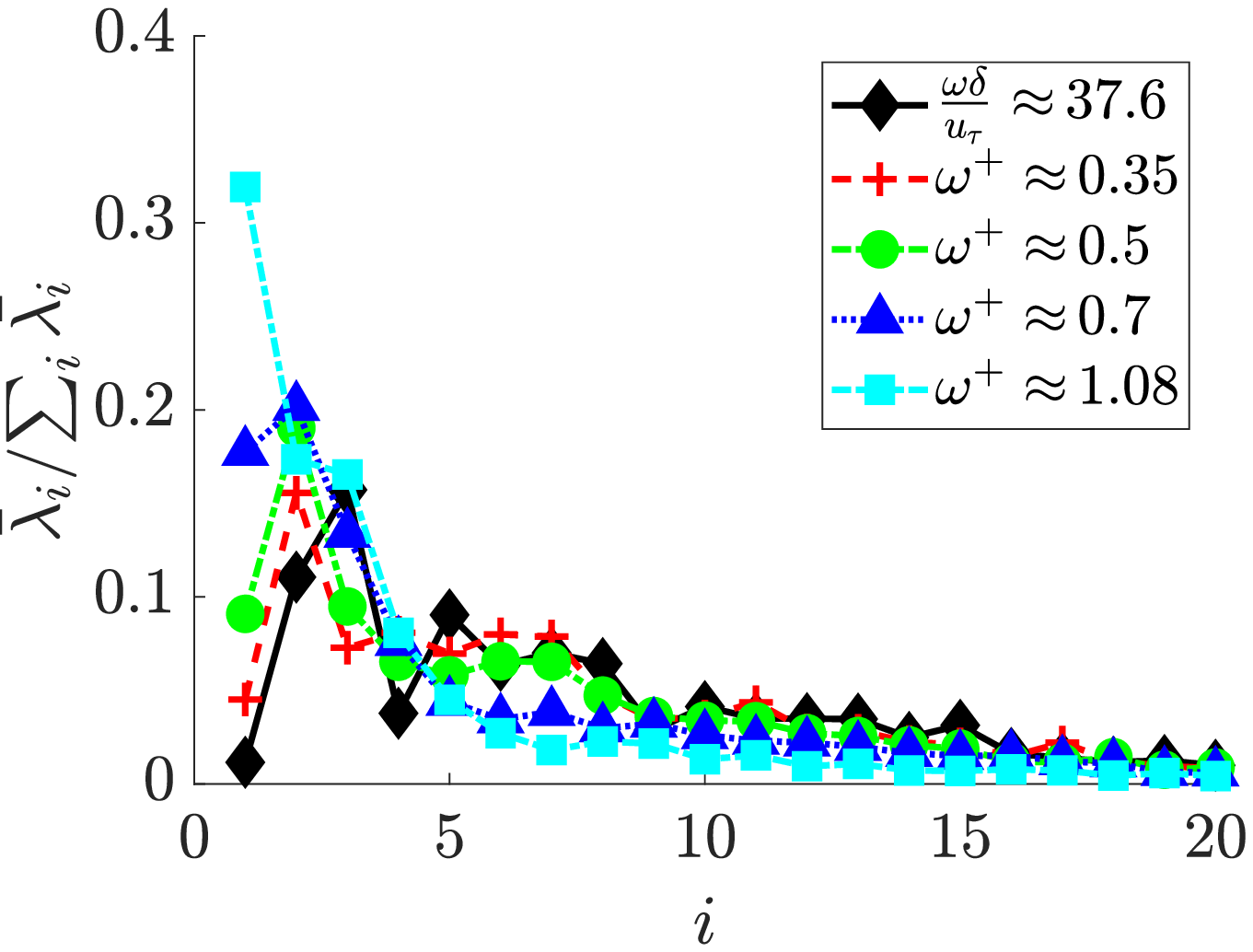}
        \put(1,65){\scriptsize$(a)$}
      \end{overpic}
    \end{subfigure}%
    \begin{subfigure}{.5\textwidth}
      \begin{overpic}[width=\linewidth]{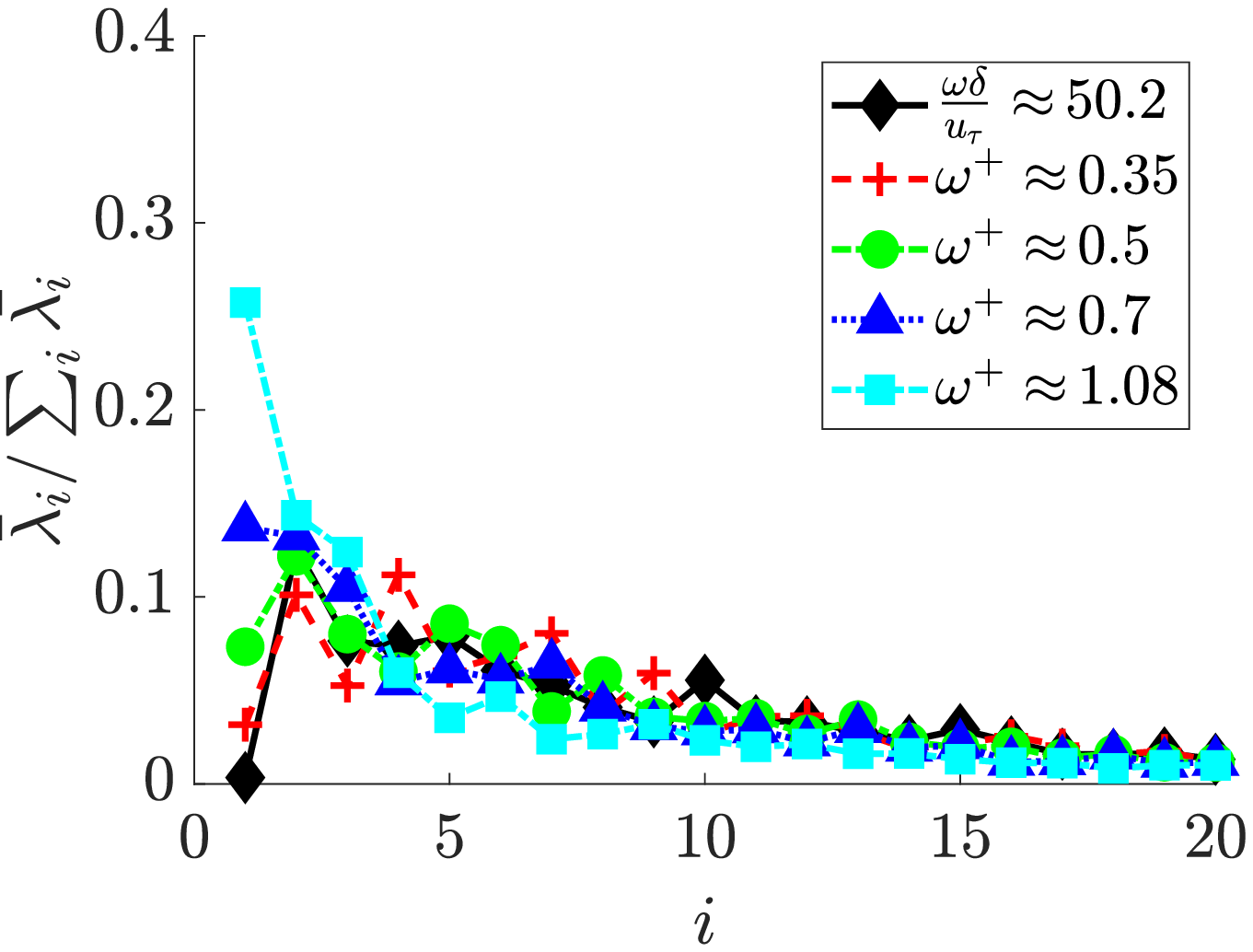}
        \put(1,65){\scriptsize$(b)$}
      \end{overpic}
    \end{subfigure}%
    \caption{Fractional contribution of the first 20 spectral POD
      modes computed using the Poisson inner product ($\beta=0.1$) to
      the integrated net source PSD for a) $Re_{\tau}=180$ and b)
      $Re_{\tau}=400$ at different frequencies.}
    \label{fig:spodeigpsdwt_wmat_alpha_0p1}
  \end{figure}

  \begin{figure}
    \centering
    \begin{subfigure}{.25\textwidth}
      \begin{overpic}[width=\linewidth]{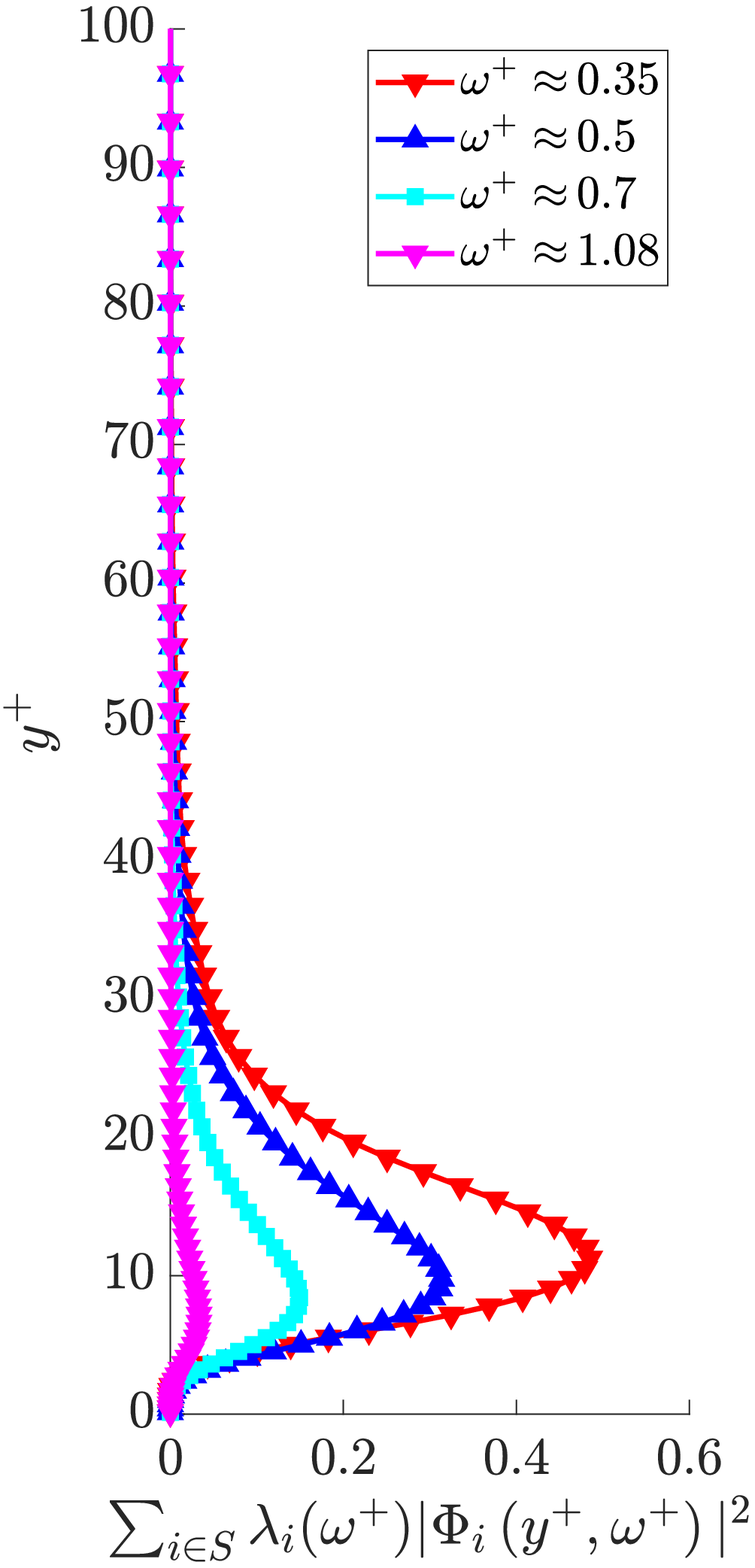}
        \put(1,95){\scriptsize$(a)$}
      \end{overpic}
    \end{subfigure}%
    \begin{subfigure}{.25\textwidth}
      \begin{overpic}[width=\linewidth]{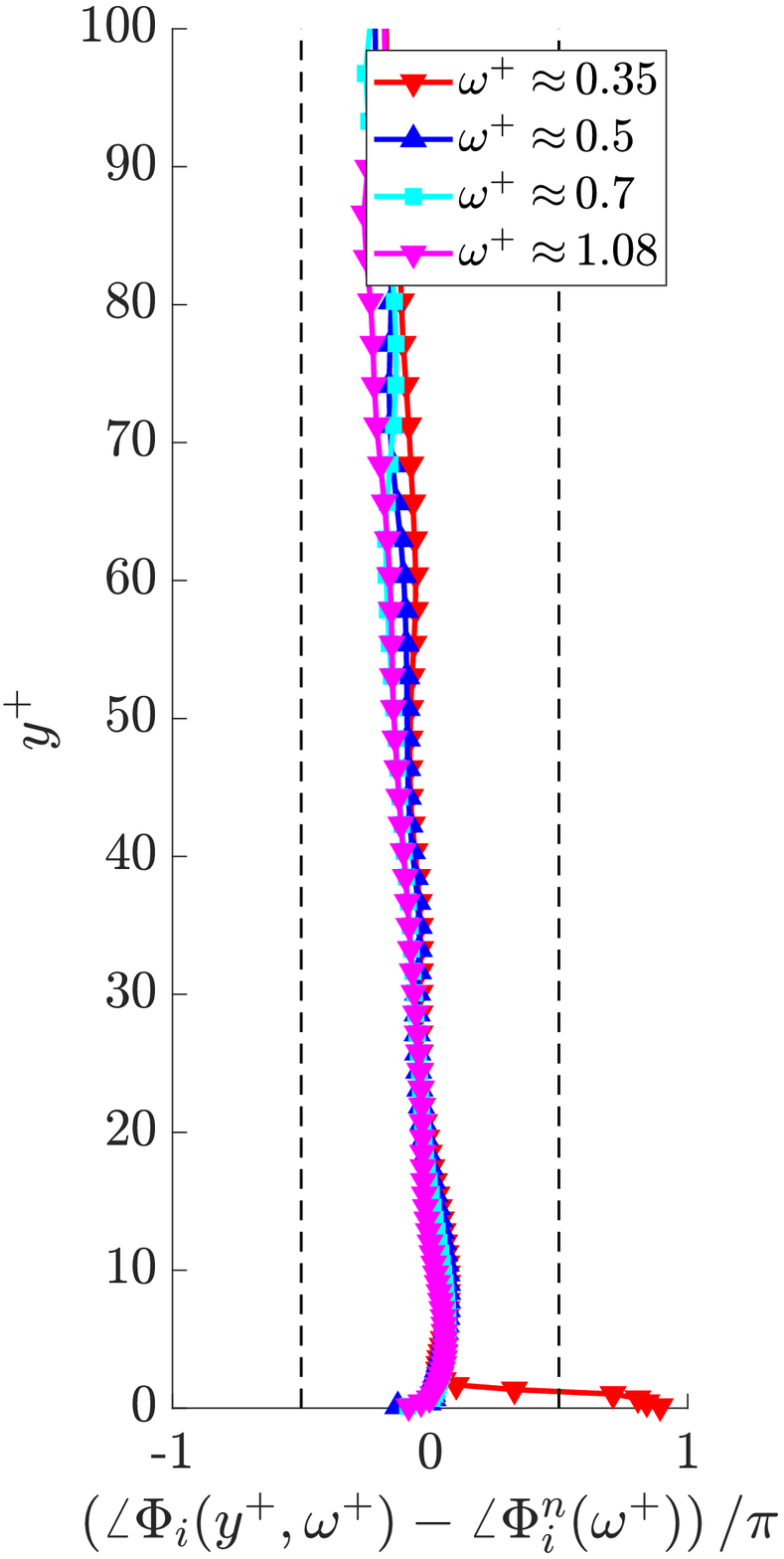}
        \put(1,95){\scriptsize$(b)$}
      \end{overpic}
    \end{subfigure}%
    \begin{subfigure}{.25\textwidth}
      \begin{overpic}[width=\linewidth]{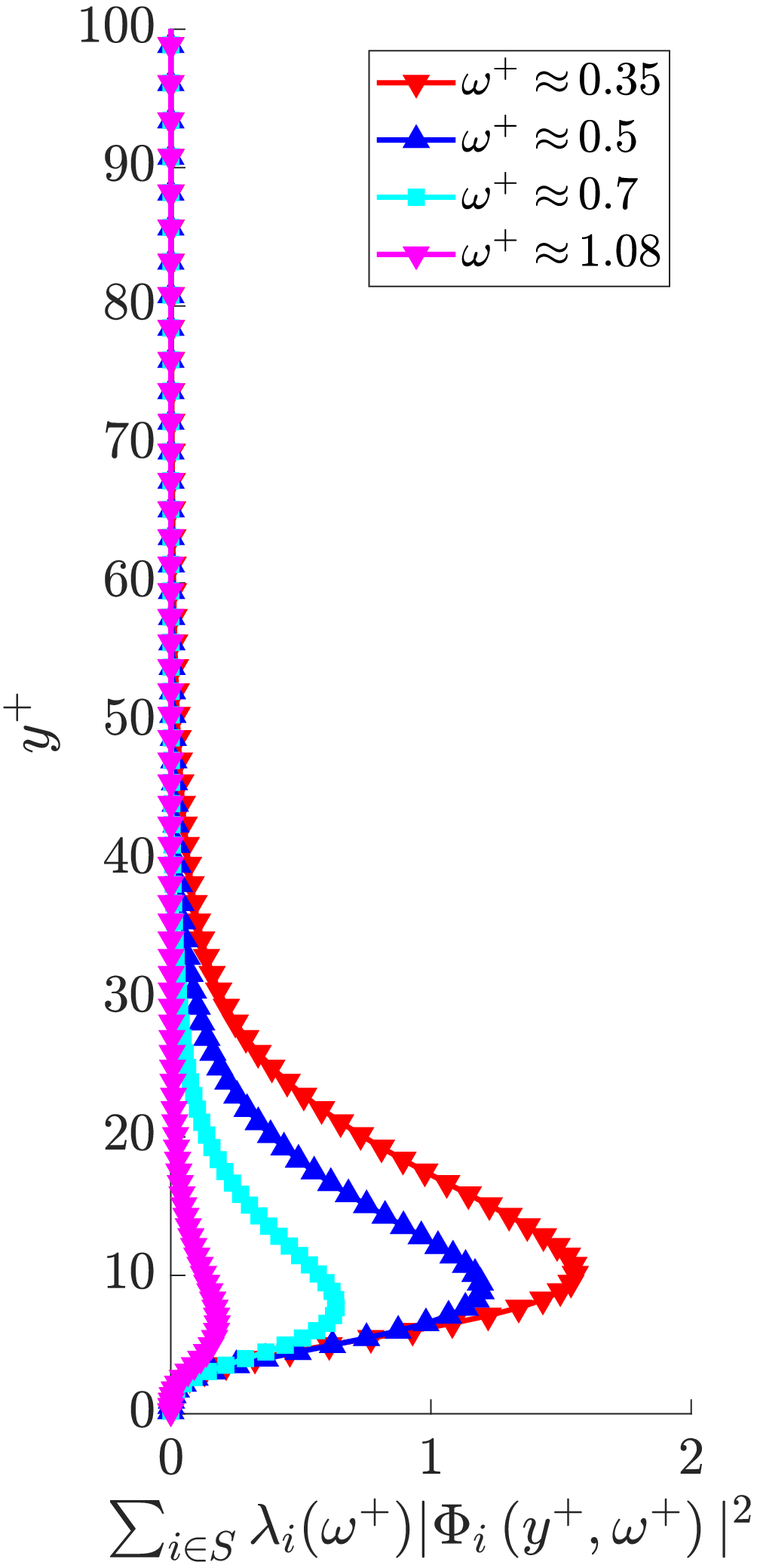}
        \put(1,95){\scriptsize$(c)$}
      \end{overpic}
    \end{subfigure}%
    \begin{subfigure}{.25\textwidth}
      \begin{overpic}[width=\linewidth]{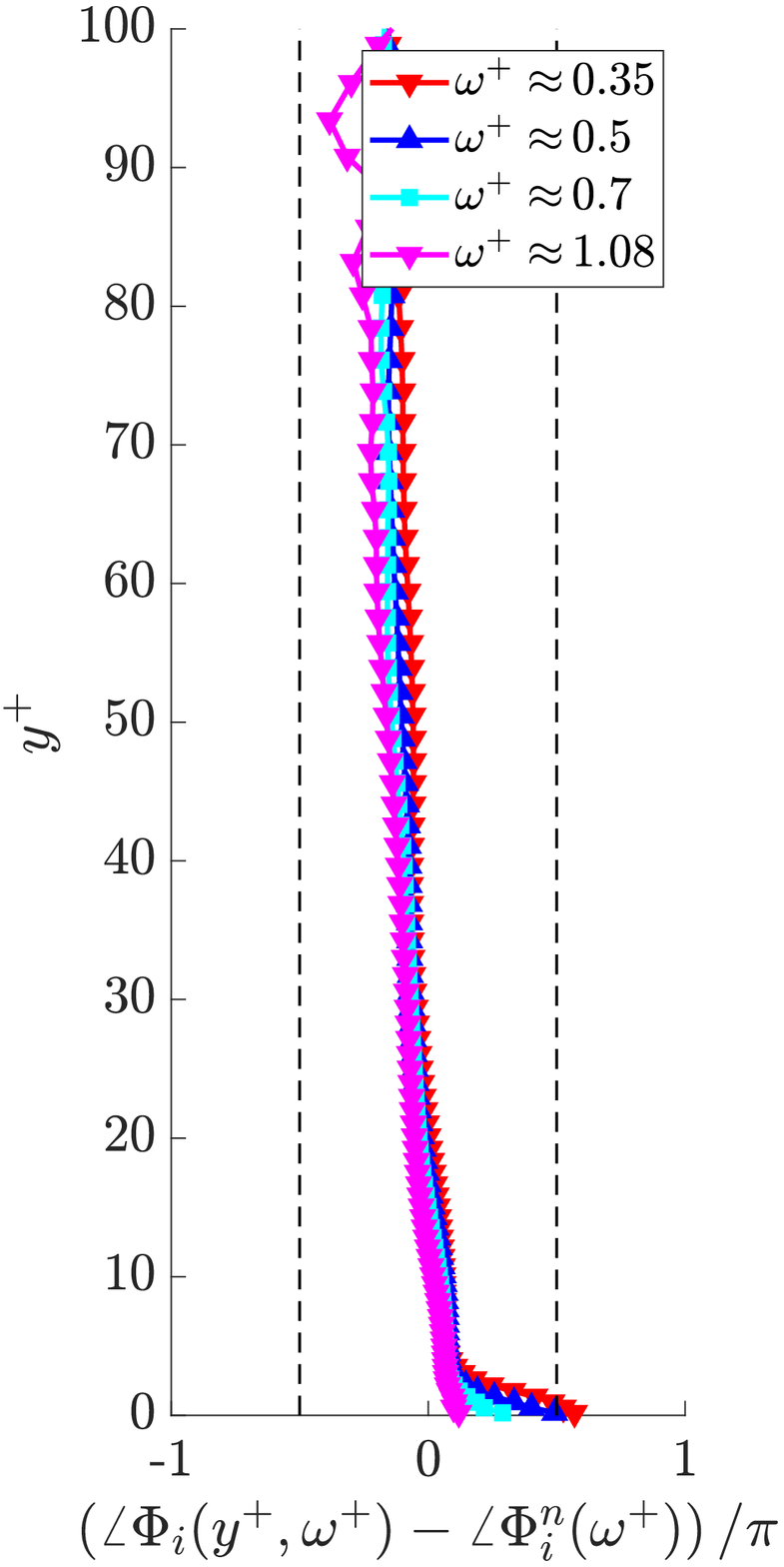}
        \put(1,95){\scriptsize$(d)$}
      \end{overpic}
    \end{subfigure}%
    \caption{Envelope (figures a, c) and phase (figures b, d) of the
      dominant spectral POD mode computed using the Poisson inner product
      ($\beta=0.1$) for few selected high frequencies. Figures a-b and c-d are
      for $Re_{\tau}=180$ and $400$, respectively. The left and right
      dashed black solid line in figures b and d indicate
      $\angle\Phi_i(y^+,\omega^+)-\angle\Phi_i^n(\omega^+)$ equal to
      $-\pi/2$ and $\pi/2$, respectively.}
    \label{fig:domspod_lowfreq_wmat_alpha_0p1}
  \end{figure}

Figure \ref{fig:domspod_lowfreq_wmat_alpha_0p1} shows the wall-normal variation of the envelope and phase of the dominant mode at the premultiplied spectra peak $\omega^+=0.35$ and a few higher frequencies $\omega^+=0.5,\,0.7$ and $1$. The dominant modes have a similar shape in inner units for both $Re_{\tau}$. Its envelope (figures \ref{fig:domspod_lowfreq_wmat_alpha_0p1}a and c) represents sources confined near the wall with intensities peaking in the buffer layer. With increasing frequency, the wall-normal location of the peak moves closer to the wall, and the width of the envelope decreases. This behavior of the dominant mode is consistent with that of the wall-pressure fluctuation - net source CSD. The phase (figures \ref{fig:domspod_lowfreq_wmat_alpha_0p1}b
 and d) of the dominant mode varies between $-\pi/2$ and $\pi/2$. Therefore, the contributions from different wall-normal locations undergo constructive interference. Further, the phase variation is almost linear with a negative slope, at least around the envelope peak. The negative slope indicates that the envelope encloses a wave traveling towards the wall.}

 {\color{black} 

   Figures \ref{fig:lowfreqpeak_spod}a and b show the magnitude and phase, respectively, of the dominant spectral POD mode at the low-frequency linear PSD peak (figure \ref{fig:wallpresk3spec}). Recall that the frequencies are $\omega\delta/u_{\tau}=37.6$ for $Re_{\tau}=180$ and $50.2$ for $Re_{\tau}=400$. The envelope of the dominant mode peaks around $y^+\approx 15$ for both $Re_{\tau}$. The phase variation does not show any noticeable slope indicating that the different wall-normal locations are in phase with each other, at least around the envelope peak. 

   In figure \ref{fig:lowfreqpeak_spod}c and d, we show the wall-normal contribution of the dominant mode to the wall-pressure PSD. Figure \ref{fig:lowfreqpeak_spod}c and d are in inner and outer units, respectively. The curves are normalized to obtain unit integral along the y-axis. The contribution peaks at $y^+\approx 15$ for both $Re_{\tau}$. Also, figure \ref{fig:lowfreqpeak_spod}c shows a negative contribution close to the wall for $Re_{\tau}=400$ that is not present for $Re_{\tau}=180$. We observe that the region $y^+>30$ contributes more for the higher Reynolds number, signifying an increase in the outer region contribution. Further, from figure \ref{fig:lowfreqpeak_spod}d, we observe that the width of this dominant source is around $0.25\delta$ since the y-coordinate is significant for $y<0.25\delta$. Overall, at the low-frequency PSD peak, the contribution from the dominant mode peaks at $y^+=15$, and its width is around $0.25\delta$.

\begin{figure}
  \centering
  \begin{subfigure}{.25\textwidth}
    \begin{overpic}[width=\linewidth]{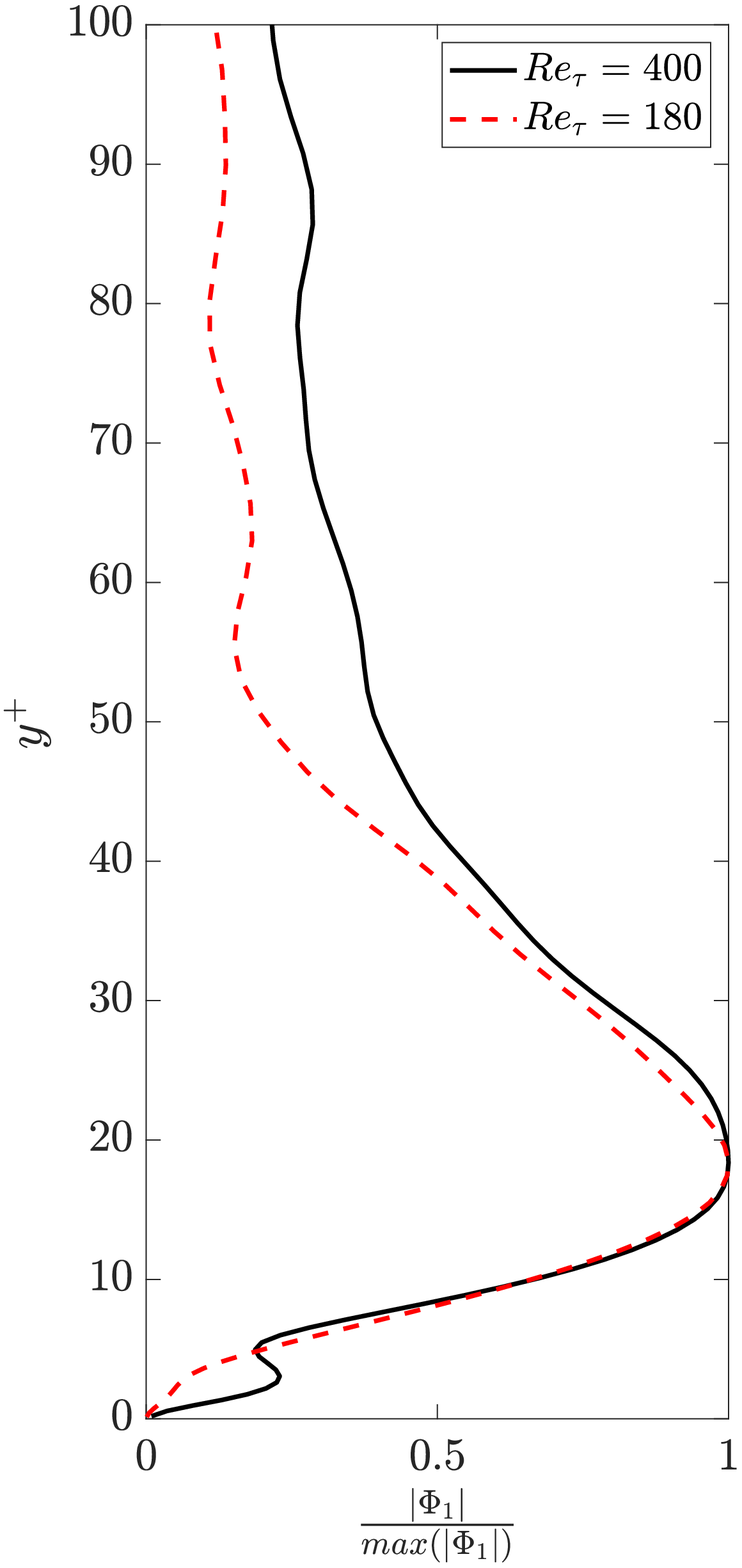}
      \put(1,95){\scriptsize$(a)$}
    \end{overpic}
  \end{subfigure}%
  \begin{subfigure}{.25\textwidth}
    \begin{overpic}[width=\linewidth]{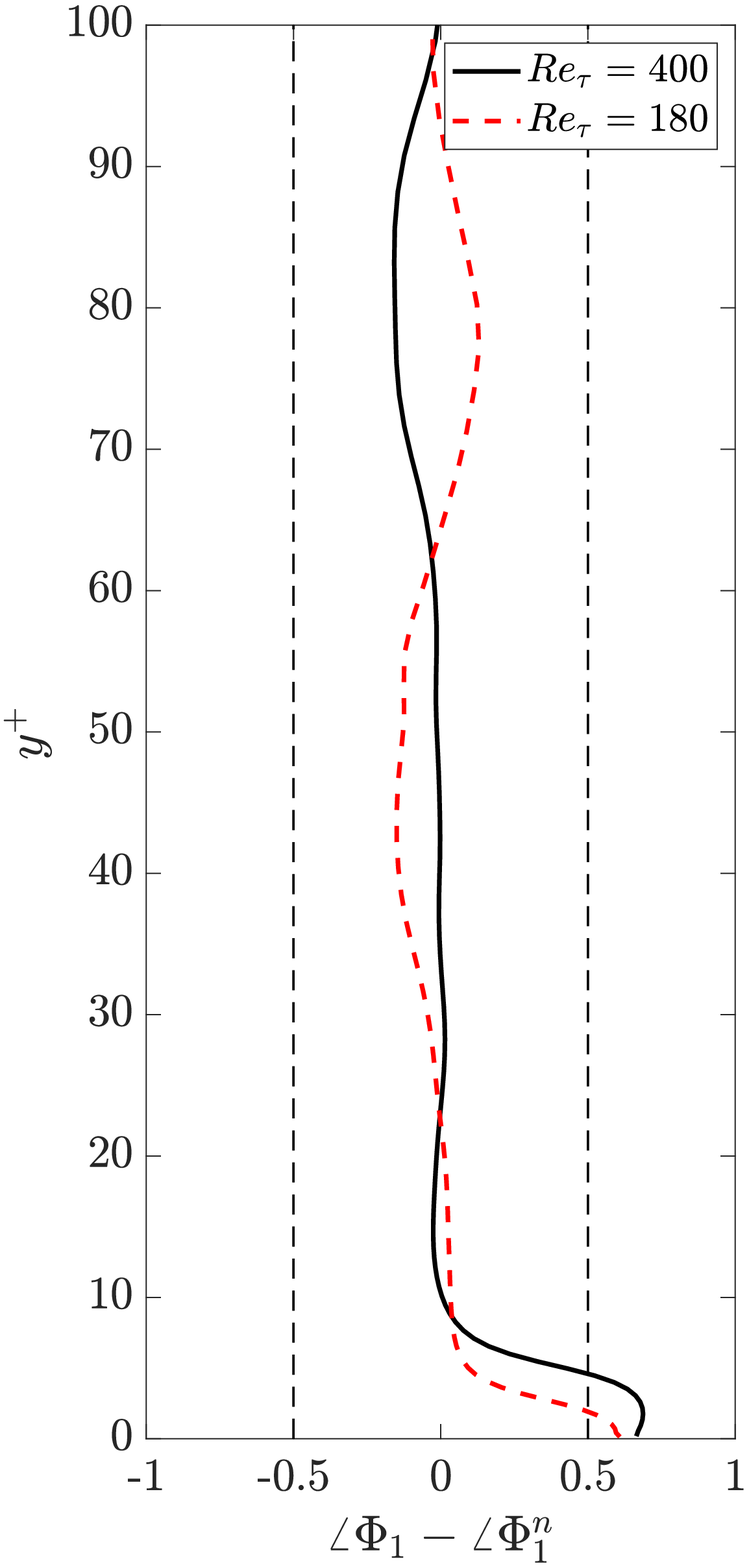}
      \put(1,95){\scriptsize$(b)$}
    \end{overpic}
  \end{subfigure}%
  \begin{subfigure}{.25\textwidth}
    \begin{overpic}[width=\linewidth]{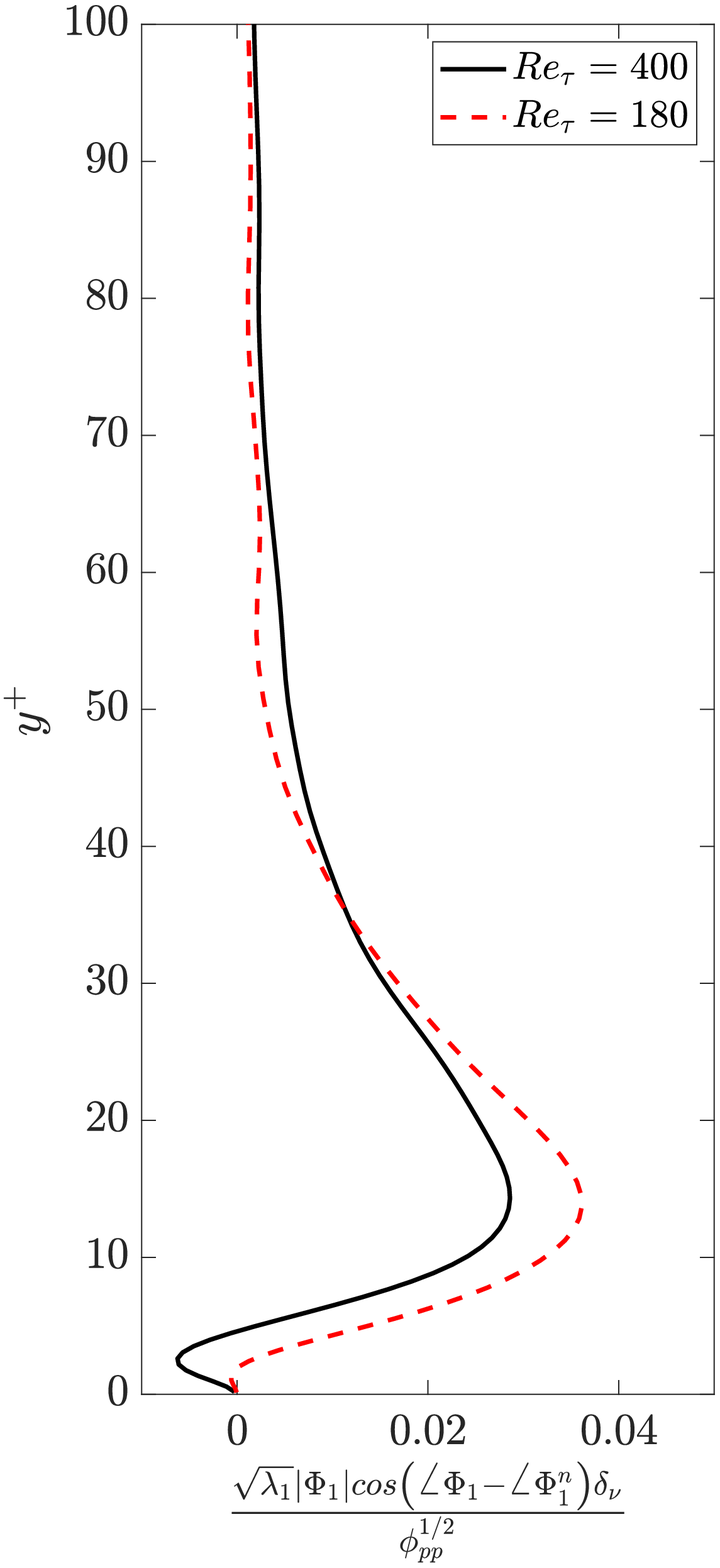}
      \put(1,95){\scriptsize$(c)$}
    \end{overpic}
  \end{subfigure}%
  \begin{subfigure}{.25\textwidth}
    \begin{overpic}[width=\linewidth]{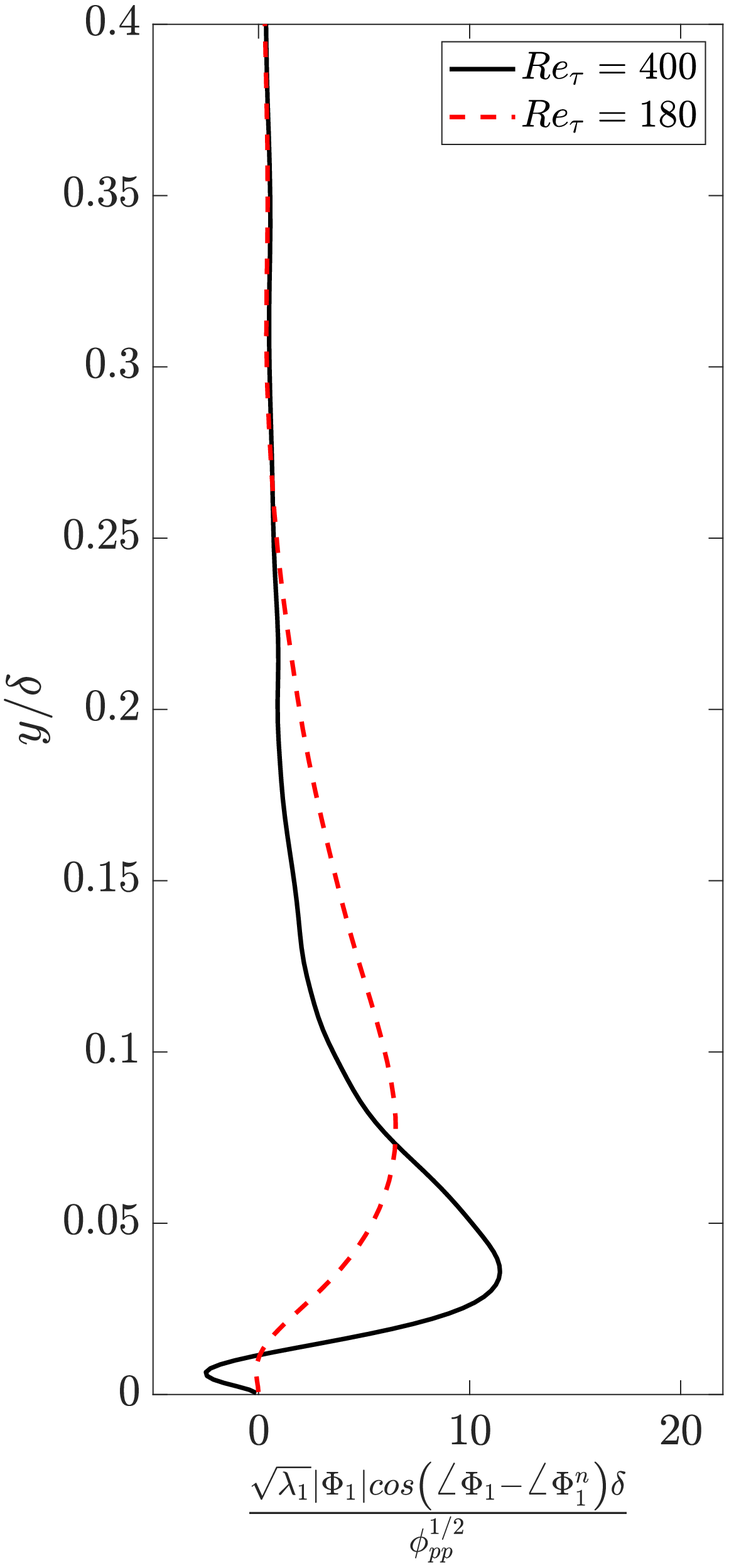}
      \put(1,95){\scriptsize$(d)$}
    \end{overpic}
  \end{subfigure}%
%
  \caption{Envelope (figure a) and phase (figure b) of the dominant spectral POD mode compute using the Poisson inner product at the low frequency linear PSD peak. Figures c and d show the normalized wall-normal contribution of the dominant mode to the wall-pressure PSD in inner and outer units, respectively.}
  \label{fig:lowfreqpeak_spod}
\end{figure}}

 {\color{black} We create a representative net source field that gives the two-dimensional structure implied by a spectral POD mode. The representative field $\tilde{f}_G(x,y,z,t)$ implied by mode $\Phi_j(y,\omega_o)$ at frequency $\omega_o$ is constructed as
  \begin{equation}
    \tilde{f}_G(x,y,z,t)=Re\left(\alpha(x,z,\omega_o)e^{-i\angle\Phi_j(y,\omega_o)}|\Phi_j(y,\omega_o)|e^{i\omega_o t}\right),
  \end{equation}
  where $Re(f)$ is the real part of $f$. Since $\alpha(x,z,\omega_o)$ is homogenous in $x$ and $z$, we decompose it using the Fourier transform. Intuitively, we expect the streamwise convective Fourier component $e^{-i\omega_ox/c_o(\omega_o)}$ of $\alpha(x,z,\omega_o)$ to be the most dominant one. Here, $c_o(\omega_o)$ is the convective velocity at frequency $\omega_o$. For simplicity, we assume no spanwise variation of the representative net source. Substituting $\alpha(x,z,\omega_o)=e^{-i\omega_ox/c_o(\omega_o)}$ in the above equation, we obtain the representative field
  \begin{equation}
    \tilde{f}_G(x,y,z,t)=Re\left(e^{-i\omega_ox/c_o(\omega_o)}e^{-i\angle\Phi_j(y,\omega_o)}|\Phi_j(y,\omega_o)|e^{i\omega_o t}\right).
  \end{equation}

  To create the representative field at a frequency $\omega_o$, we need three inputs - the mode $\Phi_j(y,\omega_o)$, the convection velocity $c_o(\omega_o)$ and the time $t$. Figure \ref{fig:rep_fg_poisson} shows the representative net source field constructed from the dominant spectral POD mode. Figure \ref{fig:rep_fg_poisson}a is at the premultiplied PSD peak frequency and \ref{fig:rep_fg_poisson}b is at the linear PSD peak frequency. We use a convection velocity defined as $c_o(\omega_o)/u_{\tau}=\left(\omega_o\delta/u_{\tau}\right)/k_p(\omega_o)\delta$, where $k_p(\omega_o)$ is the peak wavenumber coordinate at frequency $\omega_o$ in the wavenumber-frequency spectrum of wall-pressure. We choose time $t$ to be $0$.

Figures \ref{fig:rep_fg_poisson}a and b show a convecting coherent structure inclined in the downstream direction. Essentially, this is because of the negative slope in the phase of the mode. As the inclined structures convect across a fixed streamwise location $x_o^+$, the wall-normal intensity (magnitude of the field that depends on $y$ and $x_o$) propagates towards the wall as indicated by the negative slope.

Figures \ref{fig:rep_fg_poisson}c and d show the coherent structure represented by the dominant POD mode at the linear PSD peak. These structures are vertical, with almost no inclination in the downstream direction, as indicated by almost no slope in the phase of the mode. Such large scale vertical patterns with streamwise spacing of $\sim 2\delta$ have been previously observed in the instantaneous rapid pressure fields for $Re_{\tau}\sim 1000$ by \cite{abe2005dns}. They proposed that these patterns are responsible for the low-wavenumber peak in the wall-pressure spectra. The coherent structures in figures \ref{fig:rep_fg_poisson}c and d further support this case. 

\begin{figure}
  \centering
  \begin{subfigure}{\textwidth}
    \begin{overpic}[width=\linewidth]{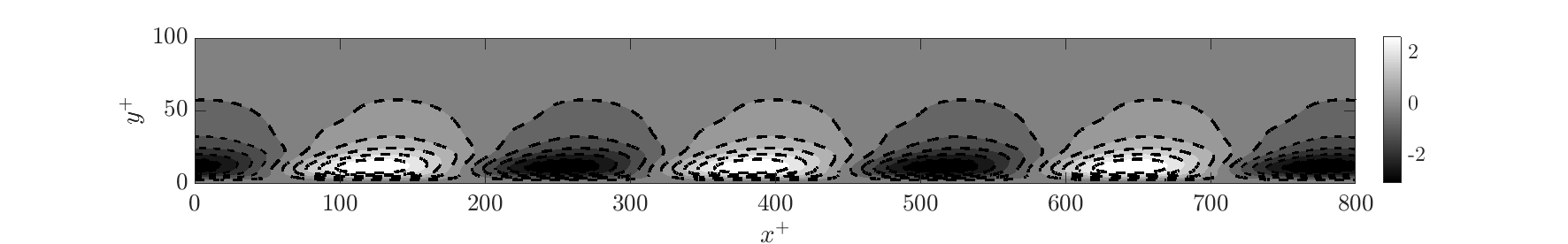}
      \put(5,13){\scriptsize$(a)$}
    \end{overpic}
  \end{subfigure}%

  \begin{subfigure}{\textwidth}
    \begin{overpic}[width=\linewidth]{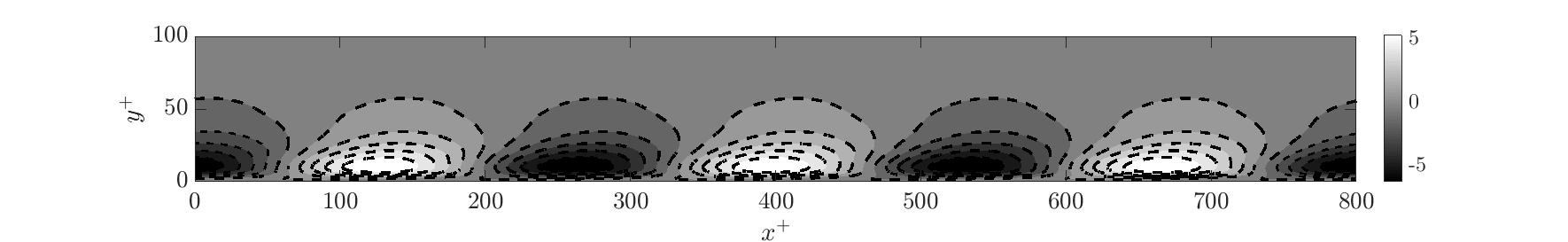}
      \put(5,13){\scriptsize$(b)$}
    \end{overpic}
  \end{subfigure}%

  \begin{subfigure}{\textwidth}
    \begin{overpic}[width=\linewidth]{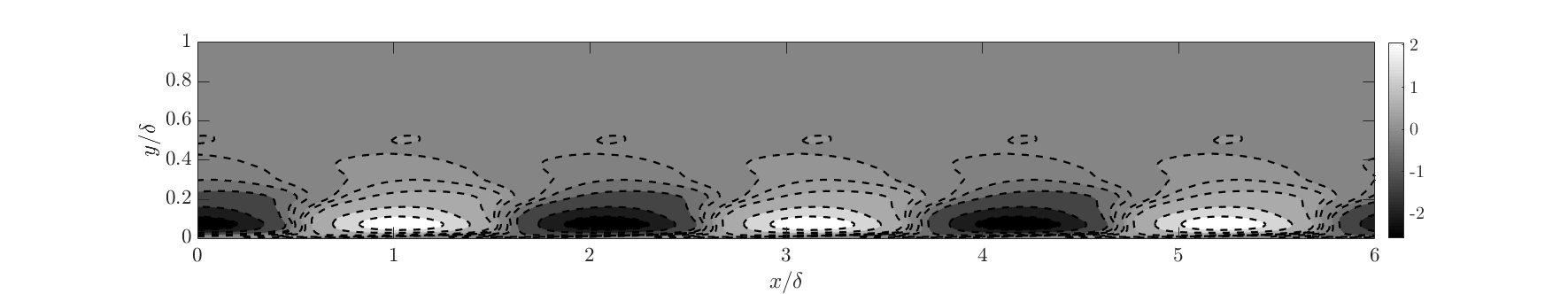}
      \put(5,15){\scriptsize$(c)$}
    \end{overpic}
  \end{subfigure}%

  \begin{subfigure}{\textwidth}
    \begin{overpic}[width=\linewidth]{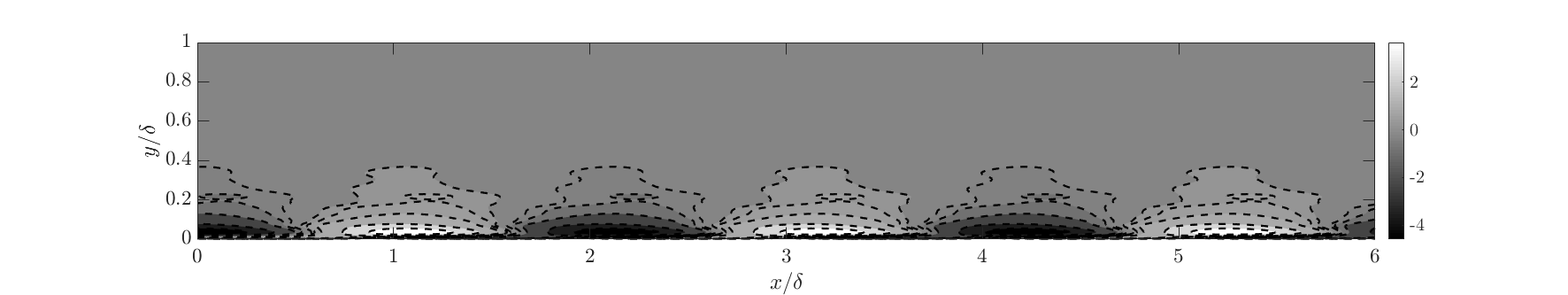}
      \put(5,15){\scriptsize$(d)$}
    \end{overpic}
  \end{subfigure}%

  \caption{Representative net source $\tilde{f}_G$ at the premultiplied spectra peak (a-b) and the linear spectrum peak (c-d). Figures a and c are for $Re_{\tau}=180$, and b and d are for $Re_{\tau}=400$. Contours in figures a and b are 10 equally spaced values between the minimum and maximum of $\tilde{f}_G$. Contours in figures c and d are $[\pm 0.05\,\pm 0.1\,\pm 0.2\,\pm 0.5\,\pm 0.8]$ times the maximum value of $\tilde{f}_G$ }
  \label{fig:rep_fg_poisson}
\end{figure}


{\color{black} Overall, the dominant spectral POD mode represents the active portion of the net source that contributes to the entire wall-pressure PSD. The remaining POD modes that comprise the inactive portion have zero contribution to the PSD. Further, the active and inactive parts are decorrelated. At high frequencies ($\omega^+\geq 0.35$), the shape of the dominant POD mode is similar in inner units for the two $Re_{\tau}$. The two-dimensional (2D) coherent structure at the premultiplied PSD peak inclines in the downstream direction. At the low-frequency  linear PSD peak, the wall-normal contribution peaks in the buffer layer at $y^+\approx 15$ with a width of $y/\delta\approx 0.25$. The corresponding 2D structure has a large scale vertical pattern similar to the previous observations of the instantaneous rapid pressure field by \cite{abe2005dns}.

  We expect the similarity of the high frequency dominant modes in inner units to continue at even higher Reynolds numbers. At the low wavenumber/frequency wall-pressure linear spectra peak, the outer region $(y^+>30)$ contributes more for $Re_{\tau}=400$ than for $Re_{\tau}=180$. This low wavenumber/frequency peak is present in the linear wall-pressure spectra up to $Re_{\tau}=5000$ \citep{panton2017correlation,abe2005dns}. With increasing Reynolds number, we expect this contribution from the outer region to grow larger. Further, at $Re_{\tau}\approx 5000$, the low and high wavenumber contributions to the premultiplied wall-pressure spectra show mild separation. We expect the spectral POD modes responsible for the low wavenumber peak to depend on outer units. Further, high Reynolds number effects like amplitude modulation \citep{tsuji2016amplitude} in the wall-pressure sources could be studied using the above spectral POD framework.}

}


\begin{figure}
  \centering
  \begin{subfigure}{.5\textwidth}
    \begin{overpic}[width=\linewidth]{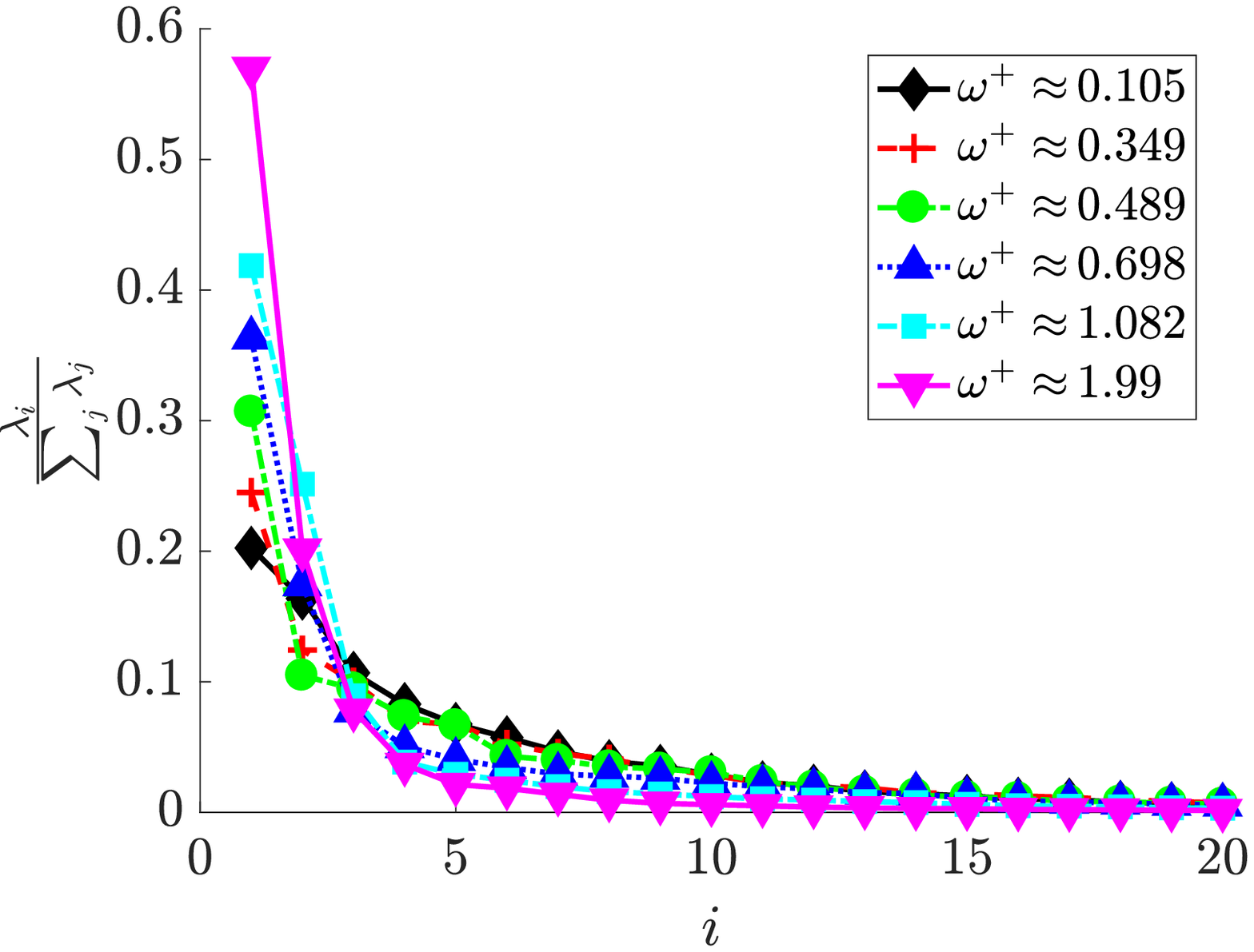}
      \put(1,65){\scriptsize$(a)$}
    \end{overpic}
  \end{subfigure}%
  \begin{subfigure}{.5\textwidth}
    \begin{overpic}[width=\linewidth]{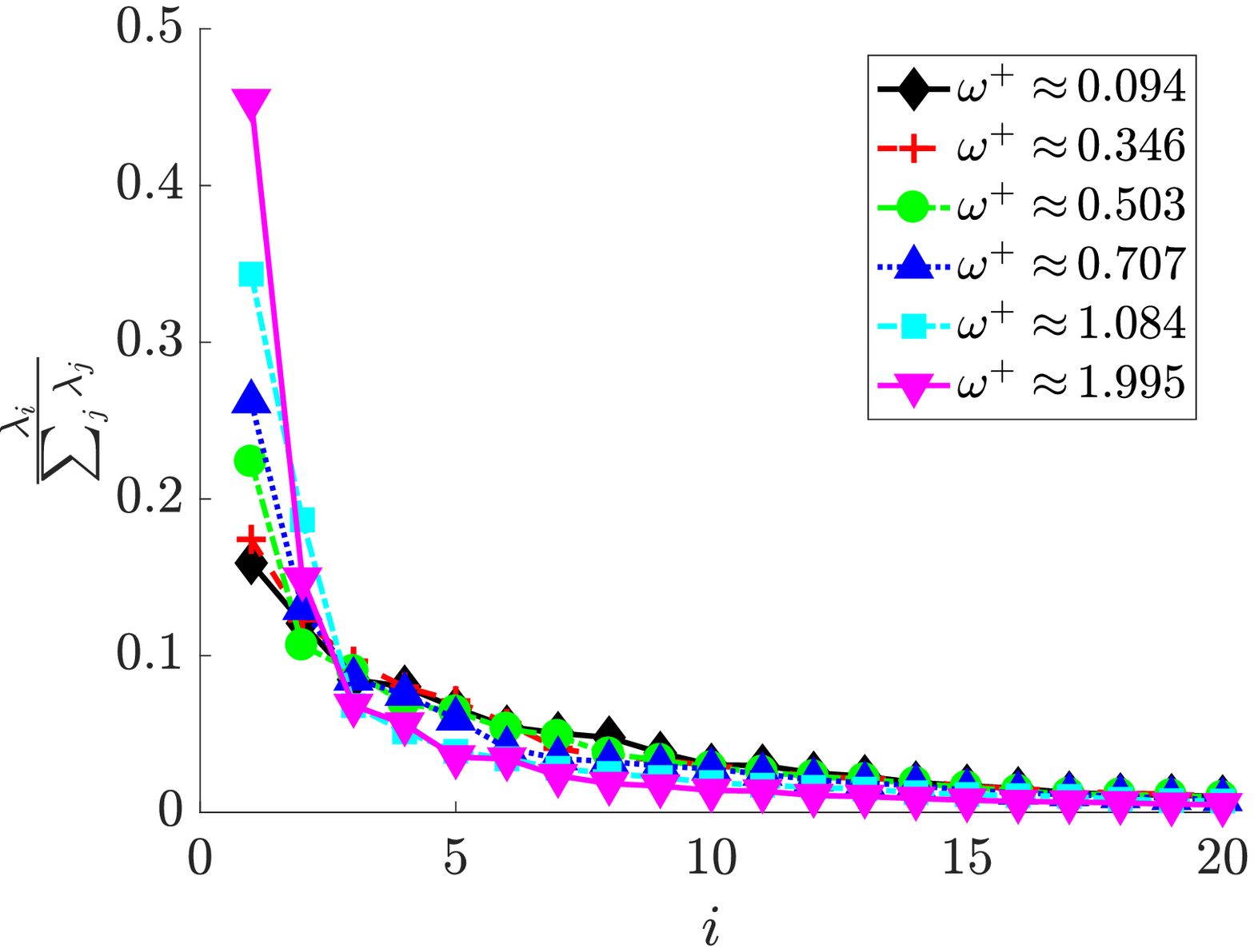}
      \put(1,65){\scriptsize$(b)$}
    \end{overpic}
  \end{subfigure}%
  \caption{Spectral POD eigenvalues computed using the $L^2$ inner product for a) $Re_{\tau}=180$ and b) $Re_{\tau}=400$ at different frequencies.}
  \label{fig:spodeig}
\end{figure}


\begin{figure}
  \centering
  \begin{subfigure}{.5\textwidth}
    \begin{overpic}[width=\linewidth]{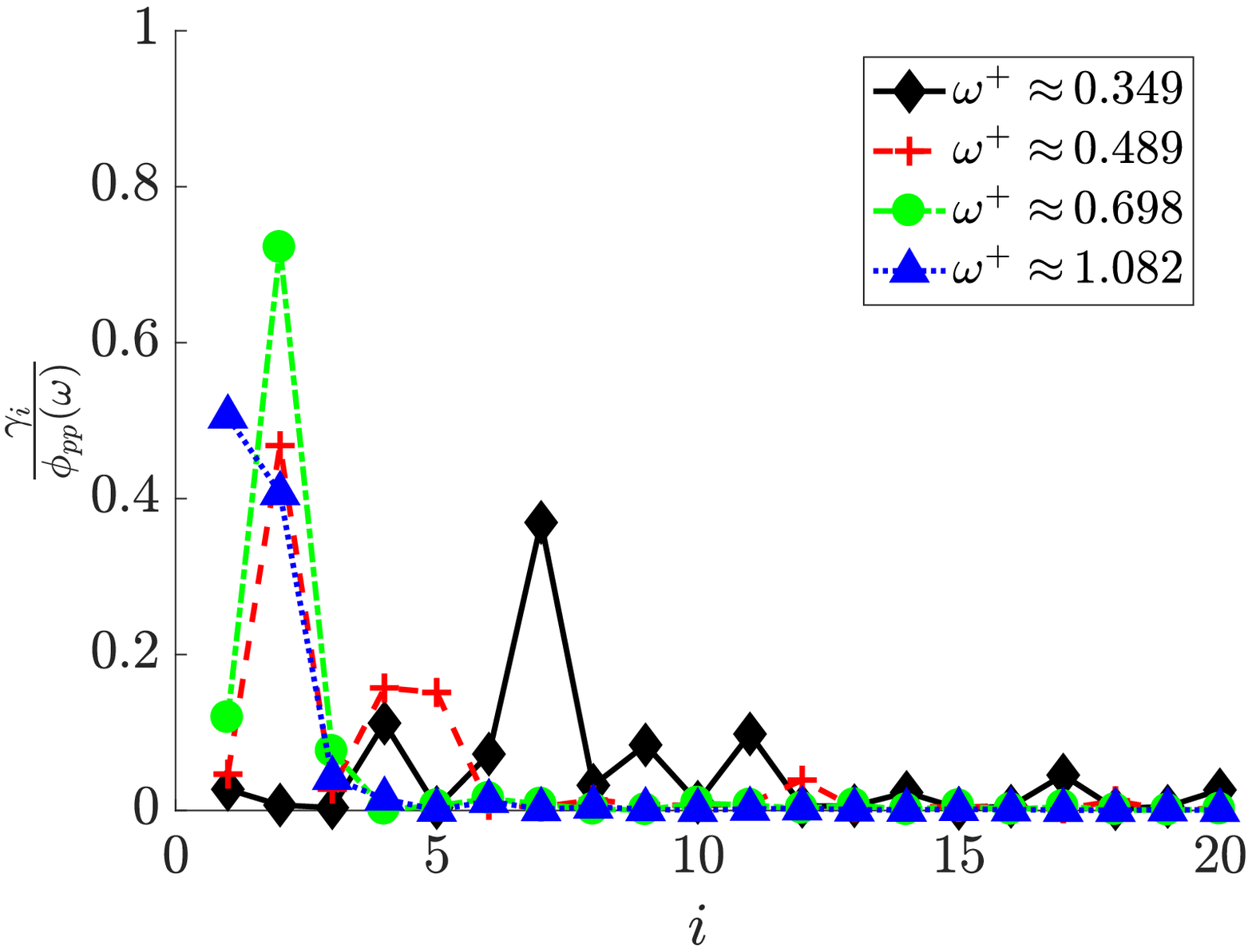}
      \put(1,65){\scriptsize$(a)$}
    \end{overpic}
  \end{subfigure}%
  \begin{subfigure}{.5\textwidth}
    \begin{overpic}[width=\linewidth]{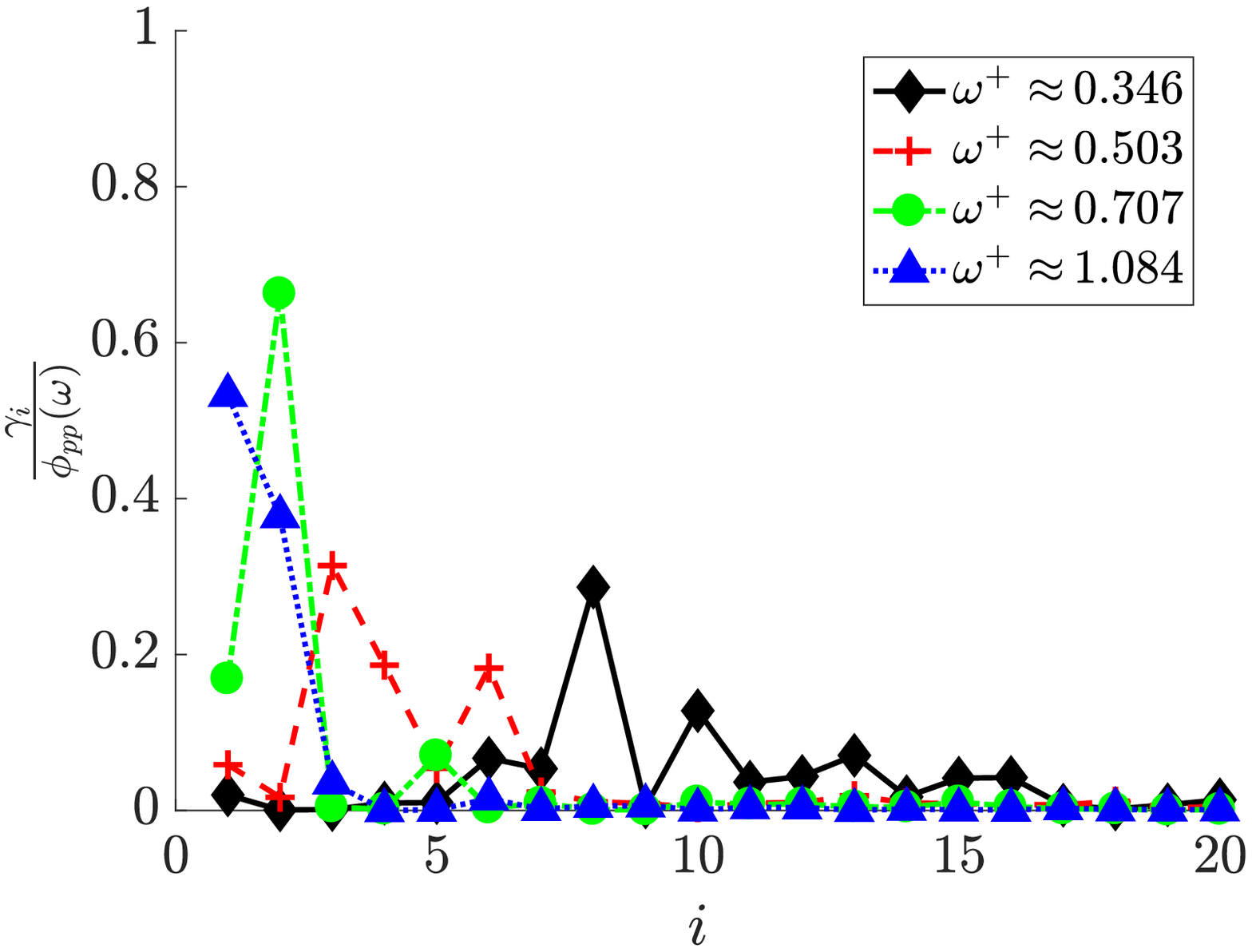}
      \put(1,65){\scriptsize$(b)$}
    \end{overpic}
  \end{subfigure}%
  \caption{Contribution of the first 20 spectral POD modes computed using the $L^2$ inner product (normalized by the wall-pressure PSD) to wall-pressure PSD for a) $Re_{\tau}=180$ and b) $Re_{\tau}=400$ at different frequencies.}
  \label{fig:spodeigwt}
\end{figure}

\begin{figure}
  \centering
  \begin{subfigure}{.5\textwidth}
    \begin{overpic}[width=\linewidth]{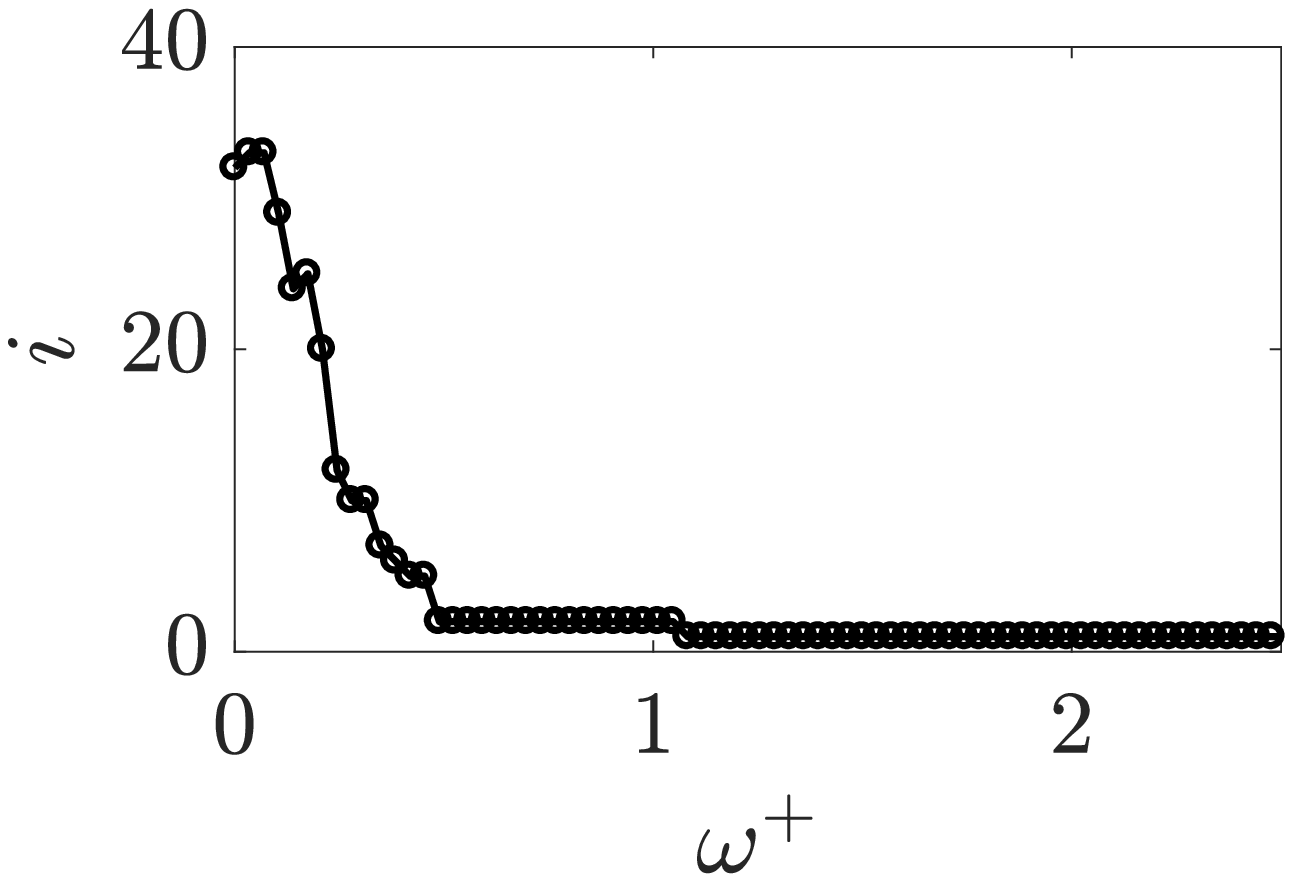}
      \put(1,60){\scriptsize$(a)$}
    \end{overpic}
  \end{subfigure}%
  \begin{subfigure}{.5\textwidth}
    \begin{overpic}[width=\linewidth]{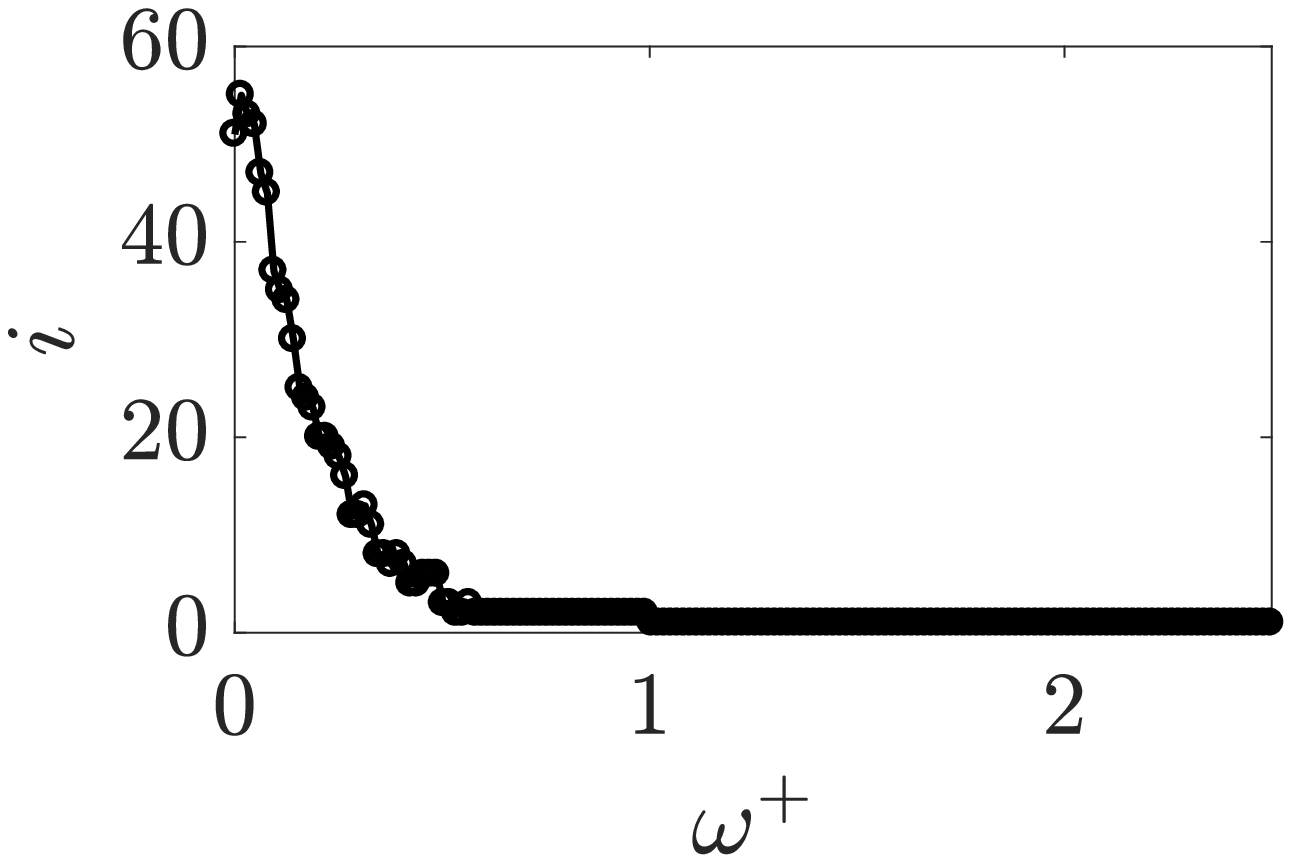}
      \put(1,60){\scriptsize$(b)$}
    \end{overpic}
  \end{subfigure}%
  \caption{Index of the spectral POD (computed using $L^2$ inner product) that contributes the most to the wall-pressure PSD for a) $Re_{\tau}=180$ and b) $Re_{\tau}=400$.}
  \label{fig:domspodmodeindex}
\end{figure}

{\color{black}\subsubsection{Remark on spectral POD with the $L^2$ inner product}

  We also performed spectral POD of the net source CSD using the $L^2$ inner product. Figure \ref{fig:spodeig} shows the obtained eigenvalues for both $Re_{\tau}$. The eigenvalues give the contribution of each POD mode to the wall-normal integral of the net source PSD. The POD modes obtained with the $L^2$ inner product, by definition, optimally decompose the integral of the net-source PSD. However, the dominant POD mode might not contribute significantly to the wall-pressure PSD. Clearly, figure \ref{fig:spodeigwt} shows this behavior for $\omega^+<1$. 

To investigate this further, we plot the index of the POD mode that contributes the most to the wall-pressure PSD as a function of frequency in figure \ref{fig:domspodmodeindex}. In the frequency ranges $0.55<\omega^+<1$ and $\omega^+>1$, the dominant wall-pressure mode (largest $\gamma_i(\omega)$) is the second and the first spectral POD mode, respectively. At low frequencies $\omega^+<0.55$, the dominant wall-pressure mode index is larger than or equal to 3. The dominant spectral POD mode is not the dominant wall-pressure mode because of destructive interference. The contributions of the dominant spectral POD mode from different wall-normal regions cancel each other out. For more details, we refer the reader to Appendix \ref{sec:destint}.

\begin{figure}
  \centering
  \begin{subfigure}{.25\textwidth}
    \begin{overpic}[width=\linewidth]{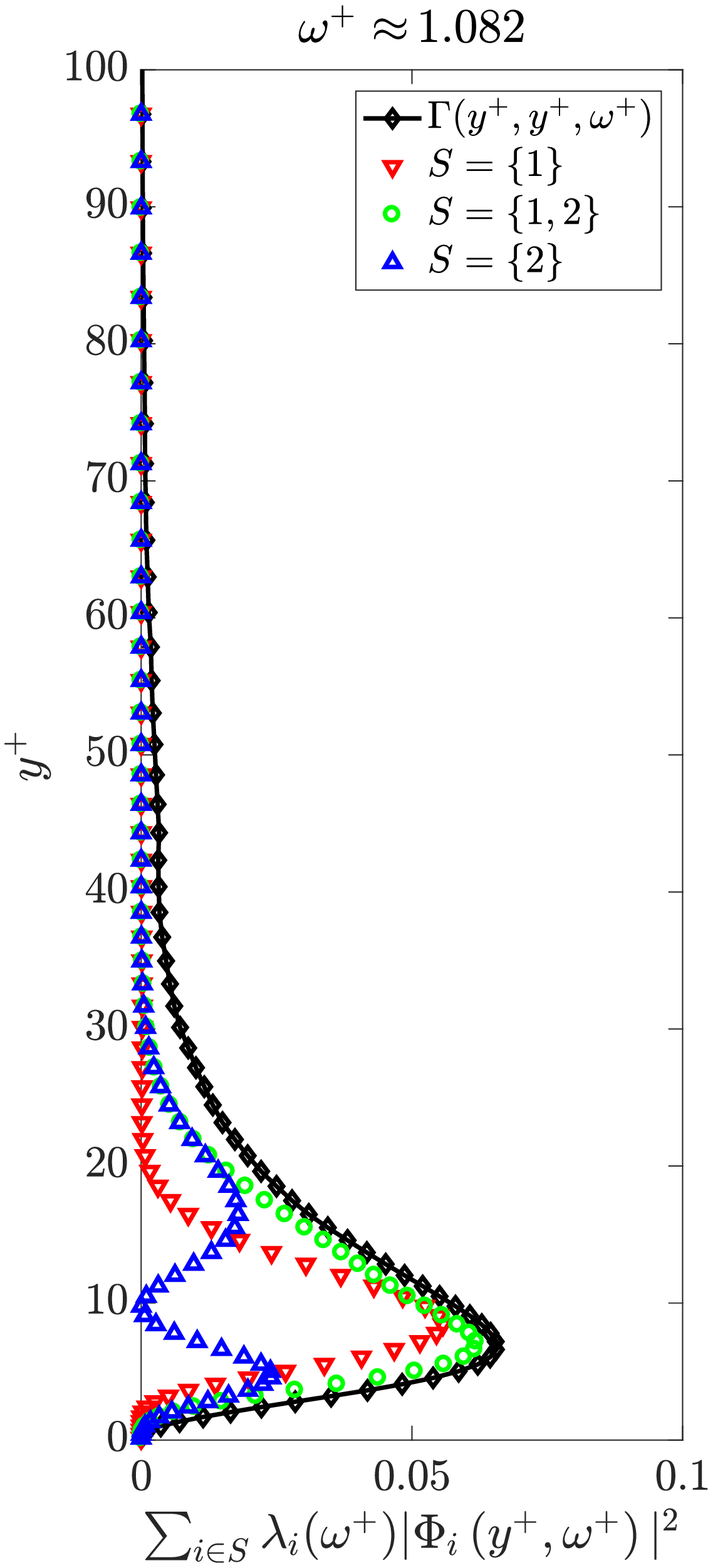}
      \put(1,95){\scriptsize$(c)$}
    \end{overpic}
  \end{subfigure}%
  \begin{subfigure}{.25\textwidth}
    \begin{overpic}[width=\linewidth]{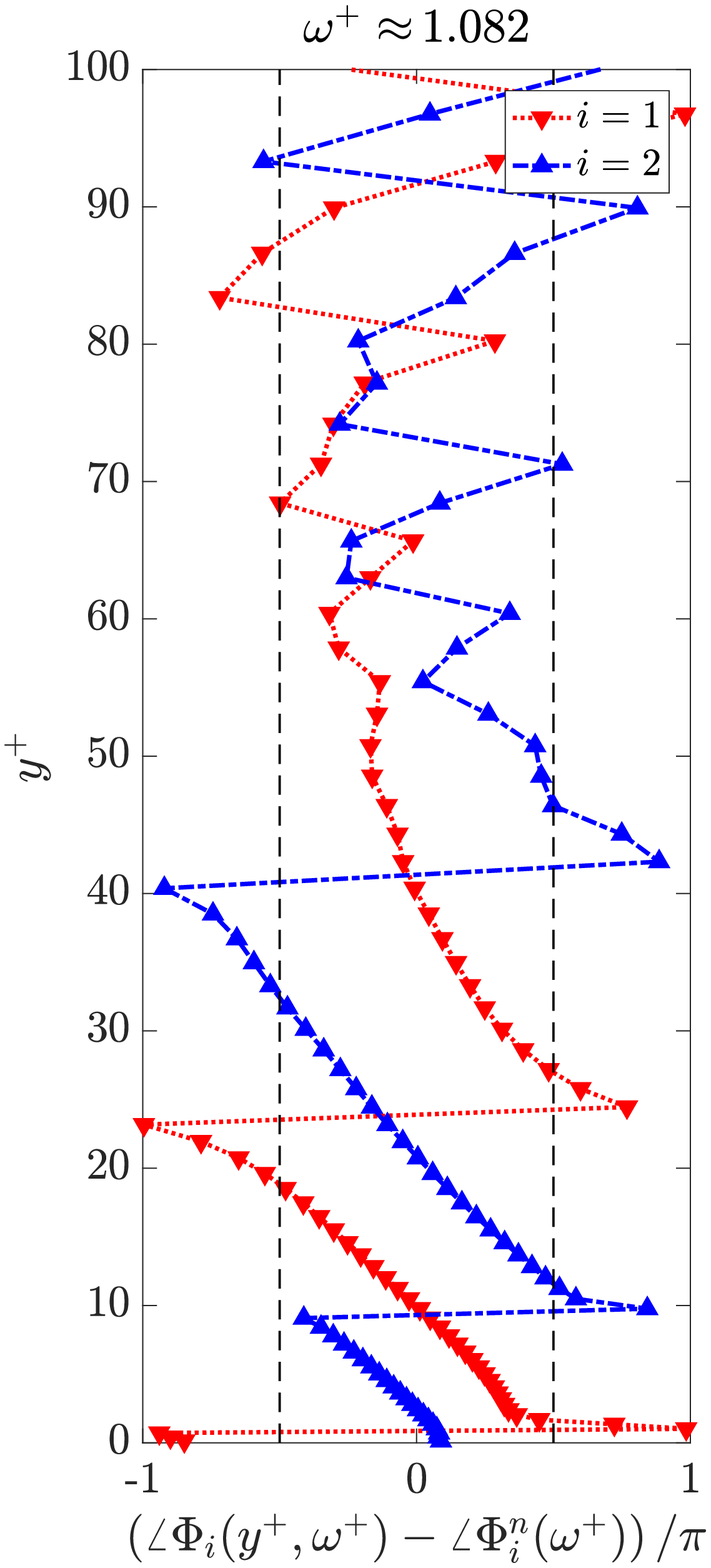}
      \put(1,95){\scriptsize$(d)$}
    \end{overpic}
  \end{subfigure}%
  \begin{subfigure}{.25\textwidth}
    \begin{overpic}[width=\linewidth]{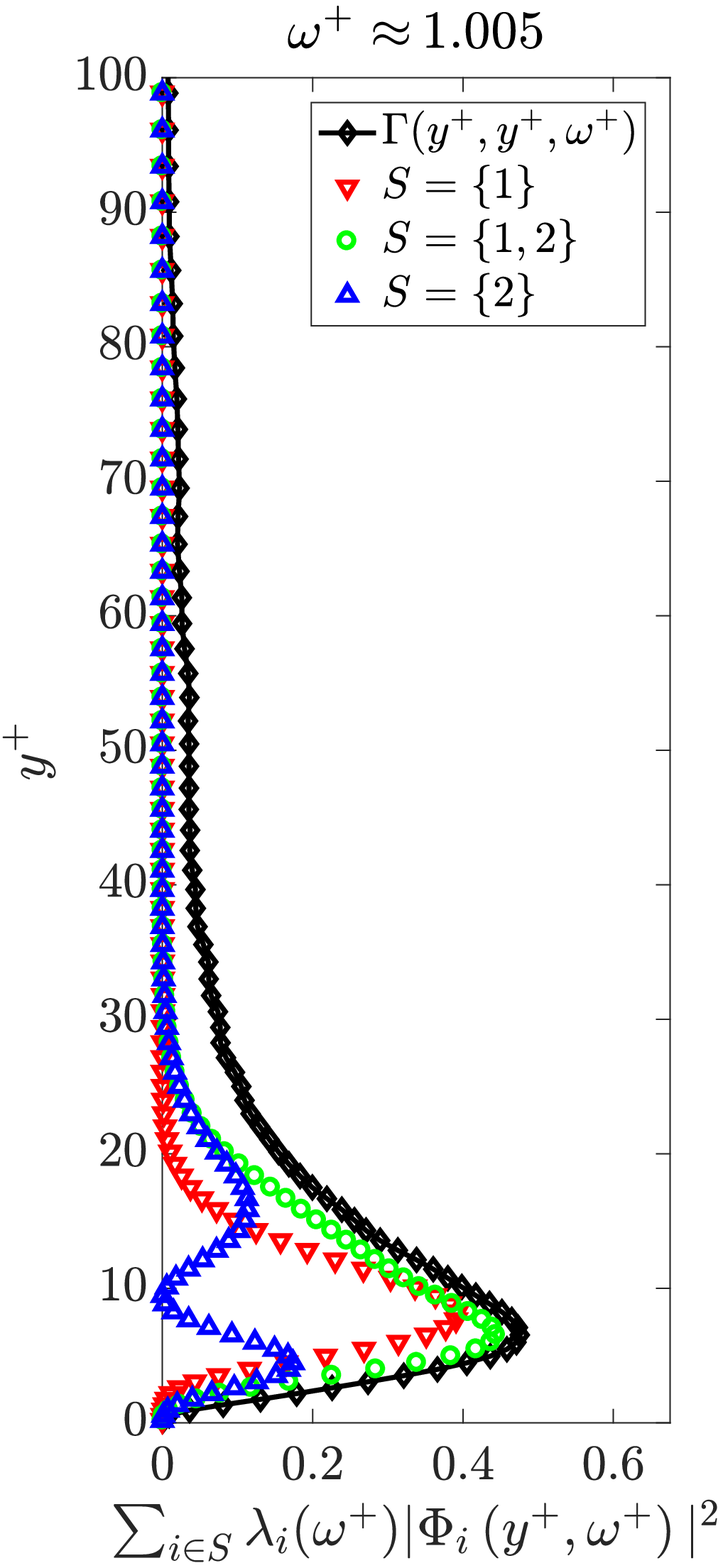}
      \put(1,95){\scriptsize$(g)$}
    \end{overpic}
  \end{subfigure}%
  \begin{subfigure}{.25\textwidth}
    \begin{overpic}[width=\linewidth]{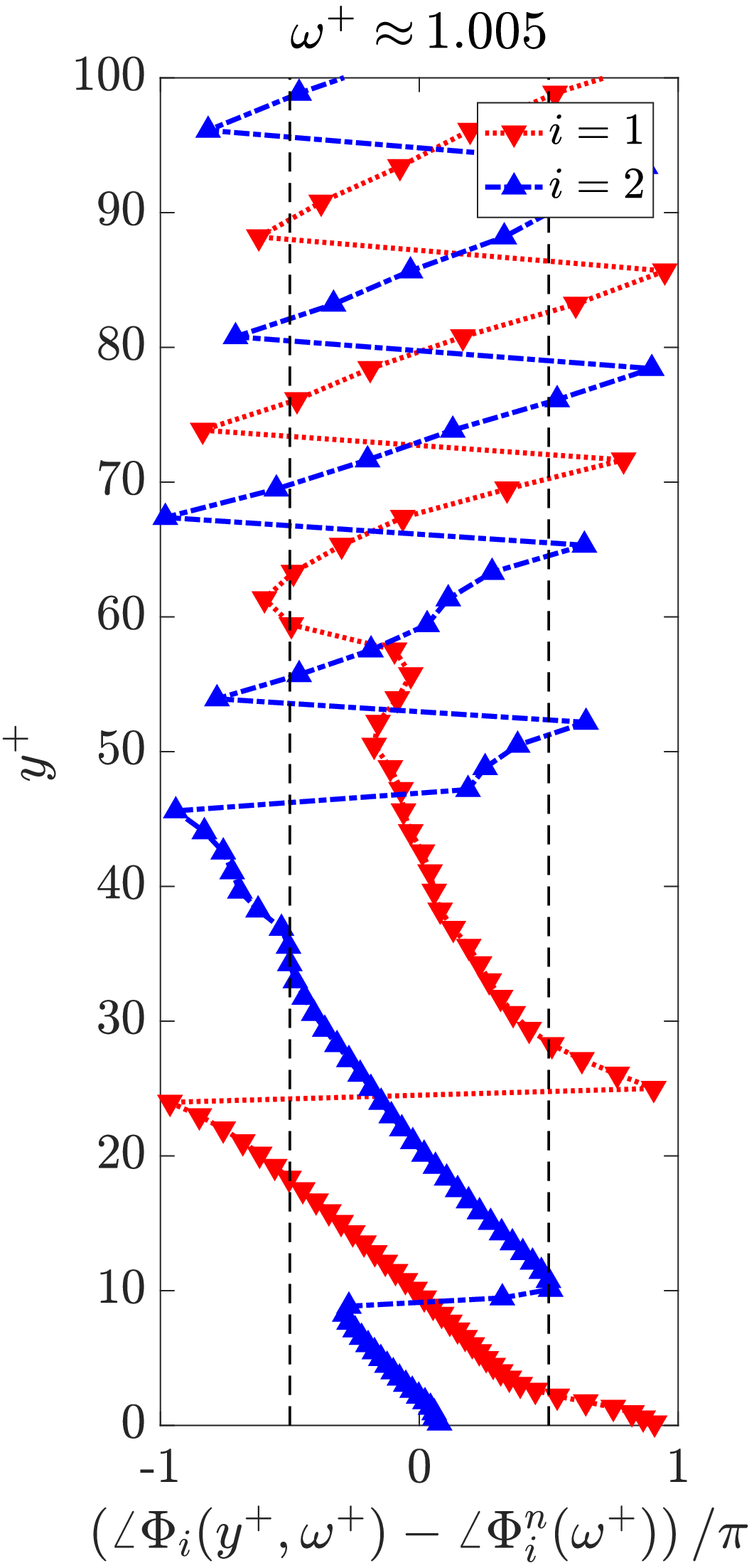}
      \put(1,95){\scriptsize$(h)$}
    \end{overpic}
  \end{subfigure}%
%
%
  \caption{Envelope (figures a, c) and phase (figures b, d) of the two dominant spectral POD modes computed using the $L^2$ inner product at different frequencies. Figures a-b and c-d are for $Re_{\tau}=180$ and $400$, respectively. The left and right dashed black solid line in figure b, d, f and h indicate $\angle\Phi_i(y^+,\omega^+)-\angle\Phi_i^n(\omega^+)$ equal to $-\pi/2$ and $\pi/2$, respectively.}

  \label{fig:highfreqspod}
\end{figure}

The magnitude and phase of the first two dominant spectral POD modes at a high frequency of $\omega^+\approx 1$ are shown in figure \ref{fig:highfreqspod}. Note that for this frequency, the dominant spectral POD and wall-pressure modes coincide. Clearly, we observe that the dominant modes resemble wavepackets. For both $Re_{\tau}$, the envelope and phase of the wavepackets have similar shape, which indicates similarity of the dominant modes at high frequencies. The envelope shows that dominant modes correspond to sources in the near wall region ($y^+<30$). The first and second dominant mode envelopes have one and two lobes respectively {\color{black}(figures \ref{fig:highfreqspod}a, c, e and g)}. Since the slope of the phase variation of both modes is negative {\color{black}near the wall} {\color{black}(figures \ref{fig:highfreqspod}b, d, f and h)}, equation \ref{eqn:wallnormalphasespeed} implies that these modes correspond to sources moving {\color{black} towards} the wall.

\begin{figure}
  \centering
  \begin{subfigure}{.25\textwidth}
    \begin{overpic}[width=\linewidth]{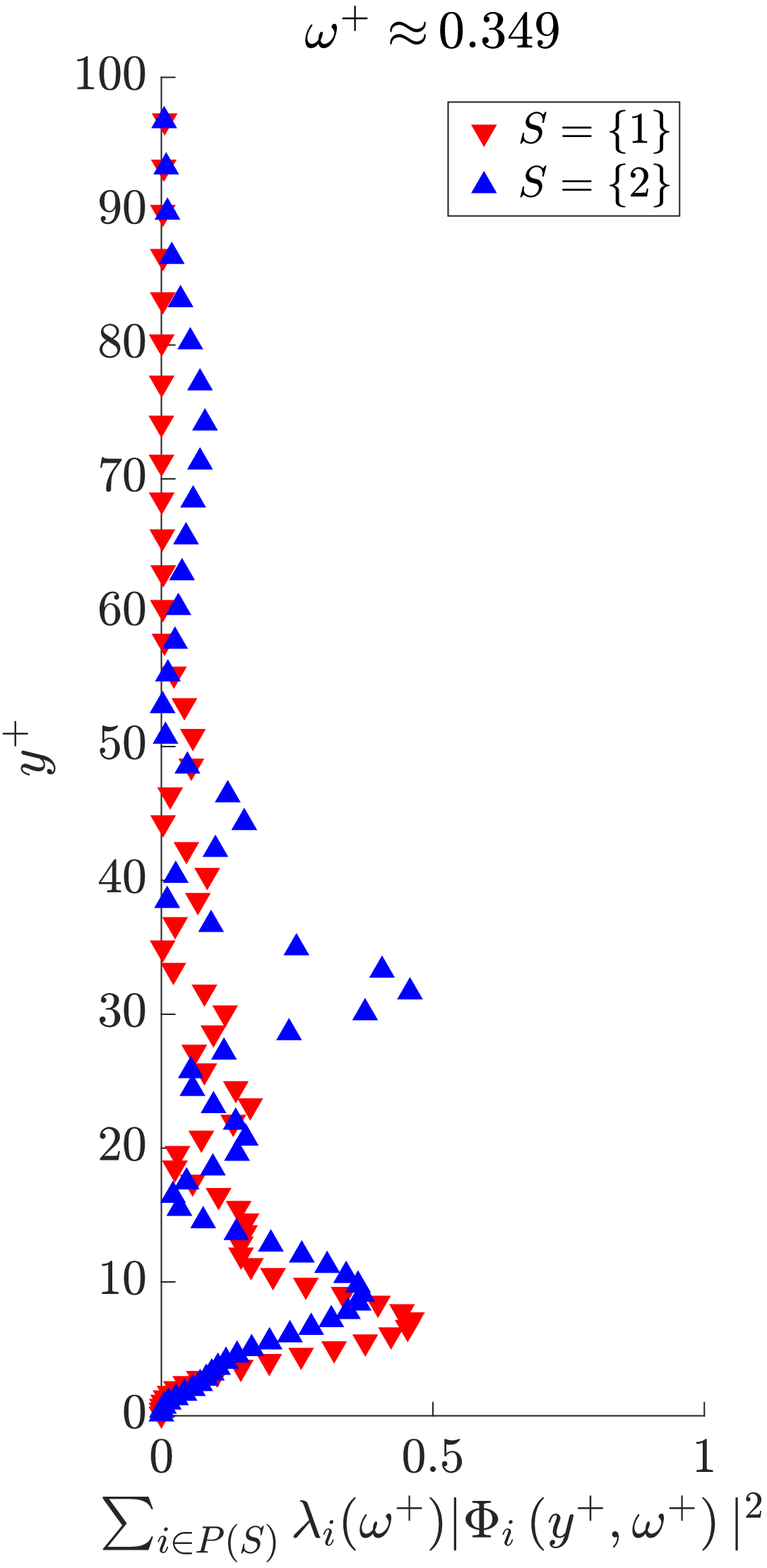}
      \put(1,95){\scriptsize$(a)$}
    \end{overpic}
  \end{subfigure}%
  \begin{subfigure}{.25\textwidth}
    \begin{overpic}[width=\linewidth]{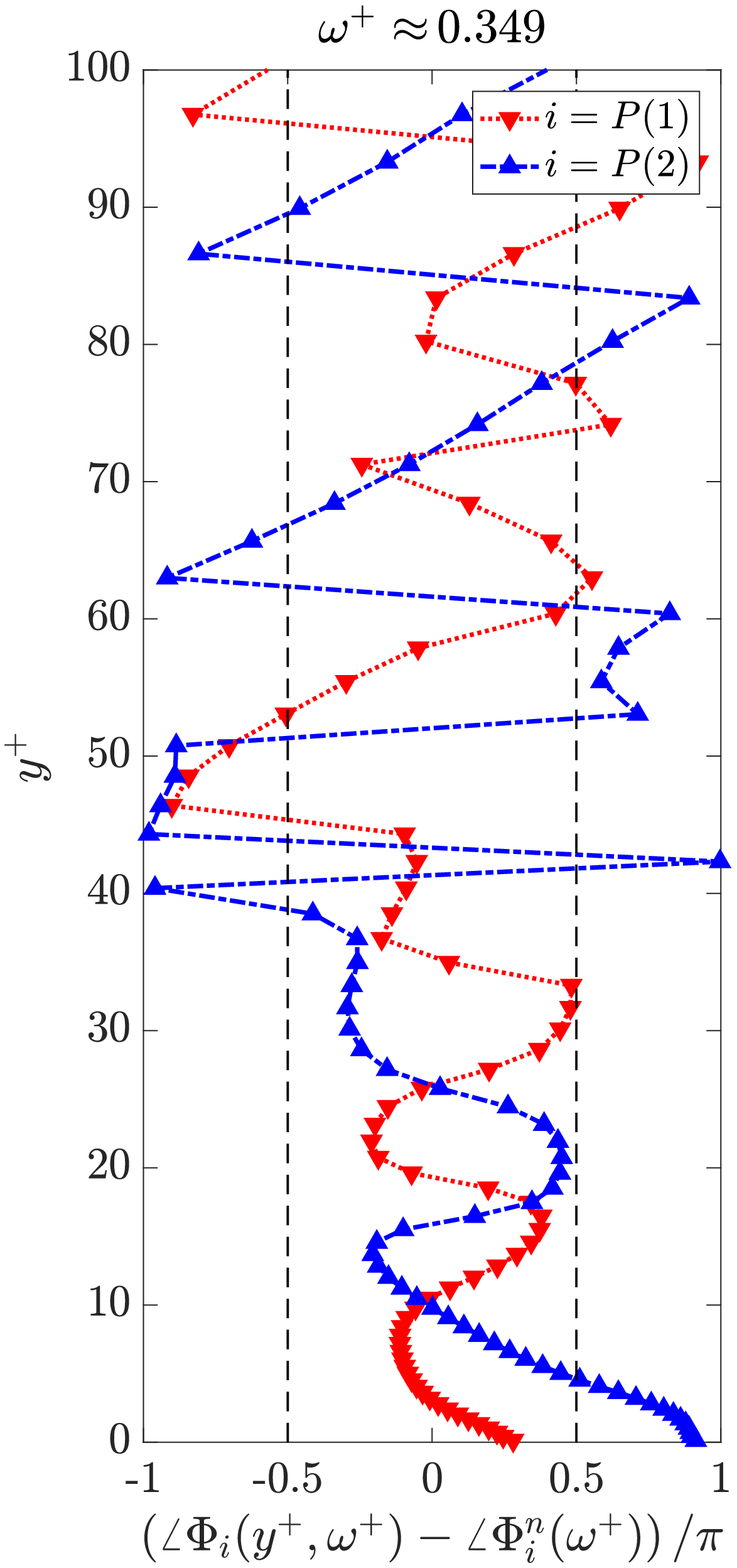}
      \put(1,95){\scriptsize$(b)$}
    \end{overpic}
  \end{subfigure}%
  \begin{subfigure}{.25\textwidth}
    \begin{overpic}[width=\linewidth]{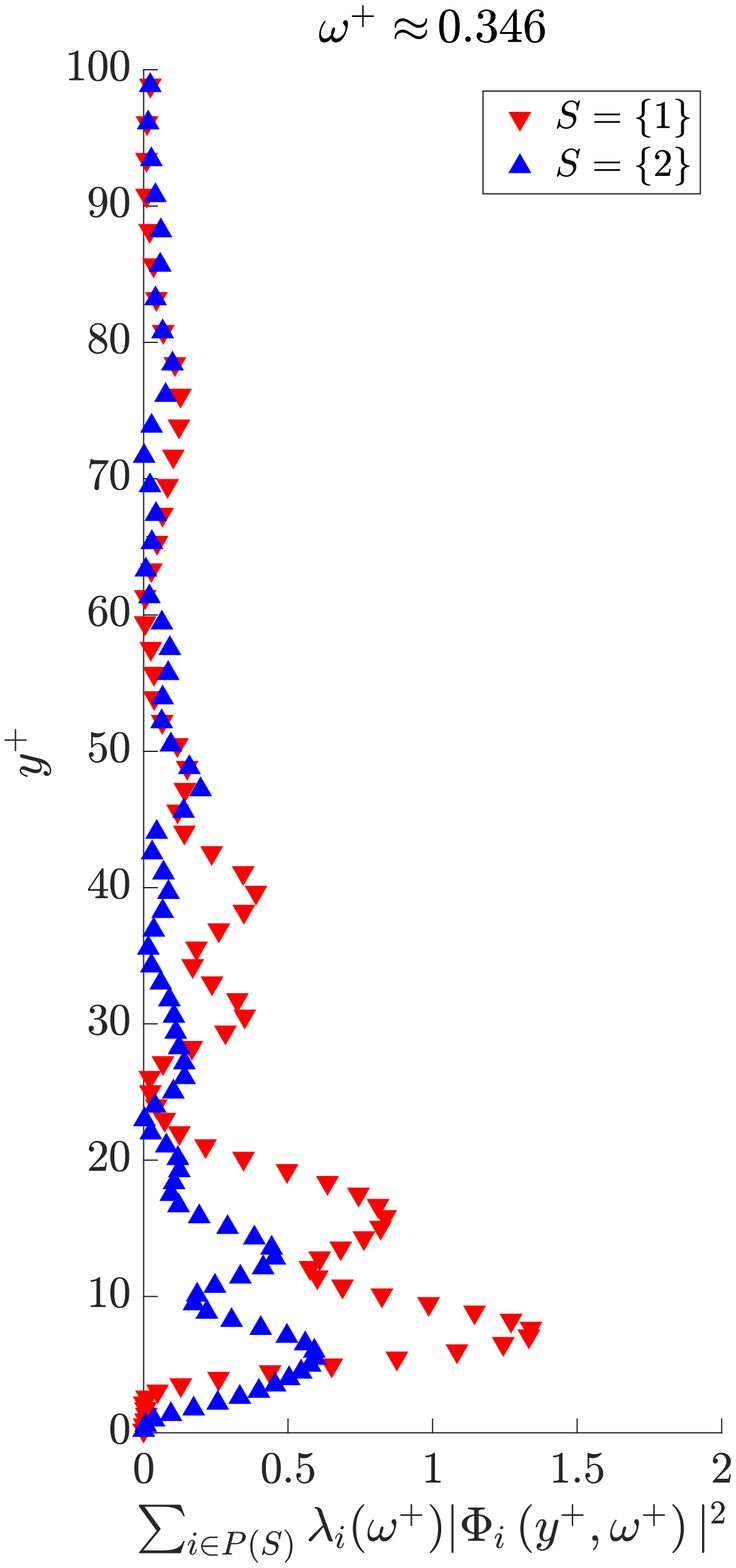}
      \put(1,95){\scriptsize$(c)$}
    \end{overpic}
  \end{subfigure}%
  \begin{subfigure}{.25\textwidth}
    \begin{overpic}[width=\linewidth]{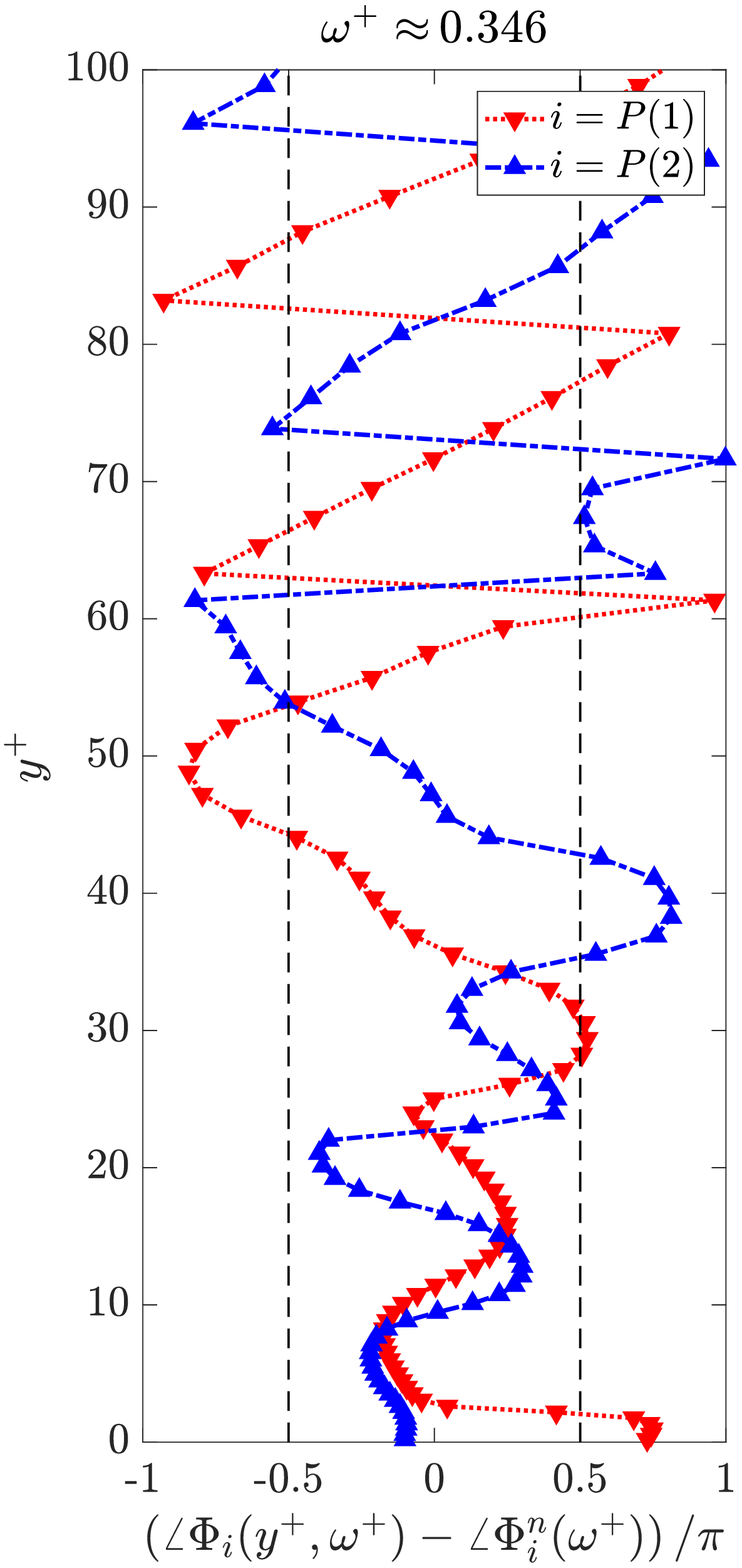}
      \put(1,95){\scriptsize$(d)$}
    \end{overpic}
  \end{subfigure}%
%
  \caption{Envelope (figures a, c) and phase (figures b, d) of the two
    dominant wall-pressure modes computed using the $L^2$ inner product at
    $\omega^+=0.35$. Figures a-b and c-d are for $Re_{\tau}=180$ and
    $400$, respectively. The left and right dashed black solid line in
    figures b and d indicate
    $\angle\Phi_i(y^+,\omega^+)-\angle\Phi_i^n(\omega^+)$ equal to
    $-\pi/2$ and $\pi/2$, respectively. $P(S)$ is the spectral POD
    mode index of the $S^{th}$ dominant wall-pressure mode.}
  \label{fig:domwallp_lowfreq}
\end{figure}

Next, we investigate the first two dominant wall-pressure modes at $\omega^+=0.35$, which {\color{black}together} contribute approximately $50\%$ to the wall-pressure PSD in figure \ref{fig:domwallp_lowfreq}. Note that the premultiplied spectra peak occurs at this frequency (figure \ref{fig:wallprespowerspec}b). The magnitude and phase variation shows that the these modes do not resemble a near-wall wavepacket. {\color{black} The envelope is not localized and the phase variation shows no sign of linear variation. Not much can be said of the pattern of these low frequency wall-pressure sources, except that} the contributions from different wall-normal regions undergo constructive interference. This is because the phase of the mode varies mostly between the two dashed lines. {\color{black} Further, several suboptimal spectral POD modes each contribute a small fraction the wall-pressure PSD at this frequency (figure \ref{fig:spodeigwt}). Thus, the individual dominant wall-pressure mode obtained using the $L^2$ inner product does not give us much information of the wall-pressure sources. However, the mode obtained using the Poisson inner product with $\beta=0.1$ (figure \ref{fig:domspod_lowfreq_wmat_alpha_0p1}) gives useful information of the wall-pressure source.

Therefore, spectral POD using the Poisson inner product performs better than the $L^2$ inner product in isolating dominant wall-pressure sources for both low and high frequencies. This is because the Poisson inner product decomposes the integral $\iint_{-\delta}^{+\delta}\frac{{G}(s,r,\frac{\beta}{1-\beta},0)}{1-\beta}\Gamma(r,s,\omega)\,\mathrm{dr}\,\mathrm{ds}$ as sum of eigenvalues. For small enough $\beta$, this integral is a good proxy for wall-pressure PSD. On the other hand, the $L^2$ inner product decomposes the integrated net source PSD ($\int_{-\delta}^{+\delta}\Gamma(r,r,\omega),\mathrm{dr}$) instead which is not a good proxy for wall-pressure PSD.

}
}

\section{Summary} \label{sec:conclusion}

We present a novel framework to analyze the sources of wall-pressure fluctuation in turbulent channel flow. A net source function $f_G(y,t)$ is defined whose integral in the wall-normal direction gives the wall-pressure fluctuation, i.e., $p(t)=\int_{-\delta}^{+\delta}f_G(y,t)\mathrm{dy}$. The spectral properties of the defined net source function are studied by computing its CSD using the generated DNS dataset at $Re_{\tau}=$ $180$ and $400$. The wall-pressure fluctuation - net source CSD shows a premultiplied peak at $\omega^+=0.35$ for both $Re_{\tau}$. The wall-normal location corresponding to the peak is $y^+=16.5$ and $18.4$ for $Re_{\tau}=180$ and $400$, respectively. Therefore, the peak in the premultiplied wall-pressure PSD at $\omega^+=0.35$ is due to the correlation with the sources in buffer layer. The wall-pressure fluctuation - net source CSD has a log-normal behavior in $y^+$ for $\omega^+>0.35$. The location of the dominant wall-parallel plane obtained from the mean of the log-normal profile varies exponentially with frequency. The wall-normal width of the dominant region obtained from the standard deviation of the log-normal profile is approximately proportional to the location of the dominant plane. At low frequencies, a dominant inner and overlap/outer region interaction is observed at both $Re_{\tau}$. 



{\color{black}We obtain the decorrelated net source patterns by performing spectral POD of the net source CSD using an inner product that has a symmetric positive definite kernel. The net source can be decomposed into active and inactive parts. The dominant spectral POD mode identified with this new inner product is active in the sense that it contributes to the entire wall-pressure PSD. The remaining portion of the net source constituted by the suboptimal POD modes is inactive in the sense that it does not contribute to wall-pressure PSD. Further, the active and inactive portions of the net source are decorrelated.

The dominant mode at the premultiplied PSD peak ($\omega^+\approx 0.35$) has a similar shape in inner units for both $Re_{\tau}$. It represents structures inclined in the downstream direction. At the low frequency linear PSD peak, the wall-normal contribution peaks at $y^+\approx 15$ and has a width of $y/\delta\approx 0.25$. The corresponding two-dimensional structure has a large scale vertical pattern similar to the observations of \cite{abe2005dns} in the instantaneous fields of rapid pressure.}

The analysis framework presented in this paper can be used to quantitatively understand the contribution of large scale coherent motions in the outer region at very high Reynolds numbers. Such contributions are believed to be the reason for the increasing low wavenumber contribution to wall-pressure RMS \citep{panton2017correlation}. The analysis has implications on wall modelled large eddy simulations (LES). The wall-pressure fluctuation net source CSD shows that sources correlated with the buffer layer are essential contributors to the premultiplied power spectra peak at $\omega^+= 0.35$. However, in wall modelled LES where the first point is in the logarithmic layer, one would not resolve the net source terms that lie in the buffer region. Hence, {\color{black} wall-modeled LES would fail} to accurately pblackict the wall-pressure spectra at high frequencies. {\color{black}These conclusions are consistent with \cite{bradshaw1967inactive} who noted the importance of buffer layer eddies to the higher frequencies in the wall-pressure spectra, and are consistent with \cite{park2016space} who attribute the errors in the high frequency slope of wall-pressure spectrum to the lack of resolution of the buffer layer eddies in their wall modeled LES.} {\color{black}Also, high Reynolds number effects like amplitude modulation of wall-pressure \citep{tsuji2016amplitude} can be studied using the above framework. The framework can also be used to quantitatively investigate the location of the sources that lead to $\omega^{-1}$ decay in the wall-pressure PSD at high Reynolds numbers.}


\section*{Acknowledgements}
This work is supported by the United States Office of Naval Research
(ONR) under grant N00014-17-1-2939 with Dr. Ki-Han Kim as the
technical monitor. The computations were made possible through
computing resources provided by the US Army Engineer Research and
Development Center (ERDC) in Vicksburg, Mississipi on the Cray X6,
Copper and Onyx of the High Performance Computing Modernization
Program.

\appendix
\section{Implementation details} \label{app:implndetails}
The theory presented in section \ref{subsec:analysistheory} consideblack
infinite domains in spanwise and streamwise directions. Here, we
present the implementation for finite periodic domains instead. The
integrals over the wavenumbers are replaced by a summation over the
discrete wavenumbers that can be represented in the periodic
domain. The wavenumber spacing is determined by the length of the
domain in each direction.

Let $N_t$ be the number of timesteps in each chunk used to compute the
Fast Fourier Transform (FFT), $N_T$ be the total number of timesteps
for which the data is acquiblack, $n_c$ be the number of chunks used for
temporal averaging the computed spectra, $T_c$ be the span of each
chunk and $p_{ovp}$ be the percentage overlap between subsequent
chunks.

The angular wavenumbers and frequencies are defined as
\begin{equation}
  \begin{split}
    k^x_l=\frac{2\pi l}{L_x}\,;\,k^z_m&=\frac{2\pi m}{L_z}\,;\,\omega_n=\frac{2\pi n}{T_c};\\
    l=-N_x/2,\dots,N_x/2-1;m&=-N_z/2,\dots,N_z/2-1;n=-N_t/2,\dots,N_t/2-1.
  \end{split}
\end{equation}

We store the source terms of the pressure Poisson equation in hard
disk from the finite volume solver. The domain in the finite volume
solver is split into multiple processors and each processor writes one
file per run containing the time history of the source terms of the
control volumes in its partition. A total of $\approx 8\,TB$ and
$\approx 30\,TB$ was requiblack to store the source terms of the
pressure Poisson equation for the $Re_{\tau}=180$ and $Re_{\tau}=400$
cases respectively.

The portion of the time series that corresponds to the current chunk
being processed is first converted to stationary frame of reference
and then written to a scratch space as wall-parallel slices. Let
$\bar{f}$ denote the four-dimensional source term array in the moving
frame of reference corrresponding to the current chunk. i.e.,
\begin{equation}
  \begin{split}
    \bar{f}&=\{\bar{f}_{i,j,k,l}\mid i=0,\dots,N_x-1,j=0,\dots,N_y-1,k=0,\dots,N_z-1,l=0,\dots,N_t-1\},\\
    \bar{f}_{i,j,k,l}&=D_{x_n}u_mD_{x_m}u_n\vert_{x_i,y_j,z_k,t_l},
  \end{split}
\end{equation}
where $D_{x_n}u_mD_{x_m}u_n$ is the discrete approximation to right
hand side of the pressure Poisson equation. The data is converted to
stationary frame of reference using Fourier interpolation and stoblack
in the source term array $f$ as
\begin{equation} \label{eqn:convtostat}
  \begin{split}
    &{f}=\{{f}_{i,j,k,l}\mid i=1,\dots,N_x,j=1,\dots,N_y,k=1,\dots,N_z,l=1,\dots,N_t\},\\
    &f_{i,j,k,l}=\sum_{m=-N_x/2}^{N_x/2-1}\tilde{f}_{m,j,k,l}\mathrm{e}^{-ik^x_{m}U_ct_{l}}\mathrm{e}^{ik^x_{m}x_{i}},\\
    &\tilde{f}_{m,j,k,l}=\frac{1}{N_x}\sum_{i=0}^{N_x-1}\bar{f}_{i,j,k,l}\mathrm{e}^{-ik^x_mx_i}.
  \end{split}
\end{equation}

Multiple processors are used to transfer the data from the Cartesian
decomposition of the solver to a wall-parallel decomposition of the
computational domain. The wall-parallel decomposition facilitates the
computation of the wavenumber frequency cross spectra of the source
terms. In order to obtain the fluctuation, the temporal mean of the
array $f$ at each spatial point is subtracted to ensure that it has
zero mean. i.e.,
\begin{equation}
  \begin{split}
    f_{i,j,k,l}&=f_{i,j,k,l}-\langle f \rangle_{i,j,k},\\
    \langle f
    \rangle_{i,j,k}&=\frac{1}{N_t}\sum_{l=0}^{N_t-1}f_{i,j,k,l}.
  \end{split}
\end{equation}

Each of the wall-parallel slices stoblack in the scratch space is then
Fourier transformed in streamwise, spanwise directions and in
time. Let $\hat{f}$ denote the Fourier transformed $f$. Then,
\begin{equation}
  \begin{split}
    &\hat{f}=\{\hat{f}_{i,j,k,l}\mid i=-N_x/2,\dots,N_x/2-1,j=0,\dots,N_y-1,\\
    &\quad \quad k=-N_z/2,\dots,N_z/2-1,l=-N_t/2,\dots,N_t/2-1\},\\
    &\hat{f}_{i,j,k,l}=\frac{1}{N_xN_zN_t}\sum_{m,n,p=1}^{N_x-1,N_z-1,N_t-1}f_{m,j,n,p}w_p\mathrm{e}^{-i\left(k^x_ix_m+k^z_kz_n+\omega_lt_p\right)},
  \end{split}
\end{equation}
where $w_p=sin^2(\pi p/N_t)$ is the Hanning Window function multiplied
with the time series in order to avoid spectral leakage. The
wall-parallel slice data is over written by its three-dimensional
Fourier transform. The processors are split in the wall-normal and
time directions to carry out the task in parallel and we use the
parallel-FFTW \citep{frigo2005design} library to carry out the Fourier
transform.

As discussed in the previous section, the memory requirement to store
the five-dimensional function $\phi_{ff}(r,s,k_1,k_3,\omega)$ is too
large. We store and append the net source cross spectral density sum
array $\Gamma^s$ (defined below) instead. The possible
$\{r_i,s_j\}_{i,j=1}^{N_y}$ pairs are split among multiple
processors. For each $(r_i,s_j)$ pair, we read the arrays
$\hat{f}_{:,i,:,:}$ and $\hat{f}_{:,j,:,:}$ from the scratch space and
update the sum $\Gamma_{i,j,:}^s$ as
\begin{equation} \label{eqn:gammasupdate}
  \begin{split}
    \Gamma^s&=\{\Gamma^s_{i,j,k}\mid i=1,\dots,N_y,j=1,\dots,N_y,k=-N_t/2,\dots,N_t/2-1\},\\
    \Gamma^s_{i,j,k}&=\Gamma^s_{i,j,k}+\frac{8}{3}\frac{T}{2\pi}\frac{L_1}{2\pi}\frac{L_3}{2\pi}\sum_{l=-N_x/2}^{N_x/2-1}\sum_{m=-N_z/2}^{N_z/2-1}\hat{f}^*_{l,i,m,k}\hat{f}_{l,j,m,k}G^*_{i,l,m}G_{j,l,m}\frac{2\pi}{L_1}\frac{2\pi}{L_3},\\
    G_{i,l,m}&=G(0,y_i,k_l,k_m).
  \end{split}
\end{equation}
The factor $8/3$ in the above equation accounts for the blackuction in
the spectral magnitude due to windowing \citep{bendat2011random}. The
update to $\Gamma^s_{i,j,:}$ given in the above equation
\ref{eqn:gammasupdate} is carried out in chunks along the frequency
dimension due to limited memory available in a cluster node. The net
source cross spectral density $\Gamma$ array is then defined by
dividing the $\Gamma^s$ array by the number of chunks $n_c$, i.e.
\begin{equation}
  \Gamma=\{\Gamma_{i,j,k}\mid \Gamma_{i,j,k}=\Gamma^s_{i,j,k}/n_c,\, i=1,\dots,N_y,j=1,\dots,N_y,k=-N_t/2,\dots,N_t/2-1\}.
\end{equation}

We store and append only half of the entire $\Gamma^s$ array since
$\Gamma_{j,i,k}=\Gamma_{i,j,k}^*$. We use $50\%$ overlap between the
chunks to increase statistical convergence. As new chunk data become
available, the net source cross spectral density $\Gamma^s$ is
updated.

Note that the Green's function had to be evaluated in quadruple
precision for $Re_{\tau}=400$ because for some wavenumbers, both the
numerator and denominator were so large that it could not be stoblack in
double precision. However, when divided, the resulting number could be
stoblack in double precision. The above post-processing methodology is
parallel, aware of the limited memory available in a supercomputer
cluster node and can be used to analyze even larger channel flow
datasets obtained for higher friction Reynolds numbers.

{\color{black} To obtain the spectral POD modes, we first obtain the eigenvalues $\{\lambda_{i,l}\}_{i=1}^{N_y}$ and the eigenvectors $\{\bar{\varphi}_{i,l}\}_{i=1}^{N_y}$ of the problem
\begin{equation}
  \begin{split}
    A_l\bar{\varphi}_{i,l}&=\lambda_{i,l}W\bar{\varphi}_{i,l}\,;\,i=1,\dots,N_y,\,l=-N_t/2,\dots,N_t/2-1,\\
    A_l&=\{A_l\in \mathbb{C}^{N_y\times N_y}\mid \{A_l\}_{m,n}=\Delta y_m \Gamma_{m,n,l} \Delta y_n \},
  \end{split}
\end{equation}
where the matrix $W$ is the finite volume discretization of the operator $\left(-(1-\alpha)\frac{\partial^2}{\partial y^2} +\alpha\right)$. The spectral POD eigenvalues are $\{\lambda_{i,l}\}_{i=1}^{N_y}$ and eigenvectors are $\{\varphi_{i,l}\}_{i=1}^{N_y}$, where $\{\varphi_{i,l}\}_{i=1}^{N_y}$ is related to $\{\bar{\varphi}_{i,l}\}_{i=1}^{N_y}$ as
\begin{equation}
\begin{split}
  \varphi_{i,l}&=D^{-1}W\bar{\varphi}_{i,l}\,;\,i=1,\dots,N_y,\,l=-N_t/2,\dots,N_t/2-1,\\
  D&=\{D\in \mathbb{C}^{N_y\times N_y}\mid \{D\}_{m,n}=\Delta y_m\delta_{mn}\}
\end{split}
\end{equation}

\begin{figure}
  \centering
  \begin{subfigure}{.5\textwidth}
    \begin{overpic}[width=\linewidth]{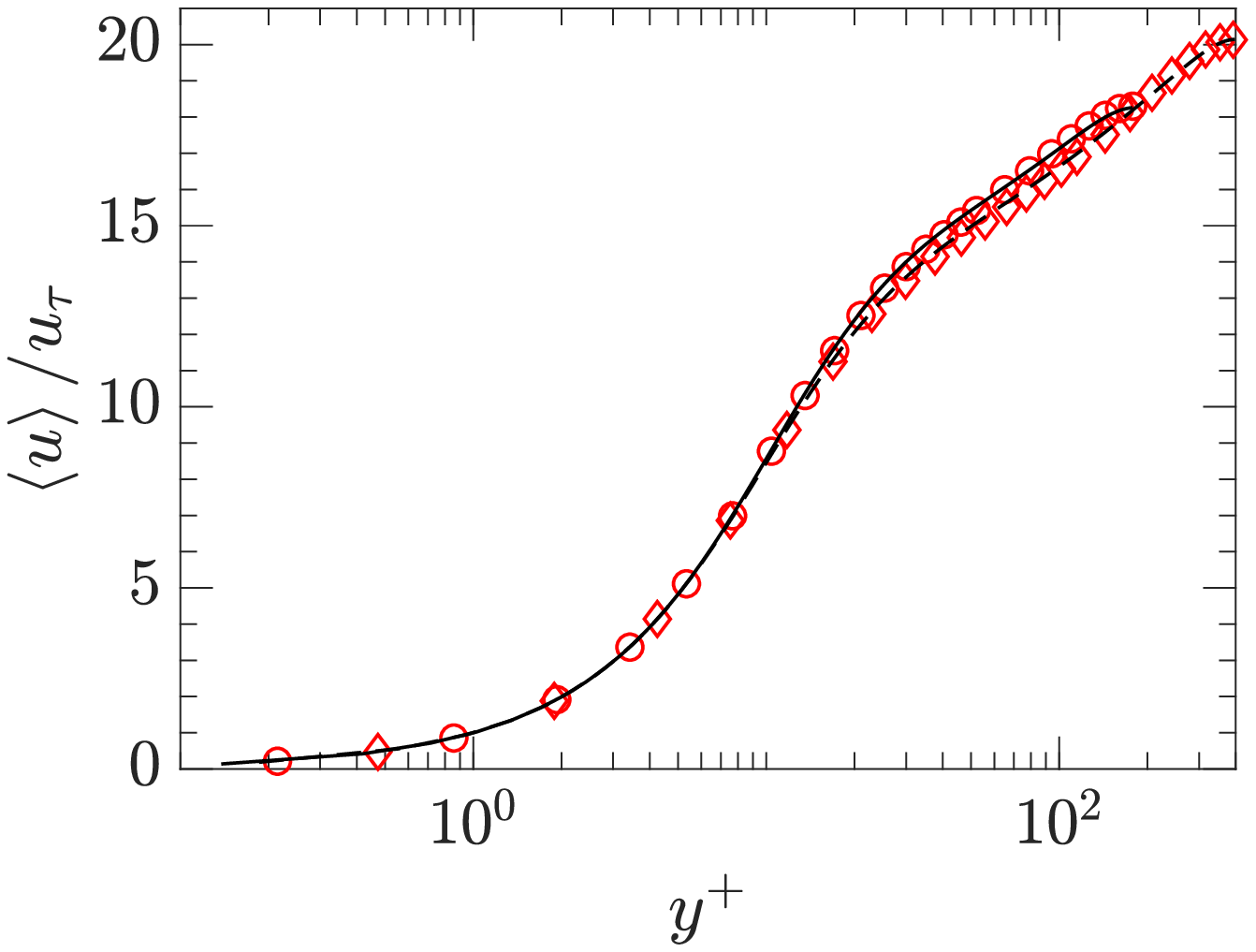}
      \put(1,70){\scriptsize$(a)$}
    \end{overpic}
  \end{subfigure}%
  \begin{subfigure}{.5\textwidth}
    \begin{overpic}[width=\linewidth]{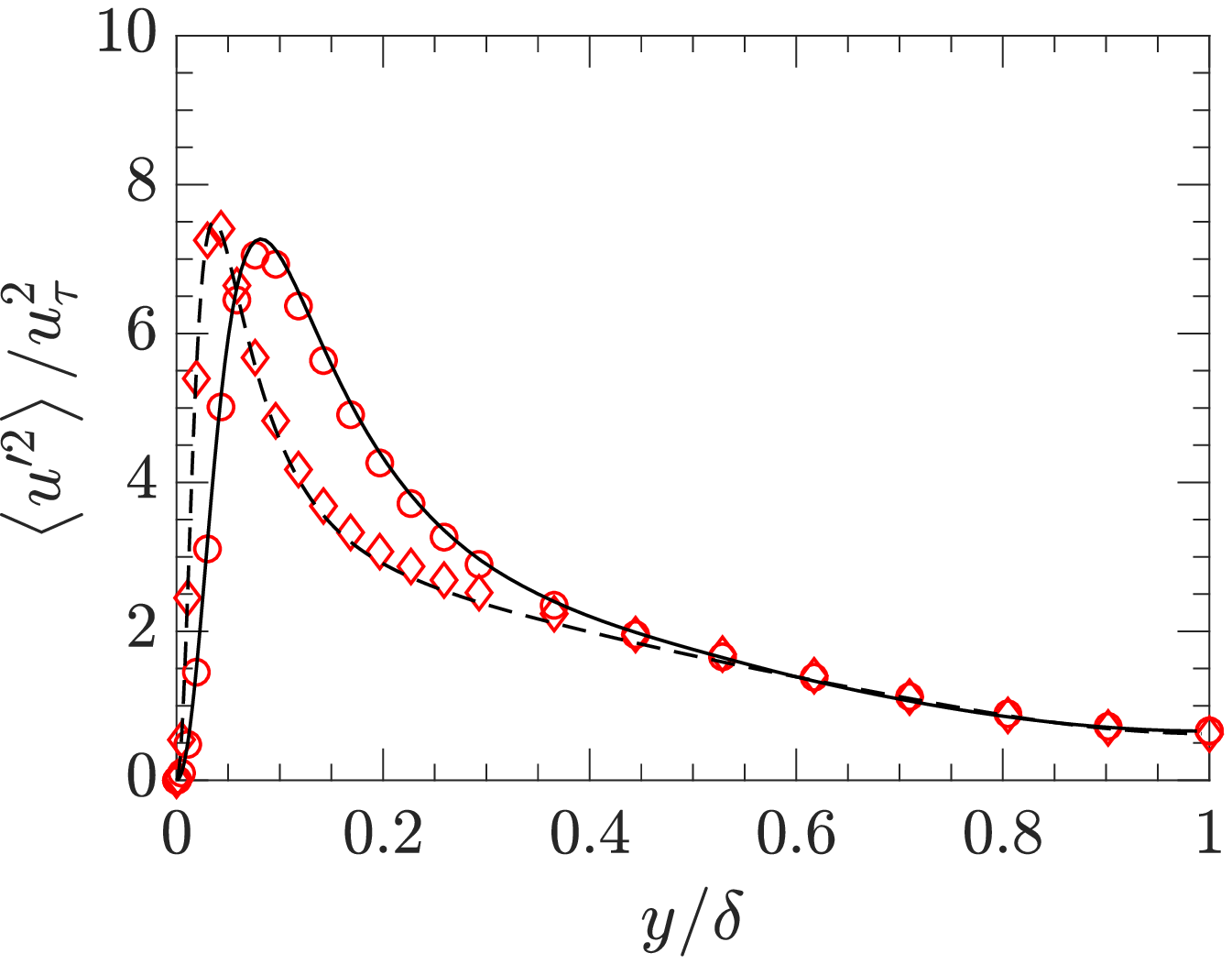}
      \put(1,70){\scriptsize$(b)$}
    \end{overpic}
  \end{subfigure}
  \begin{subfigure}{.5\textwidth}
    \begin{overpic}[width=\linewidth]{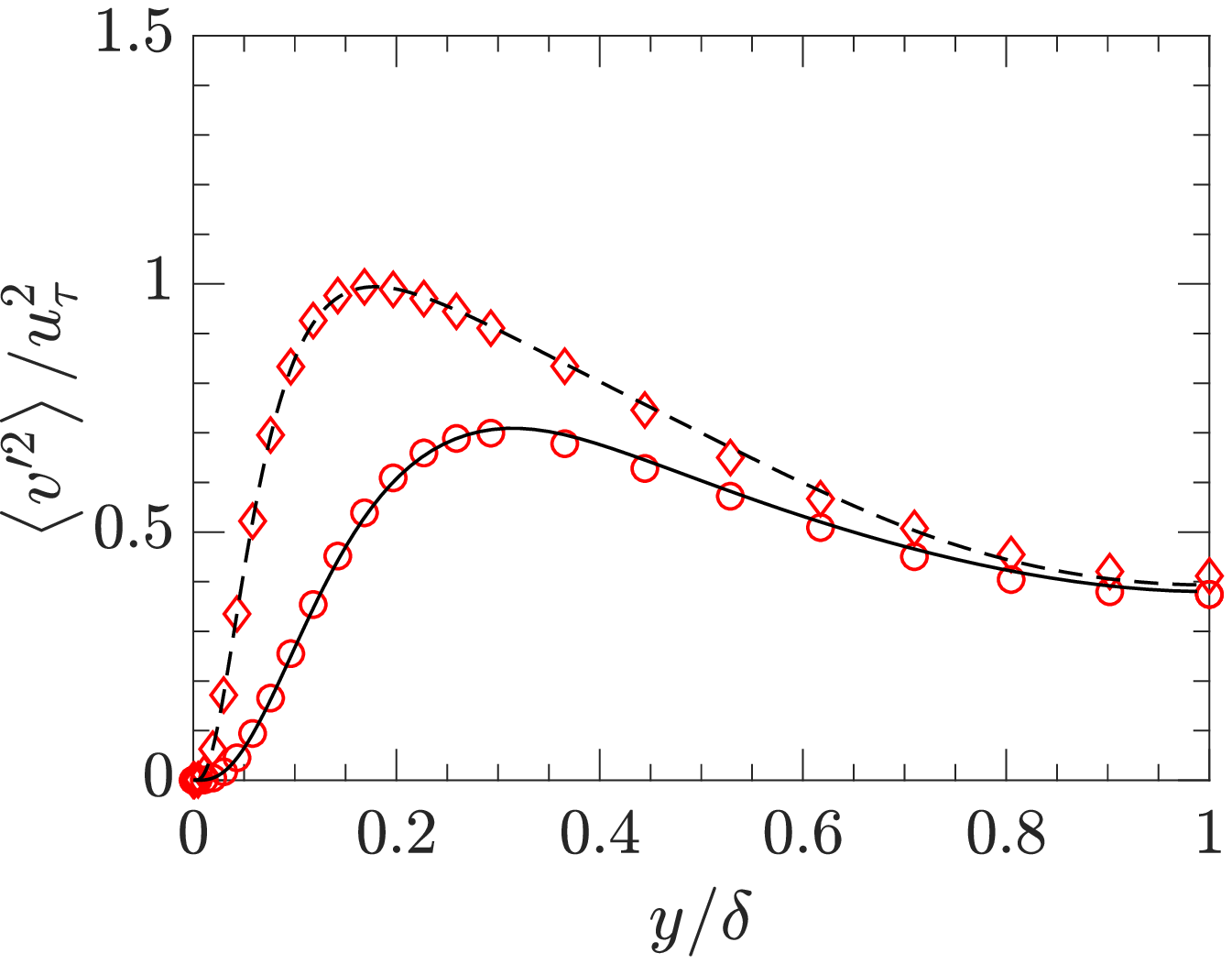}
      \put(1,70){\scriptsize$(c)$}
    \end{overpic}
  \end{subfigure}%
  \begin{subfigure}{.5\textwidth}
    \begin{overpic}[width=\linewidth]{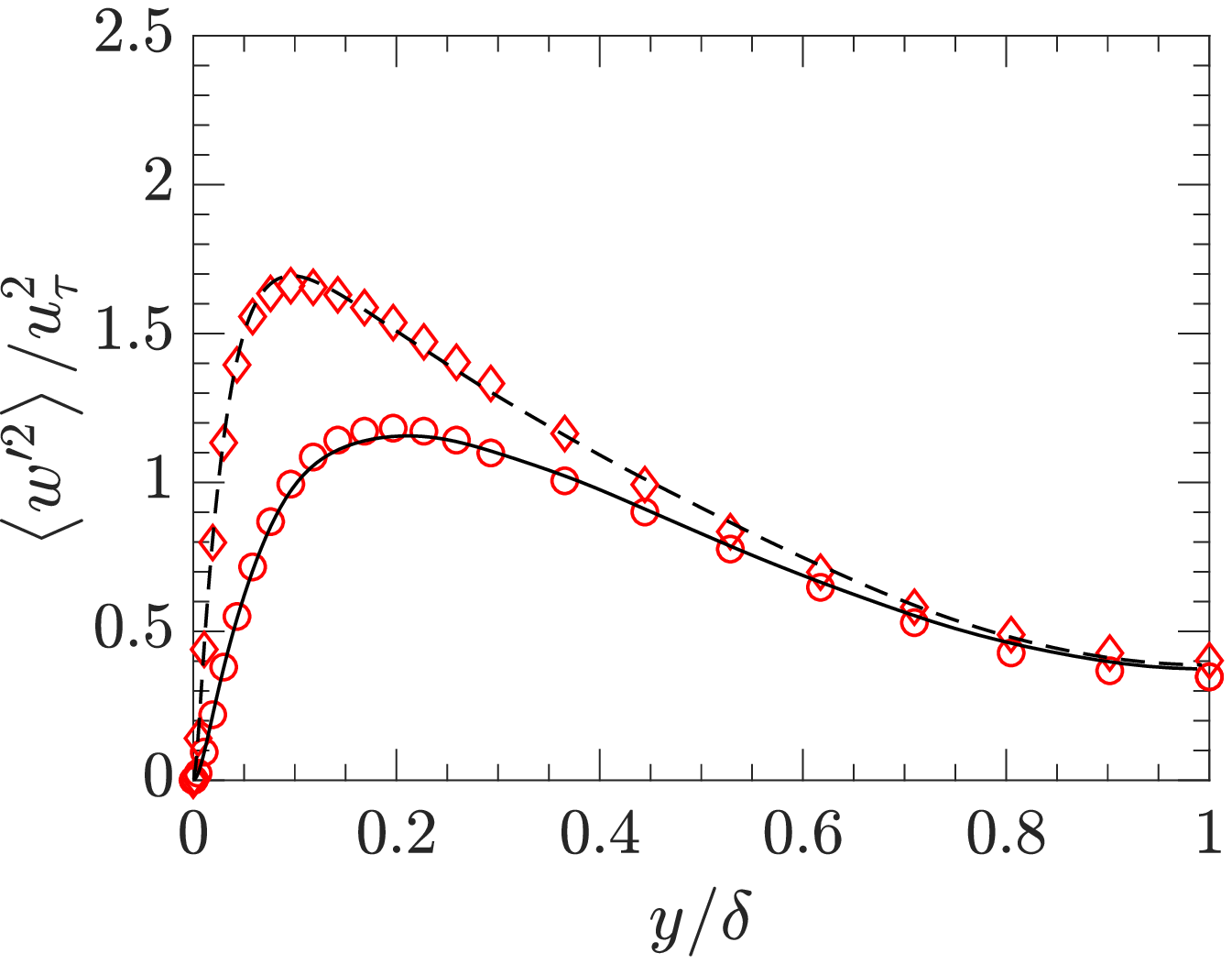}
      \put(1,70){\scriptsize$(d)$}
    \end{overpic}
  \end{subfigure}
  \begin{subfigure}{.5\textwidth}
    \begin{overpic}[width=\linewidth]{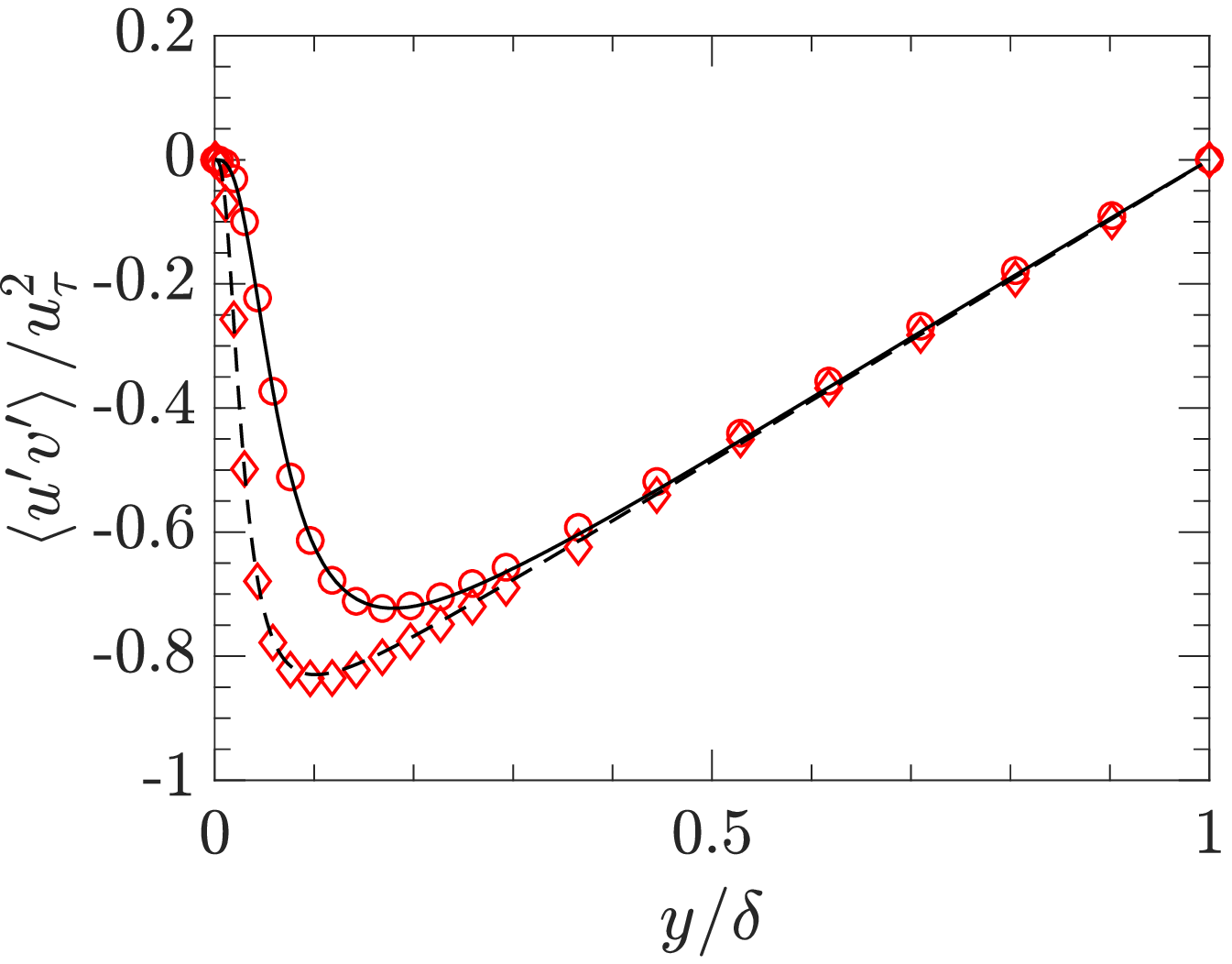}
      \put(1,70){\scriptsize$(e)$}
    \end{overpic}
  \end{subfigure}%
  \begin{subfigure}{.5\textwidth}
    \begin{overpic}[width=\linewidth]{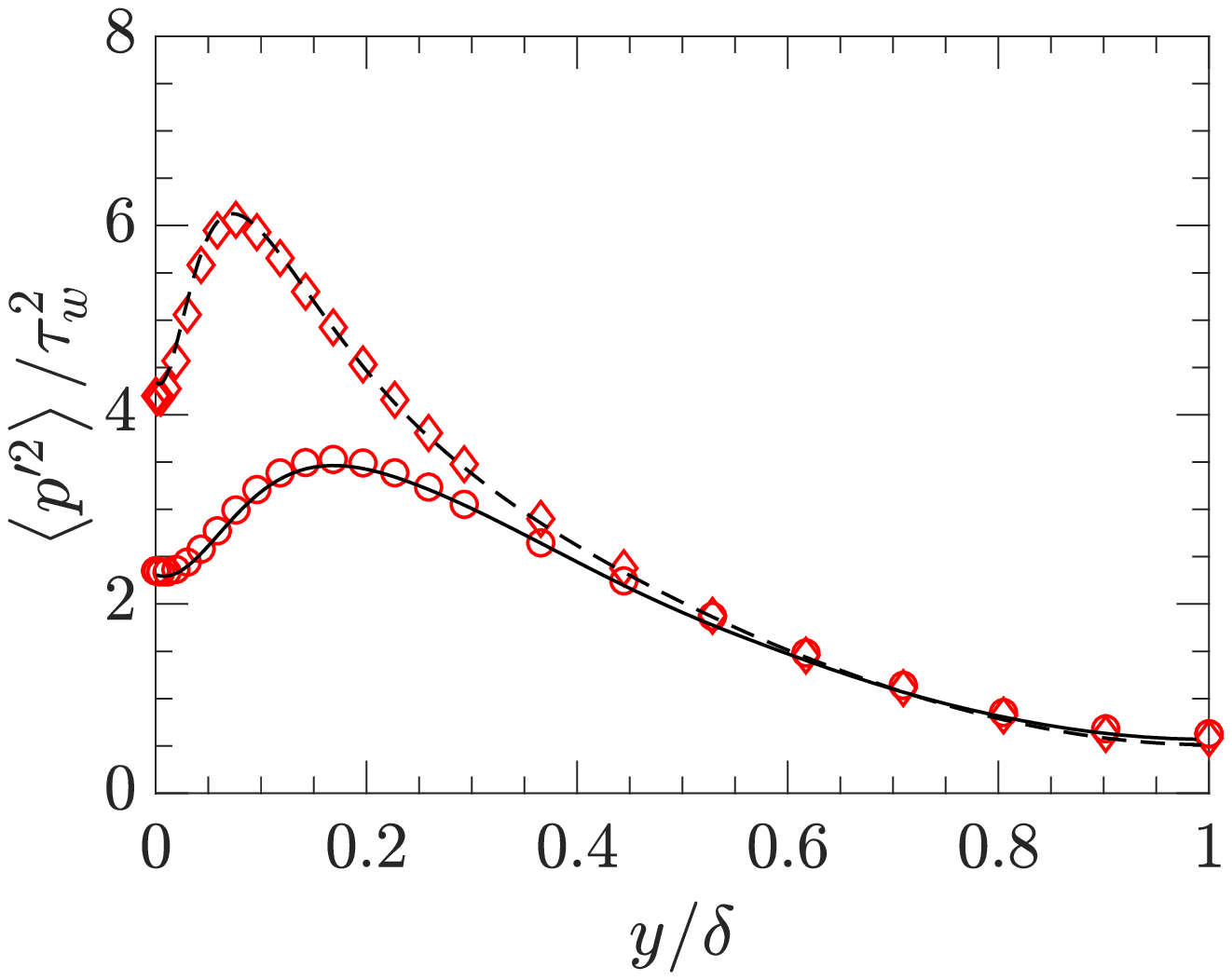}
      \put(1,70){\scriptsize$(f)$}
    \end{overpic}
  \end{subfigure}
  \caption{Velocity field and pressure fluctuation statistics. Solid and dashed lines denote the current DNS result at $Re_{\tau}=180$ and $400$, respectively. Circle and diamond symbols denote DNS data at $Re_{\tau}=182$ and $392$ from \cite{moser1999direct}. Figure a compares mean streamwise velocity, figures b, c, d compare mean-squablack streamwise, wall-normal and spanwise velocity fluctuation, respectively, and figures e and f compare mean tangential Reynolds stress and mean-squablack pressure fluctuation, respectively.}
  \label{fig:valid_stat}
\end{figure}

\begin{figure}
  \centering
  \begin{subfigure}{.33\textwidth}
    \begin{overpic}[width=\linewidth]{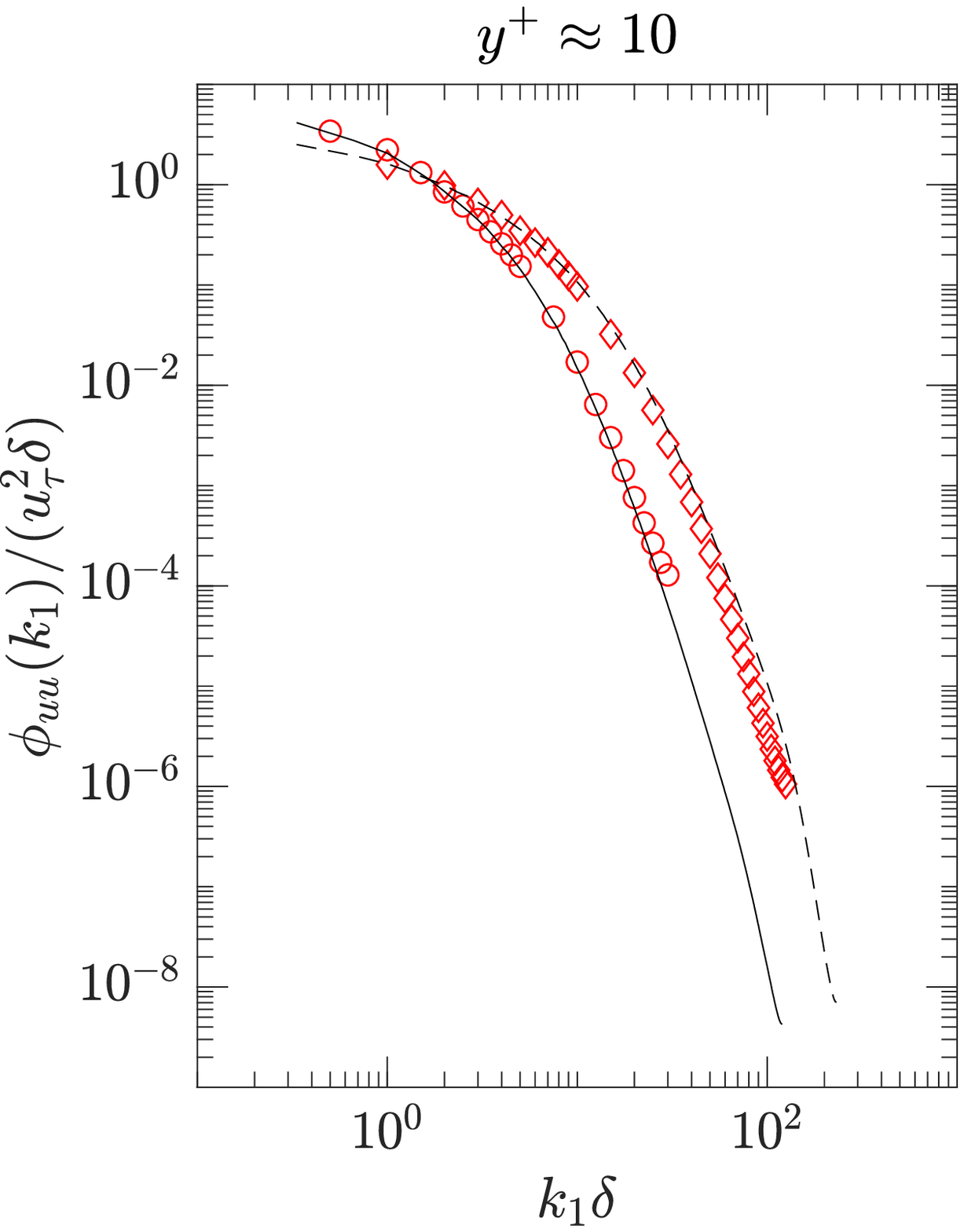}
      \put(1,90){\scriptsize$(a)$}
    \end{overpic}
  \end{subfigure}%
  \begin{subfigure}{.33\textwidth}
    \begin{overpic}[width=\linewidth]{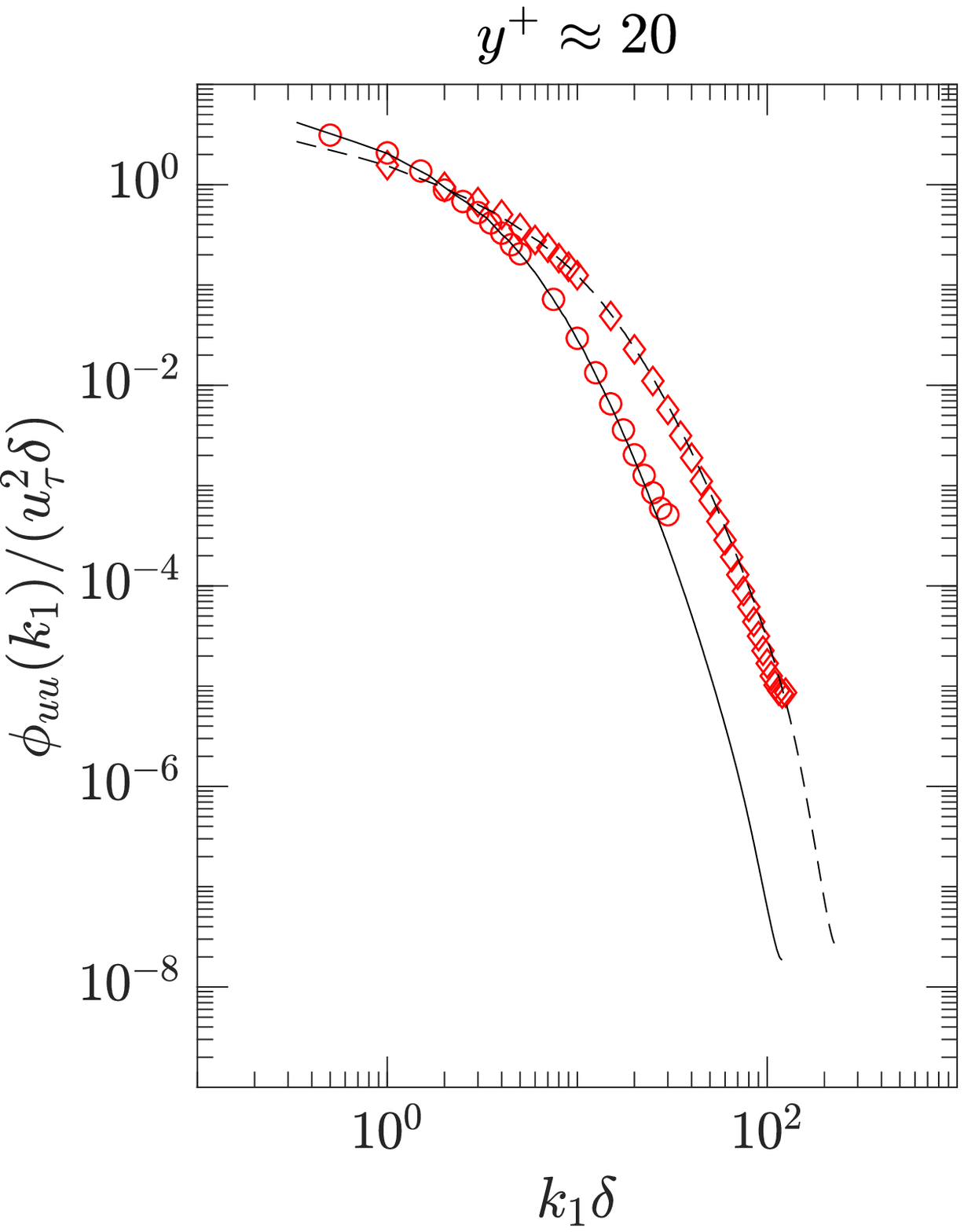}
      \put(1,90){\scriptsize$(b)$}
    \end{overpic}
  \end{subfigure}%
  \begin{subfigure}{.33\textwidth}
    \begin{overpic}[width=\linewidth]{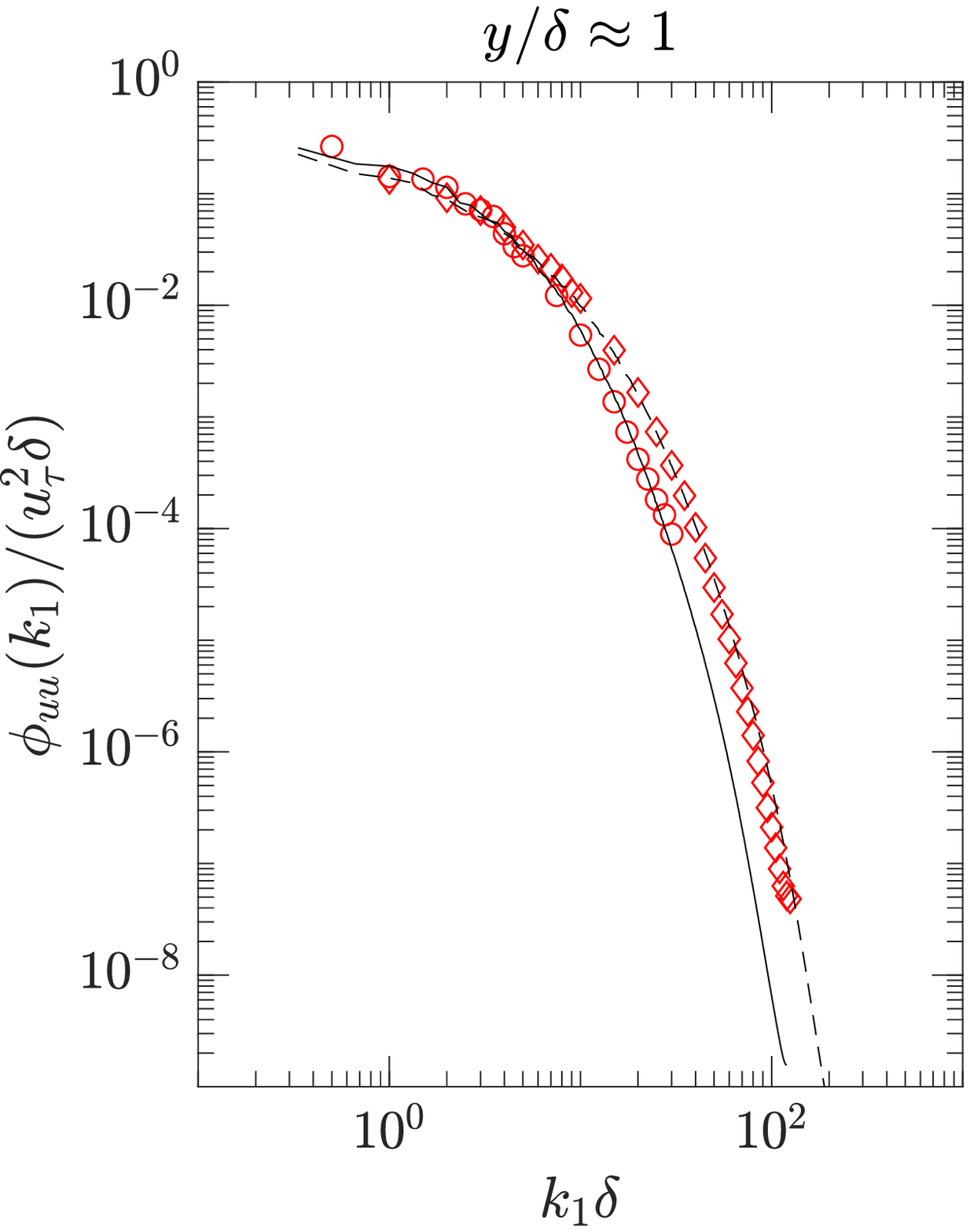}
      \put(1,90){\scriptsize$(c)$}
    \end{overpic}
  \end{subfigure}%

  \begin{subfigure}{.33\textwidth}
    \begin{overpic}[width=\linewidth]{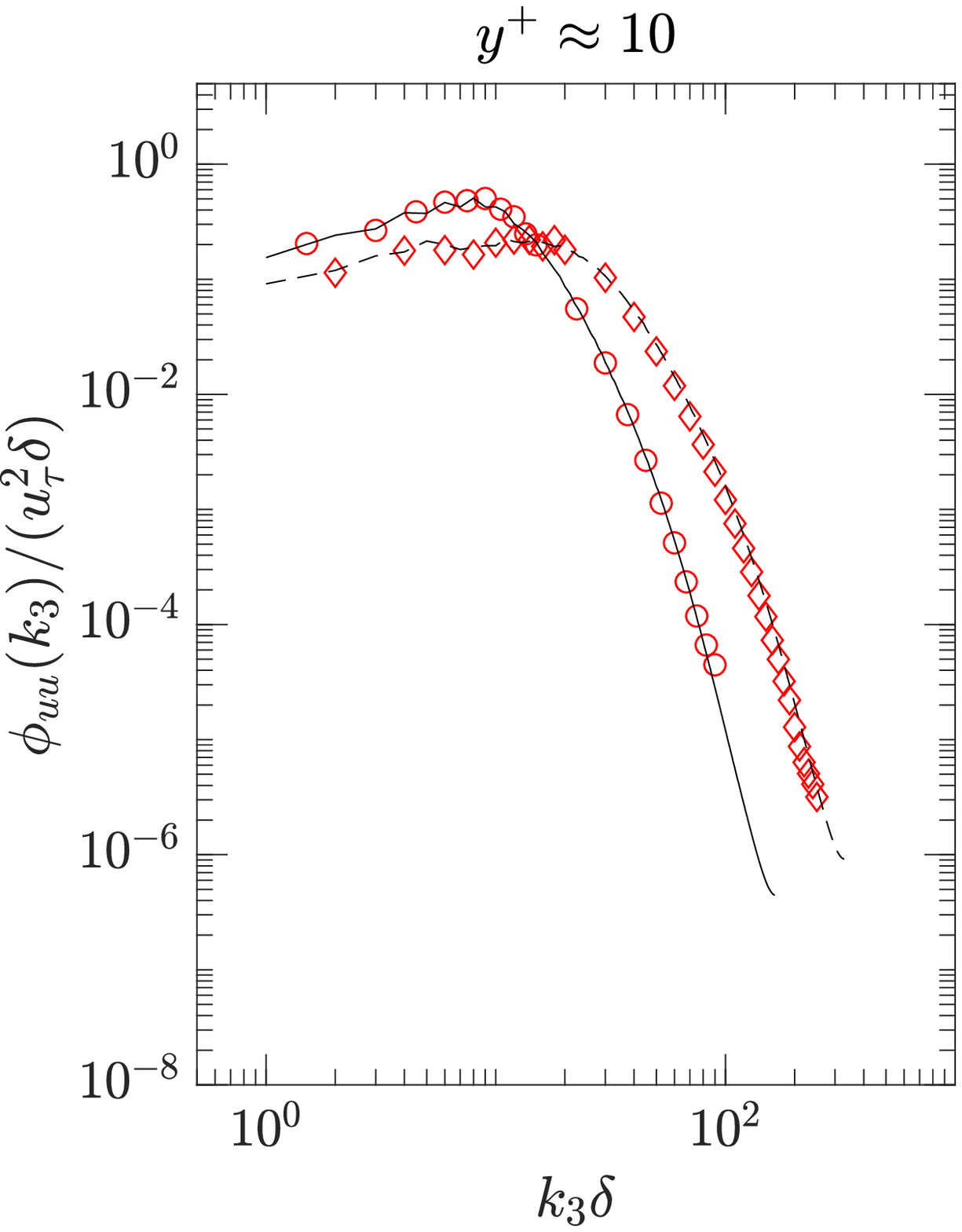}
      \put(1,90){\scriptsize$(d)$}
    \end{overpic}
  \end{subfigure}%
  \begin{subfigure}{.33\textwidth}
    \begin{overpic}[width=\linewidth]{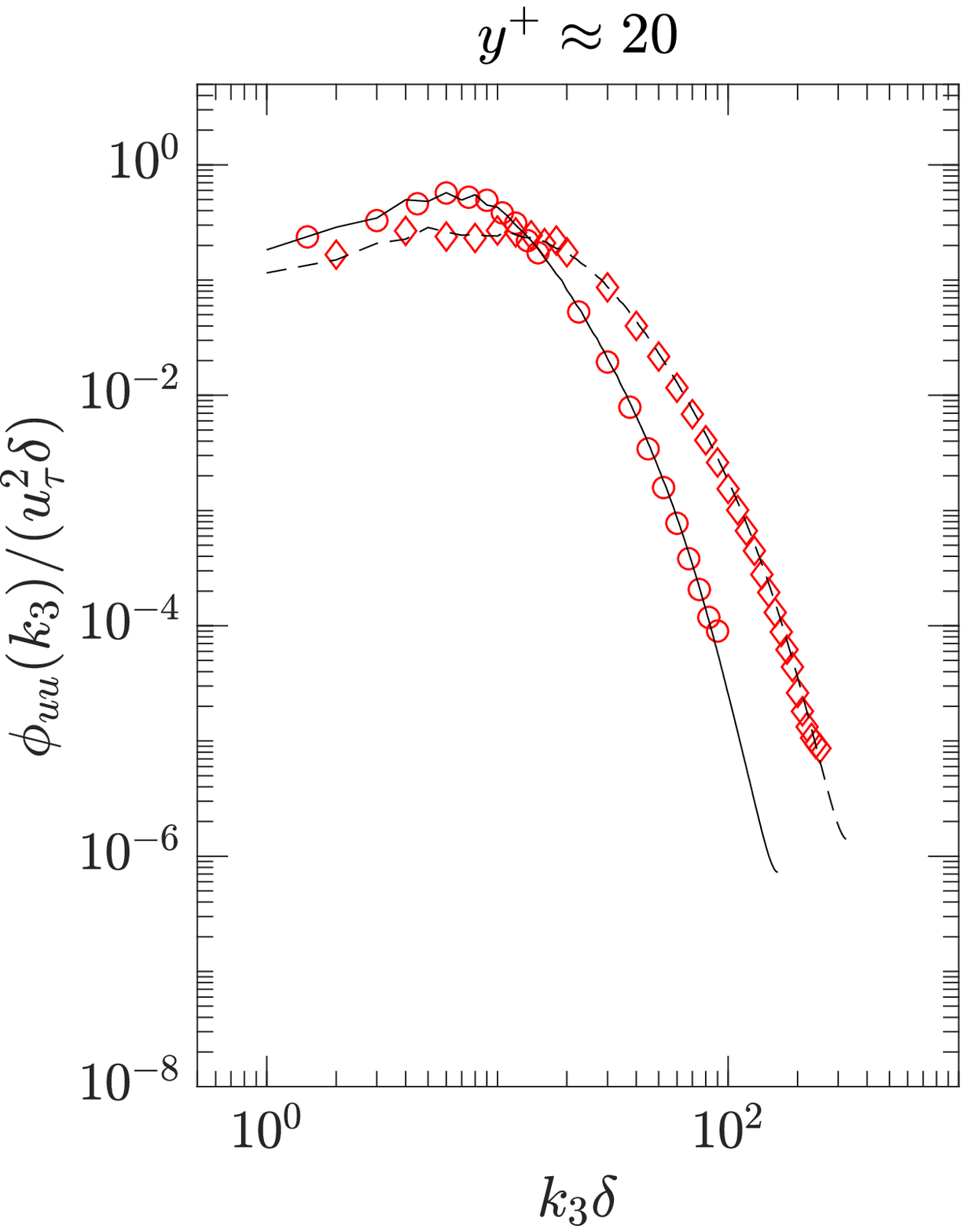}
      \put(1,90){\scriptsize$(e)$}
    \end{overpic}
  \end{subfigure}%
  \begin{subfigure}{.33\textwidth}
    \begin{overpic}[width=\linewidth]{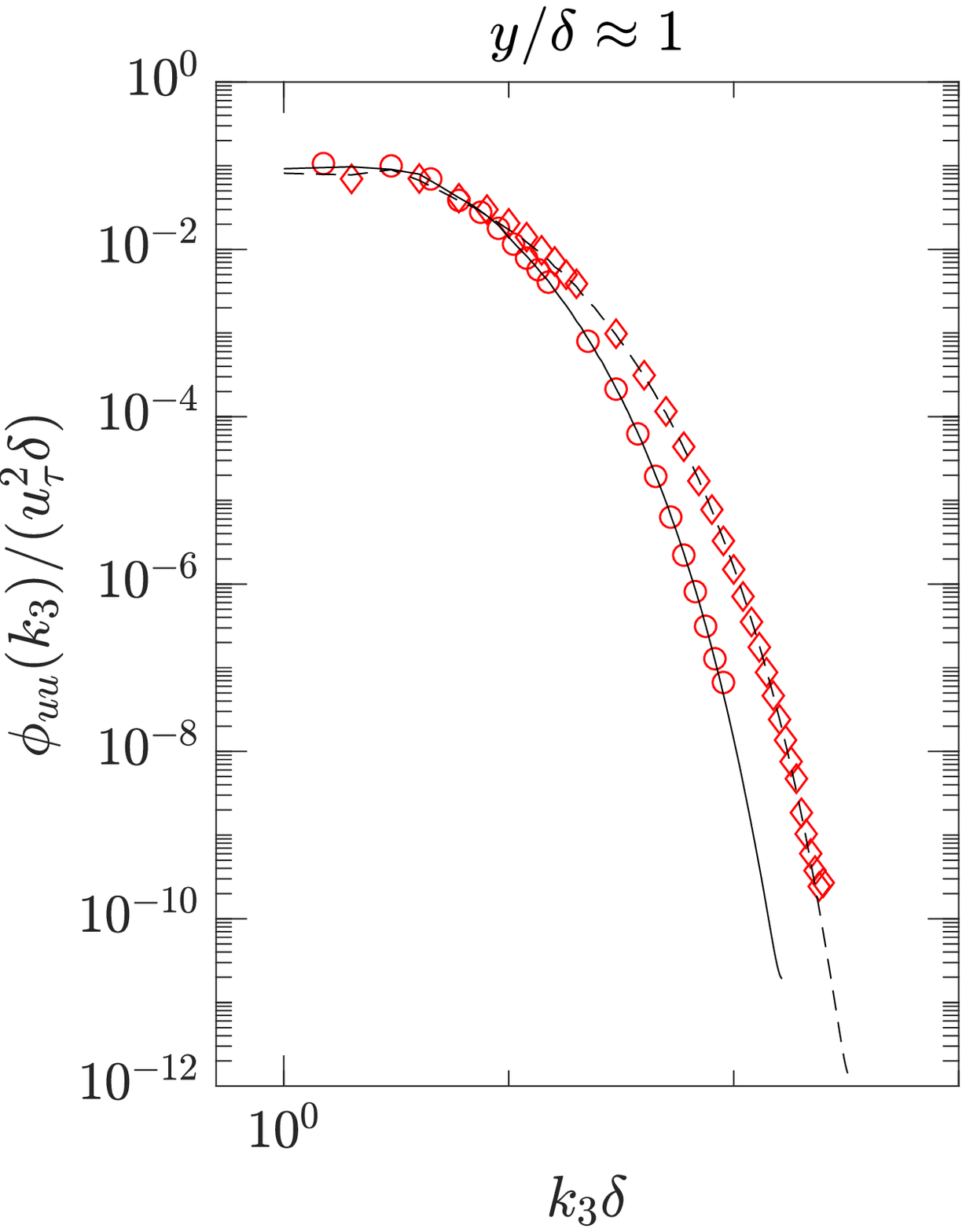}
      \put(1,90){\scriptsize$(f)$}
    \end{overpic}
  \end{subfigure}%

  \caption{Streamwise velocity fluctuation spectra. Figures a-c and d-f are streamwise and spanwise wavenumber spectra, respectively at different wall-normal locations. Solid and dashed lines denote the current DNS result at $Re_{\tau}=180$ and $400$, respectively. Circle and diamond symbols denote DNS data at $Re_{\tau}=182$ and $392$ from \cite{moser1999direct}.}
  \label{fig:spec_compare}
\end{figure}

\begin{figure}
  \centering
  \begin{subfigure}{.33\textwidth}
    \begin{overpic}[width=\linewidth]{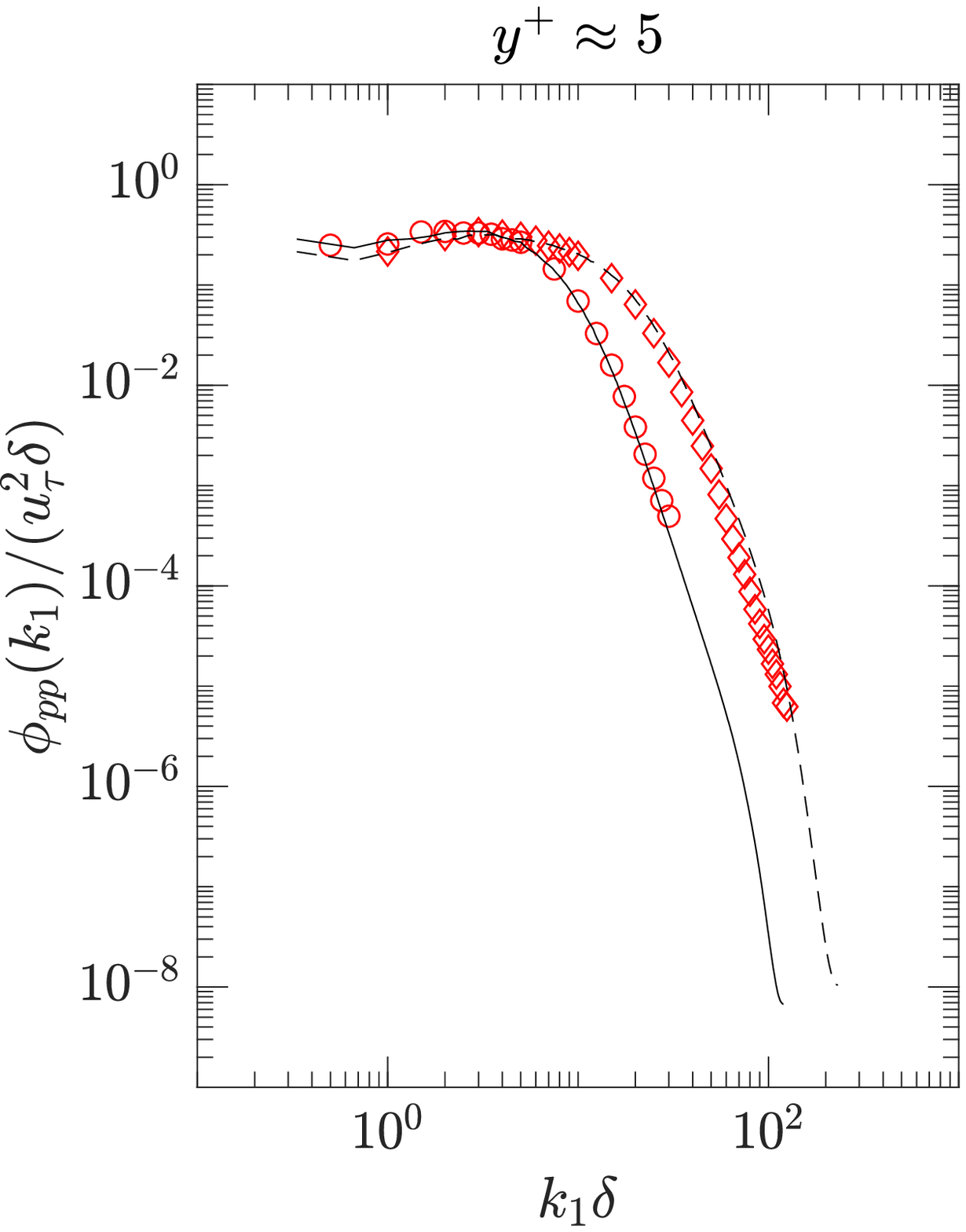}
      \put(1,90){\scriptsize$(a)$}
    \end{overpic}
  \end{subfigure}%
  \begin{subfigure}{.33\textwidth}
    \begin{overpic}[width=\linewidth]{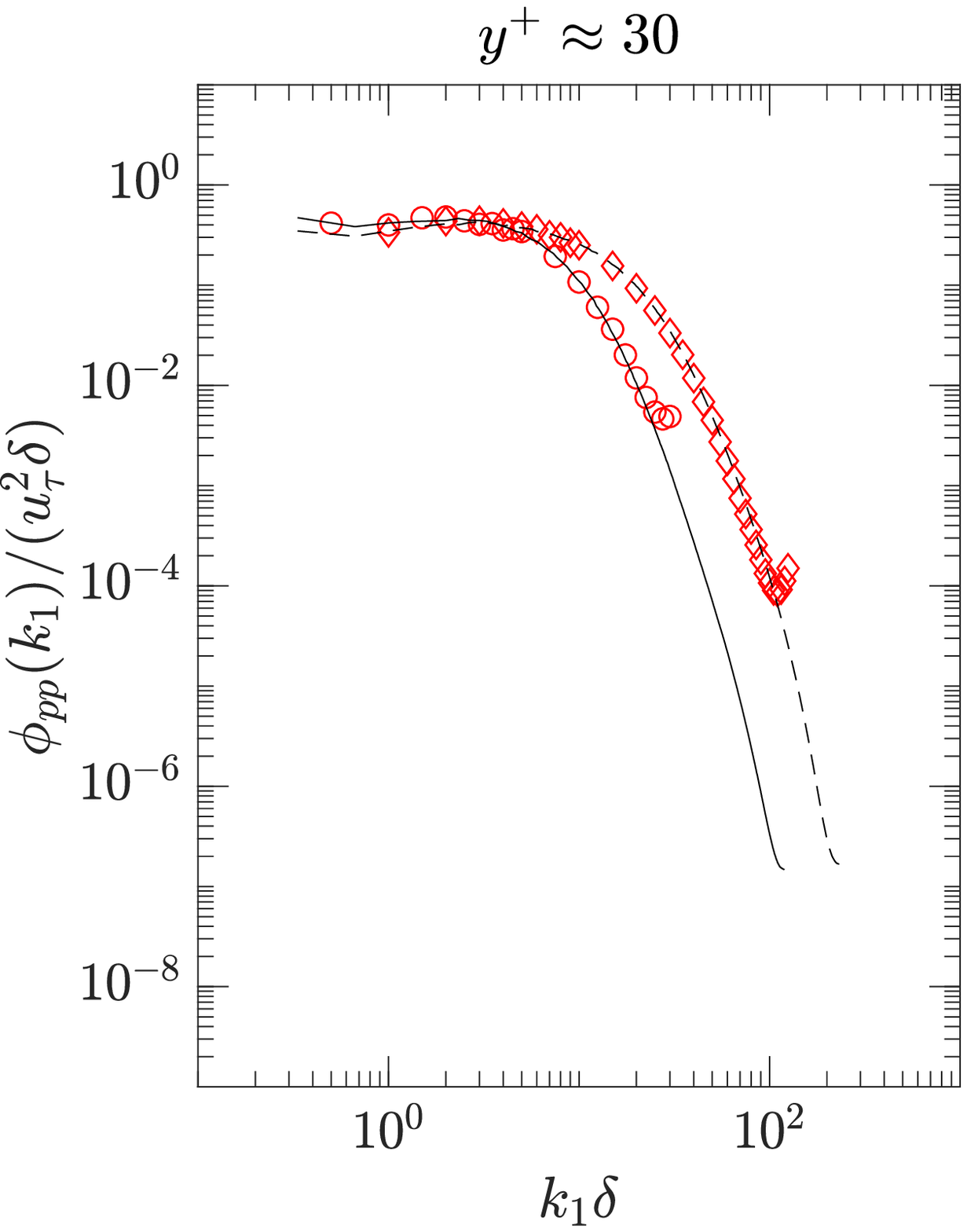}
      \put(1,90){\scriptsize$(b)$}
    \end{overpic}
  \end{subfigure}%
  \begin{subfigure}{.33\textwidth}
    \begin{overpic}[width=\linewidth]{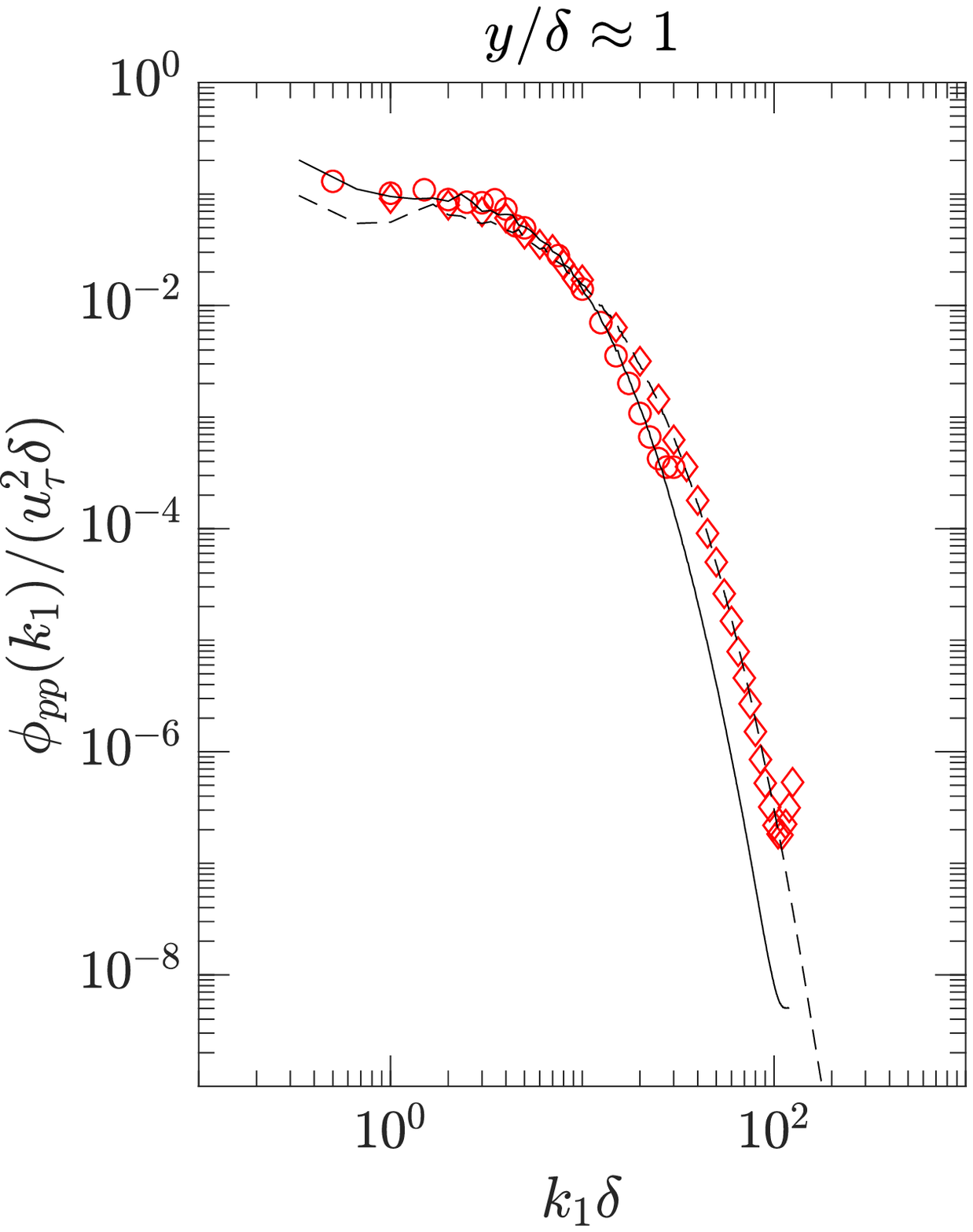}
      \put(1,90){\scriptsize$(c)$}
    \end{overpic}
  \end{subfigure}%

  \begin{subfigure}{.33\textwidth}
    \begin{overpic}[width=\linewidth]{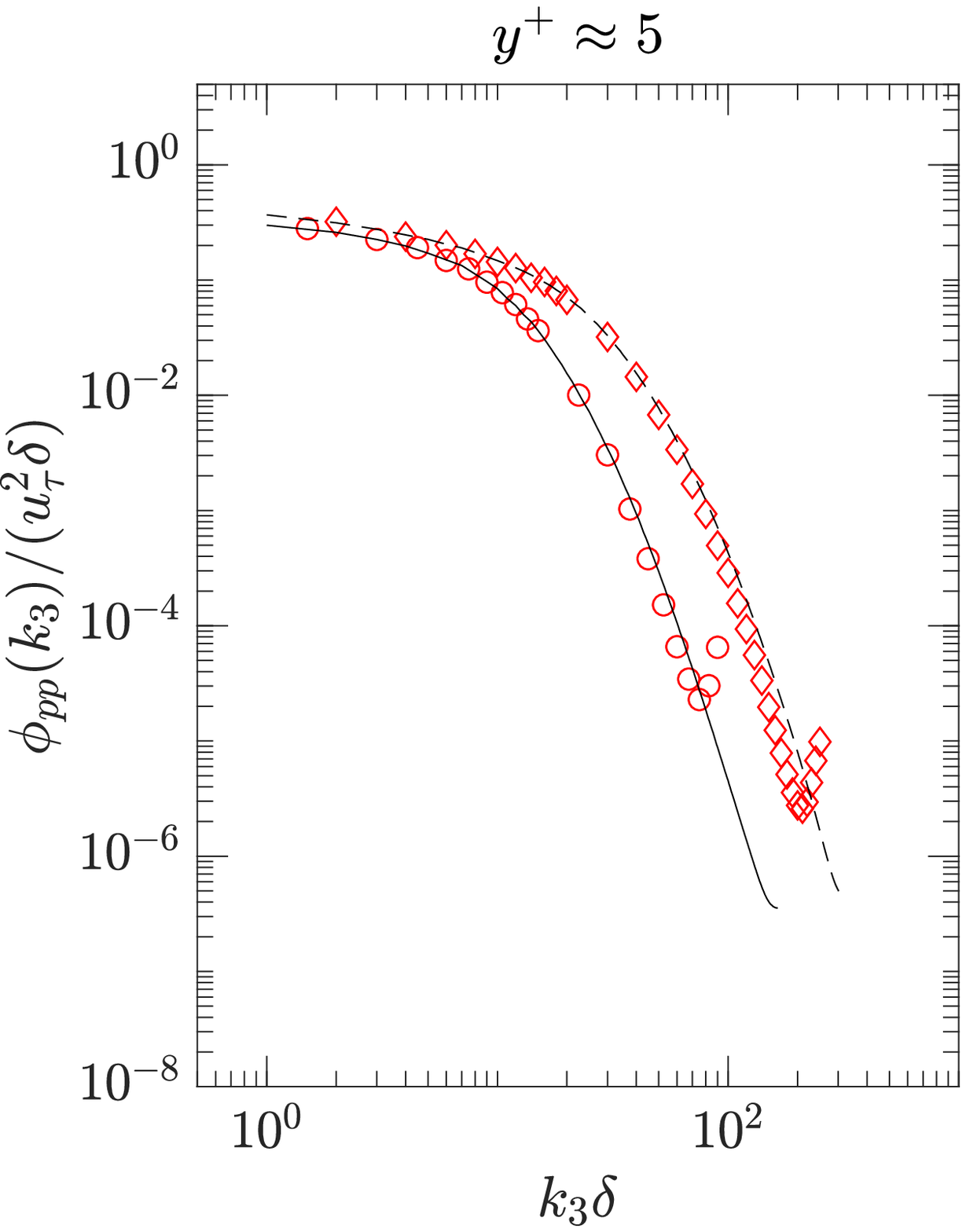}
      \put(1,90){\scriptsize$(d)$}
    \end{overpic}
  \end{subfigure}%
  \begin{subfigure}{.33\textwidth}
    \begin{overpic}[width=\linewidth]{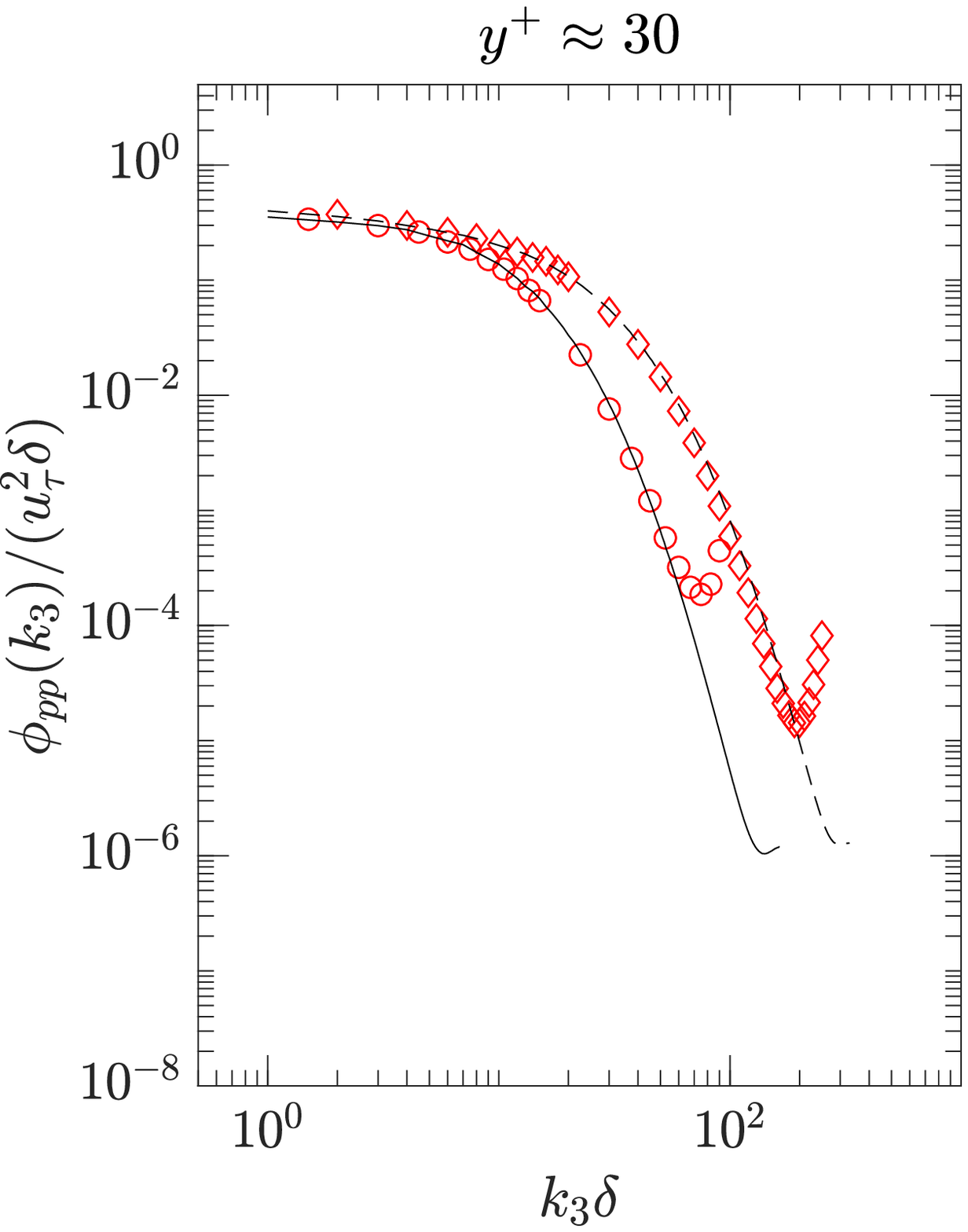}
      \put(1,90){\scriptsize$(e)$}
    \end{overpic}
  \end{subfigure}%
  \begin{subfigure}{.33\textwidth}
    \begin{overpic}[width=\linewidth]{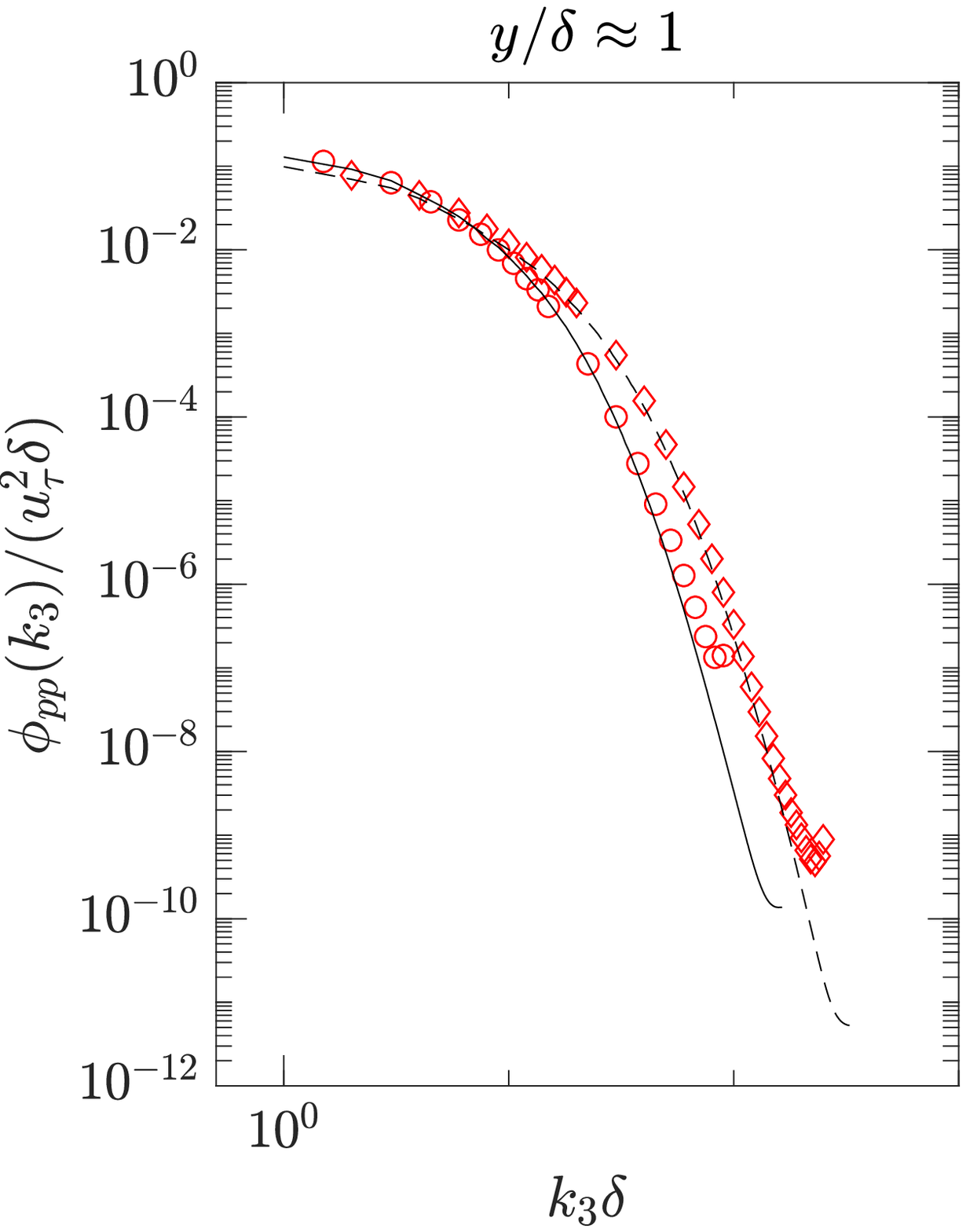}
      \put(1,90){\scriptsize$(f)$}
    \end{overpic}
  \end{subfigure}%

  \caption{Pressure fluctuation spectra. Figures a-c and d-f are streamwise and spanwise wavenumber spectra, respectively at different wall-normal locations. Solid and dashed lines denote the current DNS result at $Re_{\tau}=180$ and $400$, respectively. Circle and diamond symbols denote DNS data at $Re_{\tau}=182$ and $392$ from \cite{moser1999direct}.}
  \label{fig:spec_compare_p}
\end{figure}

\begin{figure}
  \centering
  \begin{subfigure}{.5\textwidth}
    \begin{overpic}[width=\linewidth]{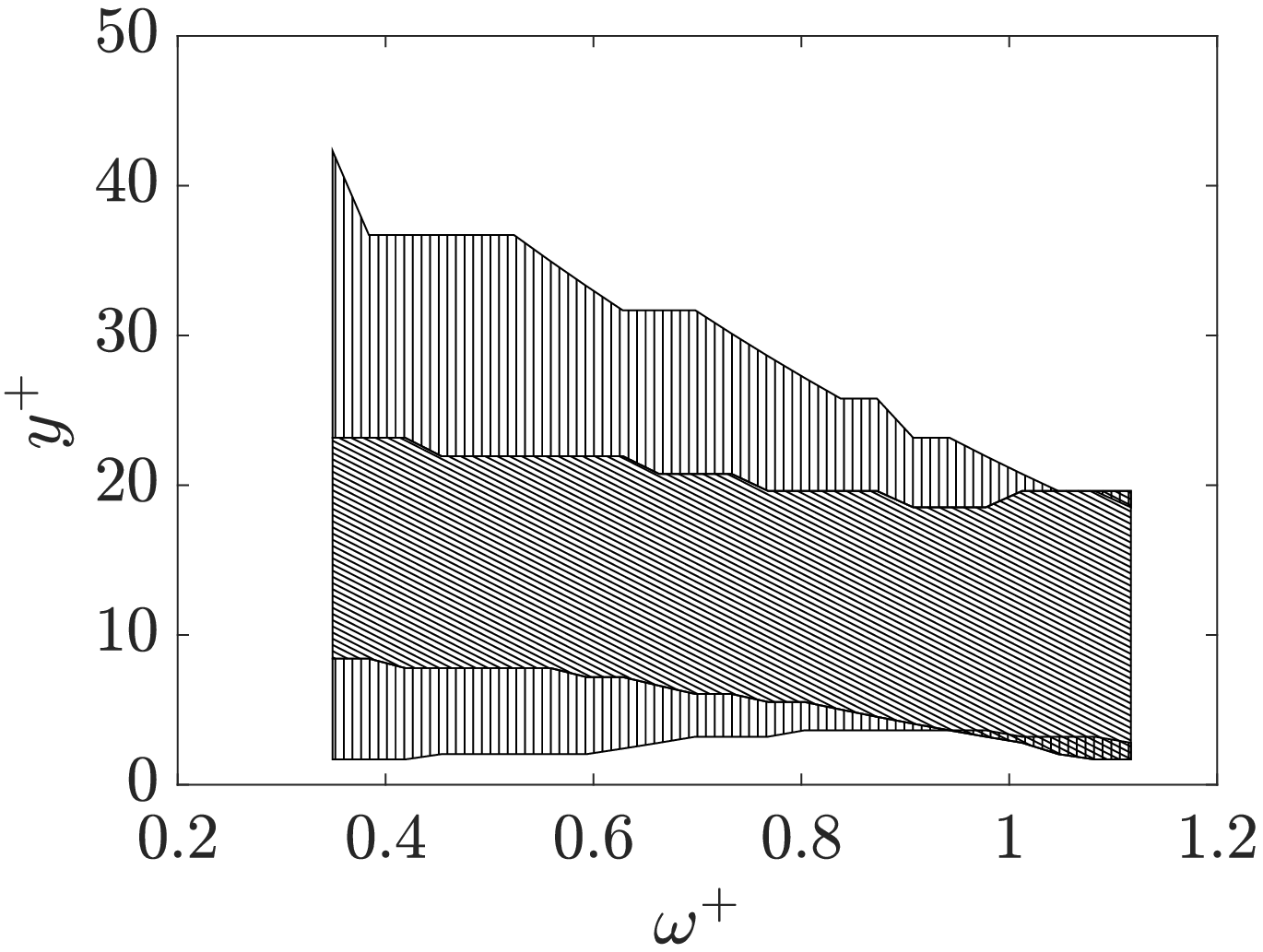}
      \put(1,60){\scriptsize$(a)$}
    \end{overpic}
  \end{subfigure}%
  \begin{subfigure}{.5\textwidth}
    \begin{overpic}[width=\linewidth]{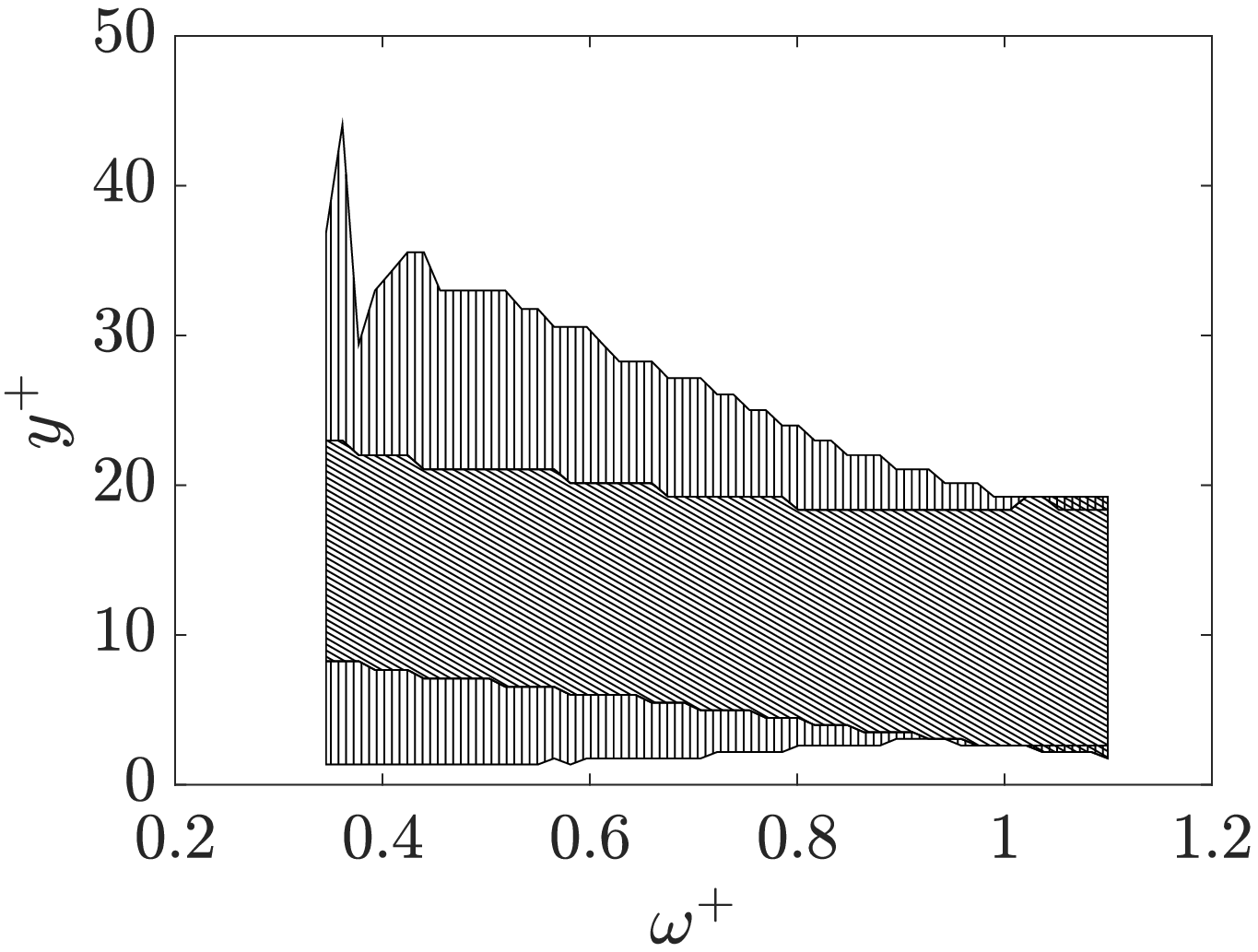}
      \put(1,60){\scriptsize$(b)$}
    \end{overpic}
  \end{subfigure}%
  \caption{Comparison of destructively interfering regions of the
    dominant spectral POD mode computed using the $L^2$ inner product
    as a function of frequency {\color{black}for a) $Re_{\tau}=180$ and
      b) $Re_{\tau}=400$}. In the cross and vertically hatched
    regions,
    $|\angle\Phi_i(y^+,\omega^+)-\angle\Phi_i^n(\omega^+)|<\pi/2$ and
    $\pi/2<|\angle\Phi_i(y^+,\omega^+)-\angle\Phi_i^n(\omega^+)|<\pi$,
    respectively.}

  \label{fig:angdommode_analysis}
\end{figure}

\section{Orthogonality of the linear transformation $C$} \label{app:orthoreln}

We prove the orthogonality relation given by equation \ref{eqn:corthoreln}. Writing the Fourier transform of the net source function as a linear combination of the set of modes $\{\hat{\Phi}_i(y,\omega)\}_{i=1}^{\infty}$ and $\{\tilde{\Phi}_i(y,\omega)\}_{i=1}^{\infty}$, we have, 
\begin{equation} \label{eqn:fhatintwobases}
  \hat{f}_G(x,y,z,\omega)=\sum_j
  \hat{\alpha}_j(x,z,\omega)\hat{\Phi}^*_j(y,\omega)=\sum_j\tilde{\alpha}_j(x,z,\omega)\tilde{\Phi}^*_j(y,\omega),
\end{equation}
where $\hat{\alpha}(\omega)$ and $\tilde{\alpha}(\omega_o)$ are the coefficients of the linear combination. For brevity, we drop the dependence of $\hat{\alpha}$ and $\tilde{\alpha}$ on $x$ and $z$. Using equation \ref{eqn:cdefn} in \ref{eqn:fhatintwobases} and equating the coefficients of $\{\tilde{\Phi}^*_j\}_{j=1}^{\infty}$, we have 
\begin{equation}
  \sum_j\hat{\alpha}_j(\omega)C_{jk}(\omega)=\tilde{\alpha}_k(\omega)
\end{equation}
Correlating the coefficients, we have
\begin{equation}
  \langle\tilde{\alpha}_k(\omega)\tilde{\alpha}^*_l(\omega_o)\rangle=\sum_j\sum_m\langle\hat{\alpha}_j(\omega)\hat{\alpha}_m(\omega_o)\rangle C_{jk}(\omega)C^*_{ml}(\omega_o).
\end{equation}
Since the coefficients are decorrelated, we obtain
\begin{equation}
  \tilde{\lambda}_k(\omega)\delta_{kl}\delta(\omega-\omega_o)=\sum_j\sum_m \hat{\lambda}_j\delta_{jm}\delta(\omega-\omega_o) C_{jk}(\omega)C^*_{ml}(\omega_o).
\end{equation}
Integrating in $\omega_o$ and expression the above relation in matrix form, we have
\begin{equation}
  \tilde{\Lambda}(\omega)=C^H(\omega)\hat{\Lambda}(\omega)C(\omega).
\end{equation}
where $\tilde{\Lambda}(\omega)$ and $\hat{\Lambda}(\omega)$ are the diagonal matrices of the eigenvalues. Since the eigenvalues are non-negative, we decompose $\hat{\Lambda}(\omega)$ as $\hat{\Lambda}^{1/2}(\omega)\hat{\Lambda}^{1/2}(\omega)$, respectively, where $\Lambda^{1/2}(\omega)$ is a diagonal matirx constructed using the set of values $\{\sqrt{\lambda_{i}(\omega)}\}_{i=1}^{\infty}$, and obtain the requiblack result
\begin{equation}
  \left({\hat{\Lambda}^{1/2}(\omega)}C(\omega)\right)^{H}{\hat{\Lambda}^{1/2}(\omega)}C(\omega)=\tilde{\Lambda}(\omega).
\end{equation}

{\color{black}
\section{DNS validation} \label{sec:valid}

We compare the mean, intensities and spectra from the current DNS to the previous reference DNS. We sample the velocity and pressure field every 50 timesteps to compute the statistics presented in this section.

Figure \ref{fig:valid_stat} shows the comparison of velocity field and pressure fluctuation statistics to the previous DNS of \cite{moser1999direct} performed at $Re_{\tau}=182$ and $392$. Figure \ref{fig:valid_stat}a compares mean streamwise velocity. Figures \ref{fig:valid_stat}b, c and d compare mean-squablack streamwise, wall-normal and spanwise velocity fluctuation, respectively. Figures \ref{fig:valid_stat}e and f compare mean tangential Reynolds stress and mean-squablack pressure fluctuation, respectively. We observe good agreement in the compablack quantities.

Figure \ref{fig:spec_compare} compares both streamwise and spanwise wavenumber spectra of the streamwise velocity fluctuations to the previous DNS of \cite{moser1999direct} at different wall-normal locations. We compare the spectra at $y^+\approx 10$ (near the buffer layer peak in the intensity), $y^+\approx 20$ and $y/\delta\approx 1$ (channel centerline). The current spectras agree well both near the wall and at the channel center for the two $Re_{\tau}$. Therefore, the DNS is well-resolved.

In figure \ref{fig:spec_compare_p}, we compare the streamwise and spanwise wavenumber spectra of the pressure fluctuations to \cite{moser1999direct} at $y^+\approx 5$ (near the wall), $y^+\approx 30$ (at the peak intensity location) and $y/\delta \approx 1$ (channel centerline). The spectra show good agreement. Also, we do not observe the spurious pile up of the spectrum levels at very high wavenumbers seen in the results of \cite{moser1999direct}.
 }

\section{Destructive interference of dominant $L^2$ inner product mode contribution to wall-pressure PSD} \label{sec:destint}

We investigate the frequency dependence of the destructive
interference of the obtained dominant spectral POD mode (computed
using the $L^2$ inner product) in figure
\ref{fig:angdommode_analysis}. The envelope and the phase of the
wavepacket are used to identify destructively interfering regions. In
the figure, the vertical and cross-hatched regions of the mode
interfere destructively. In the cross-hatched and vertically hatched
regions, the phase satisfies
$|\angle\Phi_i(y,\omega)-\angle\Phi_i^n(\omega)|<\pi/2$ and
$\pi/2<|\angle\Phi_i(y,\omega)-\angle\Phi_i^n(\omega)|<\pi$,
respectively. With increase in frequency, the ratio of the cross and
vertically hatched region increases. Therefore, the destructive
interference in the contribution from the dominant spectral POD mode
to wall-pressure PSD decreases. Hence, the dominant spectral POD mode
becomes the dominant wall-pressure mode for $\omega^+\geq 1$ (figure
\ref{fig:domspodmodeindex}). {\color{black} For small frequencies, the
  dominant spectral POD mode does not resemble a wall-normal
  wavepacket. Therefore, we would not obtain a continuously
  (continuous in frequency) varying interface between the
  destructively interfering region. Therefore, we do not include the
  frequencies below $\omega^+=0.35$.}}

\bibliographystyle{jfm}
\bibliography{papers}

\begin{thebibliography}{38}
\expandafter\ifx\csname natexlab\endcsname\relax\def\natexlab#1{#1}\fi
\def\au#1{#1} \def\ed#1{#1} \def\yr#1{#1}\def\at#1{#1}\def\jt#1{\textit{#1}}
  \def\bt#1{#1}\def\bvol#1{\textbf{#1}} \def\vol#1{#1} \def\pg#1{#1}
  \def\publ#1{#1}\def\arxiv#1{#1}\def\org#1{#1}\def\st#1{\textit{#1}}

\bibitem[Abe {\em et~al.\/}(2005)Abe, Matsuo \& Kawamura]{abe2005dns}
{\sc \au{Abe, Hiroyuki}, \au{Matsuo, Yuichi} \& \au{Kawamura, Hiroshi}}
  \yr{2005} A dns study of reynolds-number dependence on pressure fluctuations
  in a turbulent channel flow.  \bt{In {\em Fourth International Symposium on
  Turbulence and Shear Flow Phenomena\/}}. Begel House Inc.

\bibitem[Bendat \& Piersol(2011)]{bendat2011random}
{\sc \au{Bendat, J.~S.} \& \au{Piersol, A.~G.}} \yr{2011} {\em Random data:
  analysis and measurement procedures\/}.  \publ{John Wiley \& Sons}.

\bibitem[Bernardini {\em et~al.\/}(2013)Bernardini, Pirozzoli, Quadrio \&
  Orlandi]{bernardini2013turbulent}
{\sc \au{Bernardini, M.}, \au{Pirozzoli, S.}, \au{Quadrio, M.} \& \au{Orlandi,
  P.}} \yr{2013}  \at{Turbulent channel flow simulations in convecting
  reference frames}.  \jt{J.~Comp.\ Phys.}  \bvol{232},  \pg{1--6}.

\bibitem[Blake(1970)]{blake1970turbulent}
{\sc \au{Blake, W.~K.}} \yr{1970}  \at{Turbulent boundary-layer wall-pressure
  fluctuations on smooth and rough walls}.  \jt{J.~Fluid Mech.}  \bvol{44},
  \pg{637--660}.

\bibitem[Blake(2017)]{blake2017mechanics}
{\sc \au{Blake, W.~K.}} \yr{2017} {\em Mechanics of Flow-Induced Sound and
  Vibration, Volume 1 and 2\/}.  \publ{Academic Press}.

\bibitem[Bradshaw(1967)]{bradshaw1967inactive}
{\sc \au{Bradshaw, P}} \yr{1967}  \at{‘inactive’motion and pressure
  fluctuations in turbulent boundary layers}.  \jt{Journal of Fluid Mechanics}
  \bvol{30}~(2),  \pg{241--258}.

\bibitem[Bull(1996)]{bull1996wall}
{\sc \au{Bull, M.~K.}} \yr{1996}  \at{Wall-pressure fluctuations beneath
  turbulent boundary layers: some reflections on forty years of research}.
  \jt{J.~Sound Vib.}  \bvol{190},  \pg{299--315}.

\bibitem[Chang~III {\em et~al.\/}(1999)Chang~III, Piomelli \&
  Blake]{chang1999relationship}
{\sc \au{Chang~III, P.~A.}, \au{Piomelli, U.} \& \au{Blake, W.~K.}} \yr{1999}
  \at{Relationship between wall pressure and velocity-field sources}.
  \jt{Phys. of Fluids}  \bvol{11},  \pg{3434--3448}.

\bibitem[Choi \& Moin(1990)]{choi1990space}
{\sc \au{Choi, H.} \& \au{Moin, P.}} \yr{1990}  \at{On the space-time
  characteristics of wall-pressure fluctuations}.  \jt{Phys.~Fluids}  \bvol{2},
   \pg{1450--1460}.

\bibitem[Corcos(1964)]{corcos1964structure}
{\sc \au{Corcos, G.~M.}} \yr{1964}  \at{The structure of the turbulent pressure
  field in boundary-layer flows}.  \jt{J.~Fluid Mech.}  \bvol{18},
  \pg{353--378}.

\bibitem[De~Graaff \& Eaton(2000)]{de2000reynolds}
{\sc \au{De~Graaff, D.~B.} \& \au{Eaton, J.~K.}} \yr{2000}  \at{Reynolds-number
  scaling of the flat-plate turbulent boundary layer}.  \jt{J.~Fluid Mech.}
  \bvol{422},  \pg{319--346}.

\bibitem[Del~{\'A}lamo \& Jim{\'e}nez(2003)]{del2003spectra}
{\sc \au{Del~{\'A}lamo, J.~C.} \& \au{Jim{\'e}nez, J.}} \yr{2003}  \at{Spectra
  of the very large anisotropic scales in turbulent channels}.
  \jt{Phys.~Fluids}  \bvol{15},  \pg{L41--L44}.

\bibitem[Farabee \& Casarella(1991)]{farabee1991spectral}
{\sc \au{Farabee, T.~M.} \& \au{Casarella, M.~J.}} \yr{1991}  \at{Spectral
  features of wall pressure fluctuations beneath turbulent boundary layers}.
  \jt{Phys.~Fluids}  \bvol{3},  \pg{2410--2420}.

\bibitem[Frigo \& Johnson(2005)]{frigo2005design}
{\sc \au{Frigo, M.} \& \au{Johnson, S.~G.}} \yr{2005}  \at{The design and
  implementation of {FFTW3}}.  \jt{Proc. IEEE}  \bvol{93},  \pg{216--231}.

\bibitem[Gerolymos {\em et~al.\/}(2013)Gerolymos, S{\'e}n{\'e}chal \&
  Vallet]{gerolymos2013wall}
{\sc \au{Gerolymos, GA}, \au{S{\'e}n{\'e}chal, D} \& \au{Vallet, I}} \yr{2013}
  \at{Wall effects on pressure fluctuations in turbulent channel flow}.
  \jt{Journal of Fluid Mechanics}  \bvol{720},  \pg{15--65}.

\bibitem[Gravante {\em et~al.\/}(1998)Gravante, Naguib, Wark \&
  Nagib]{gravante1998characterization}
{\sc \au{Gravante, S.~P.}, \au{Naguib, A.~M.}, \au{Wark, C.~E.} \& \au{Nagib,
  H.~M.}} \yr{1998}  \at{Characterization of the pressure fluctuations under a
  fully developed turbulent boundary layer}.  \jt{AIAA journal}  \bvol{36},
  \pg{1808--1816}.

\bibitem[Hoyas \& Jim{\'e}nez(2006)]{hoyas2006scaling}
{\sc \au{Hoyas, Sergio} \& \au{Jim{\'e}nez, Javier}} \yr{2006}  \at{Scaling of
  the velocity fluctuations in turbulent channels up to re $\tau$= 2003}.
  \jt{Physics of fluids}  \bvol{18}~(1),  \pg{011702}.

\bibitem[Hu {\em et~al.\/}(2006)Hu, Morfey \& Sandham]{hu2006wall}
{\sc \au{Hu, Z.}, \au{Morfey, C.~L.} \& \au{Sandham, N.~D.}} \yr{2006}
  \at{Wall pressure and shear stress spectra from direct numerical simulations
  of channel flow}.  \jt{AIAA Journal}  \bvol{44},  \pg{1541--1549}.

\bibitem[Huang \& Shen(2014)]{huang2014hilbert}
{\sc \au{Huang, N.~E.} \& \au{Shen, S.~P.}} \yr{2014} {\em Hilbert-Huang
  transform and its applications\/}.  \publ{World Scientific}.

\bibitem[Jimenez \& Hoyas(2008)]{jimenez2008turbulent}
{\sc \au{Jimenez, Javier} \& \au{Hoyas, Sergio}} \yr{2008}  \at{Turbulent
  fluctuations above the buffer layer of wall-bounded flows}.  \jt{Journal of
  Fluid Mechanics}  \bvol{611},  \pg{215--236}.

\bibitem[Kim(1989)]{kim1989structure}
{\sc \au{Kim, J.}} \yr{1989}  \at{On the structure of pressure fluctuations in
  simulated turbulent channel flow}.  \jt{J.~Fluid Mech.}  \bvol{205},
  \pg{421--451}.

\bibitem[Kim \& Hussain(1993)]{kim1993propagation}
{\sc \au{Kim, John} \& \au{Hussain, Fazle}} \yr{1993}  \at{Propagation velocity
  of perturbations in turbulent channel flow}.  \jt{Physics of Fluids A: Fluid
  Dynamics}  \bvol{5}~(3),  \pg{695--706}.

\bibitem[Kim {\em et~al.\/}(1987)Kim, Moin \& Moser]{kim1987turbulence}
{\sc \au{Kim, J.}, \au{Moin, P.} \& \au{Moser, R.}} \yr{1987}  \at{Turbulence
  statistics in fully developed channel flow at low {R}eynolds number}.
  \jt{J.~Fluid Mech.}  \bvol{177},  \pg{133--166}.

\bibitem[Klewicki {\em et~al.\/}(2008)Klewicki, Priyadarshana \&
  Metzger]{klewicki2008statistical}
{\sc \au{Klewicki, J.~C.}, \au{Priyadarshana, P. J.~A.} \& \au{Metzger, M.~M.}}
  \yr{2008}  \at{Statistical structure of the fluctuating wall pressure and its
  in-plane gradients at high {R}eynolds number}.  \jt{J.~Fluid Mech.}
  \bvol{609},  \pg{195--220}.

\bibitem[Lumley(2007)]{lumley2007stochastic}
{\sc \au{Lumley, J.~L.}} \yr{2007} {\em Stochastic tools in turbulence\/}.
  \publ{Courier Corporation}.

\bibitem[Mahesh {\em et~al.\/}(2004)Mahesh, Constantinescu \&
  Moin]{mahesh2004numerical}
{\sc \au{Mahesh, K.}, \au{Constantinescu, G.} \& \au{Moin, P.}} \yr{2004}
  \at{A numerical method for large-eddy simulation in complex geometries}.
  \jt{J.~Comp. Phys.}  \bvol{197},  \pg{215--240}.

\bibitem[Morrison(2007)]{morrison2007interaction}
{\sc \au{Morrison, J.~F.}} \yr{2007}  \at{The interaction between inner and
  outer regions of turbulent wall-bounded flow}.  \jt{Phil. Trans. R. Soc.
  Lond.}  \bvol{365},  \pg{683--698}.

\bibitem[Moser {\em et~al.\/}(1999)Moser, Kim \& Mansour]{moser1999direct}
{\sc \au{Moser, Robert~D}, \au{Kim, John} \& \au{Mansour, Nagi~N}} \yr{1999}
  \at{Direct numerical simulation of turbulent channel flow up to
  $\mathrm{Re}_{\tau}$= 590}.  \jt{Physics of fluids}  \bvol{11}~(4),
  \pg{943--945}.

\bibitem[Panton {\em et~al.\/}(2017)Panton, Lee \&
  Moser]{panton2017correlation}
{\sc \au{Panton, R.~L.}, \au{Lee, M.} \& \au{Moser, R.~D.}} \yr{2017}
  \at{Correlation of pressure fluctuations in turbulent wall layers}.
  \jt{Phys. Rev. Fluids}  \bvol{2},  \pg{094604}.

\bibitem[Park \& Moin(2016)]{park2016space}
{\sc \au{Park, G.~I.} \& \au{Moin, P.}} \yr{2016}  \at{Space-time
  characteristics of wall-pressure and wall shear-stress fluctuations in
  wall-modeled large eddy simulation}.  \jt{Phys. Rev. Fluids}  \bvol{1},
  \pg{024404}.

\bibitem[Pope(2001)]{pope2001turbulent}
{\sc \au{Pope, Stephen~B}} \yr{2001} {\em Turbulent flows\/}.  \publ{Cambridge
  University Press}.

\bibitem[Schmidt {\em et~al.\/}(2018)Schmidt, Towne, Rigas, Colonius \&
  Br{\`e}s]{schmidt2018spectral}
{\sc \au{Schmidt, O.~T.}, \au{Towne, A.}, \au{Rigas, G.}, \au{Colonius, T.} \&
  \au{Br{\`e}s, G.~A.}} \yr{2018}  \at{Spectral analysis of jet turbulence}.
  \jt{J.~Fluid Mech.}  \bvol{855},  \pg{953--982}.

\bibitem[Sillero {\em et~al.\/}(2013)Sillero, Jim{\'e}nez \&
  Moser]{sillero2013one}
{\sc \au{Sillero, J.~A.}, \au{Jim{\'e}nez, J.} \& \au{Moser, R.~D.}} \yr{2013}
  \at{One-point statistics for turbulent wall-bounded flows at {R}eynolds
  numbers up to $\delta^+\approx$ 2000}.  \jt{Phys. Fluids}  \bvol{25},
  \pg{105102}.

\bibitem[Towne {\em et~al.\/}(2018)Towne, Schmidt \&
  Colonius]{towne2018spectral}
{\sc \au{Towne, A.}, \au{Schmidt, O.~T.} \& \au{Colonius, T.}} \yr{2018}
  \at{Spectral proper orthogonal decomposition and its relationship to dynamic
  mode decomposition and resolvent analysis}.  \jt{J.~Fluid Mech.}  \bvol{847},
   \pg{821--867}.

\bibitem[Tsuji {\em et~al.\/}(2007)Tsuji, Fransson, Alfredsson \&
  Johansson]{tsuji2007pressure}
{\sc \au{Tsuji, Y.}, \au{Fransson, J. H.~M.}, \au{Alfredsson, P.~H.} \&
  \au{Johansson, A.~V.}} \yr{2007}  \at{Pressure statistics and their scaling
  in high-{R}eynolds-number turbulent boundary layers}.  \jt{J.~Fluid Mech.}
  \bvol{585},  \pg{1--40}.

\bibitem[Tsuji {\em et~al.\/}(2016)Tsuji, Marusic \&
  Johansson]{tsuji2016amplitude}
{\sc \au{Tsuji, Yoshiyuki}, \au{Marusic, Ivan} \& \au{Johansson, Arne~V}}
  \yr{2016}  \at{Amplitude modulation of pressure in turbulent boundary layer}.
   \jt{International Journal of Heat and Fluid Flow}  \bvol{61},  \pg{2--11}.

\bibitem[Willmarth(1975)]{willmarth1975pressure}
{\sc \au{Willmarth, W.~W.}} \yr{1975}  \at{Pressure fluctuations beneath
  turbulent boundary layers}.  \jt{Ann. Rev. Fluid Mech.}  \bvol{7},
  \pg{13--36}.

\bibitem[Willmarth \& Wooldridge(1962)]{willmarth1962measurements}
{\sc \au{Willmarth, W.~W.} \& \au{Wooldridge, C.~E.}} \yr{1962}
  \at{Measurements of the fluctuating pressure at the wall beneath a thick
  turbulent boundary layer}.  \jt{J.~Fluid Mech.}  \bvol{14},  \pg{187--210}.

\end{thebibliography}

\end{document}